\documentclass[twocolumn, usenatbib]{mnras}
\usepackage{savesym}
\usepackage{graphicx}
\usepackage{longtable}
\usepackage{changepage}

\usepackage{subfig}
\usepackage{amsmath}
\usepackage{amssymb}
\usepackage{verbatim}
\usepackage[yyyymmdd,hhmmss]{datetime}
\usepackage{array}
\usepackage{times}
\usepackage{xcolor}
\usepackage{xspace}

\newcommand\rhobar {{\left\langle\rho\right\rangle}}
\newcommand\Xbar {{\left\langle X \right\rangle}}

\newcommand\mumetal{{\citep{Mummery24Plunge}}}

\title[3D disc structure in the plunging region]{ The three-dimensional structure of black hole accretion flows within the plunging region }
\author [Andrew Mummery, James M. Stone]{Andrew Mummery$^1$\thanks{E-mail:
andrew.mummery@physics.ox.ac.uk}, James M. Stone$^2$
\\
$^1$Oxford Theoretical Physics, Beecroft Building,  Clarendon Laboratory, Parks Road, Oxford, OX1 3PU, United Kingdom\\
$^2$School of Natural Sciences, Institute for Advanced Study, 1 Einstein Drive, Princeton, NJ 08540 USA }

\date{}

\begin{document}

\pagerange{\pageref{firstpage}--\pageref{lastpage}} \pubyear{2023}

\maketitle

\label{firstpage}

\begin{abstract} 
We analyse, using new analytical models and numerical general relativistic magnetohydrodynamic simulations, the three-dimensional properties  of accretion flows inside the plunging region of black hole spacetimes (i.e., at radii smaller than the innermost stable circular orbit). These simulations are of thick discs, with aspect ratios of order unity $h/r \sim 1$, and with a magnetic field geometry given by the standard low-magnetization ``SANE'' configuration. This work represents the first step in a wider analysis of this highly relativistic region. We show that analytical expressions derived in the ``thin disc'' limit describe the numerical results remarkably well, despite the large aspect ratio of the flow.  We further demonstrate that accretion within this region is typically mediated via spiral arms, and that the geometric properties of these spiral structures can be understood with a simple analytical model. These results highlight how accretion within the plunging region is fundamentally two dimensional in character, which may have a number of observational implications. We derive a modified theoretical description of the pressure within the plunging region which accounts for turbulent heating and may be of use to black hole image modelling. 
\end{abstract}

\noindent

\section{Introduction}
The accretion of material onto a black hole is a physical process of fundamental interest, as it has the potential to be leveraged as a probe of the physics of the strong-field regime of gravity and the nature of astrophysical compact objects, but also because of the role accretion has in wider astrophysics, such as the effect feedback from active galactic nuclei has on the evolution of their host galaxies.

Despite the long history of the field, and its importance for interpreting a broad range of astrophysical observations and systems, physical theories describing the black hole accretion process are yet to be fully understood in all regimes. In particular, while numerical simulations of black hole accretion naturally evolve the disc fluid throughout the entirety of the black hole's spacetime, analytical theories are typically curtailed at the so-called innermost stable circle orbit \citep[or ISCO, e.g.,][]{SS73, NovikovThorne73, PageThorne74}. The ISCO radius represents a boundary at which the dynamical nature of fluid orbits change, with circular motion thereafter unstable to inwards perturbations which result in a rapidly increasing radial velocity. While this transition undoubtedly modifies the dynamics and thermodynamics of the flow, it is known from numerous simulations that this region can contribute non-negligibly to the total emission sourced from black hole systems \citep[e.g.,][]{Noble10, Penna10, Zhu12, Schnittman16, Lancova19, Wielgus22}, and neglecting this region {\it in its entirety} will likely result in systematic errors in data analysis as we enter the era of high precision observatories. 

In a recent series of papers, \cite{MummeryBalbus2023, MummeryMori24} an analytical theory extending accretion solutions across the ISCO was developed, based on the assumption that the fluid thereafter underwent a gravitationally dominated plunge. This theory was recently shown to reproduce observations of the Galactic X-ray binary MAXI J1820+070 \mumetal, which could not be explained by conventional models which neglect intra-ISCO emission \citep{Fabian20}. While this is promising, detailed tests of this theory are perhaps best performed via comparison with numerical simulations of black hole accretion, where a much finer level of detail can be probed. 

The numerical study of black hole accretion, usually performed via the medium of general relativistic megnetohydrodynamic (GRMHD) simulations \citep[although global particle in cell techniques are being developed, e.g.,][]{Galishnikova23}, has a long history. From the original numerical studies of the magnetorotational instability \citep{BalbusHawley91}, to the analysis of warped discs \citep[e.g.,][]{White19, Liska21},  and the interpretation of event horizon telescope data \citep[e.g.,][]{EHT19, EHT22}, much of what is currently understood about black hole accretion theory has its roots in numerical simulations. 

The conventional approach involves initialising a torus which is in a relativistic hydrostatic equilibrium \citep[such as a][torus]{Fishbone76}, to which a weak (or potentially strong) seed magnetic field is added. These magnetic fields rapidly initiate the magneto-rotational instability \citep[or MRI;][]{BalbusHawley91}, rendering the flow turbulent. The subsequent turbulent evolution of the flow loses, after a sufficient period of time, memory of the initial condition whereafter analytical models of accretion flows may be tested. Numerous  codes for studying such systems have been developed over the last two decades \citep[e.g.,][]{McKinney06, Stone08, White16, Porth17, Liska18, Jiang19, White23}, and have been used to tackle a wide variety of problems in accretion physics. 

In this paper we perform and analyse {\tt AthenaK}  \citep{White23} simulations of black hole accretion flows, with a focus on the behaviour of the disc density of these systems in the plunging region. As a first step, we restrict ourselves to thick discs (aspect ratios of order unity $h/r \sim 1$), as these systems are simplest to simulate with the resolution required to capture the physics of the  magnetorotational instability, while simultaneously being the relevant regime of parameter space for comparison with the discs surrounding the supermassive black holes located in the centre of our own Galaxy (Sagittarius A$^*$) and in the centre of the giant elliptical galaxy M87. Both of these sources have of course been the subject of Event Horizon Telescope (EHT) observations \citep[e.g.,][]{EHT19, EHT22}, and analytical descriptions of the disc behaviour within the plunging region  may well be of use in future studies of these systems.  We focus on the magnetic field geometry of so-called standard and normal (``SANE'') discs, as the analytical models of \cite{MummeryBalbus2023} neglect the contribution of magnetic pressure to the dynamics (and so are not necessarily expected to well describe so-called magnetically arrested ``MAD'' discs). 

The results of this paper consist of two key findings.  We show, perhaps surprisingly, that analytical expressions derived in the ``thin disc'' limit describe the numerical results of these GRMHD simulations remarkably well, despite the large aspect ratio of the flow. In addition we highlight that accretion within the plunging region is principally mediated via over-dense (or under-dense)  spiral arms. This observation is not new, indeed spiral structures have been noticed before in magnetohydrodynamic simulations of accretion \citep[see e.g.,][for spiral features in simulations in a pseudo-Newtonian potential]{Machida03}. We show however that these structures have a simple physical origin, and  result from what we call {\it geodesic shearing}, whereby local turbulent over (or under) densities are rapidly sheared out into spiral arms by the pronounced gravitational acceleration of the intra-ISCO region. The geometric properties of these spirals are simple to describe analytically, and we derive models of spiral {arm} formation which reproduce the numerical simulations in detail. 

Conventional analytical models  of black hole accretion (i.e., those which are typically fit to data) assume azimuthal symmetry, and compute quantities as a function of radius only.  We argue that existing approaches should be extended to include asymmetric spiral features, as found in numerical simulations, as these spirals are a signature of the highly relativistic regime of gravity, and may well carry strong-field signatures which can be leveraged to probe in detail the properties of astrophysical compact objects. 

The layout of the remainder of this paper is as follows. In section \ref{rzsec} we review the analytical models of \cite{MummeryBalbus2023}, before extending these models in section \ref{spiralsec} to include the asymmetric $r-\phi$ spiral structure of the intra-ISCO region. In section \ref{numsec} we present our {\tt AthenaK} simulations, before discussing their observational implications in section \ref{obssec}. In section \ref{pressec} we present a modified theory of the intra-ISCO pressure which includes the effects of turbulent heating. We conclude in section \ref{conc}, with some results presented in Appendices. 

\section{ The vertical and radial structure of the intra-ISCO disc density  }\label{rzsec}
The conventional analytical approach to modelling accretion flows within the ISCO is to assume that the density of the flow within this region is zero, and therefore (by mass conservation) that fluid elements move infinitely quickly across the event horizon once they have crossed the ISCO. 

Of course, while fluid elements do undergo a rapid acceleration upon crossing the ISCO, their ultimate radial motion has finite velocity, and the disc density is non-zero. Recently, \cite{MummeryBalbus2023, MummeryMori24} derived analytical expressions for various thermodynamic quantities, including the disc density, within the plunging region. In this section we briefly recap these derivations, highlighting the physical and mathematical assumptions made.  

We begin by assuming that there exists a radial mass flux which is independent of radius within the ISCO 
\begin{equation}
    \dot M = 2\pi r \int \rho U^r \, {\rm d}z = {\rm constant}. 
\end{equation}
This expression describes the conservation of mass in the limit in which none of the flow variables depends on time or azimuthal angle $\phi$ (this can and will be generalised in later sections). In this expression $\rho$ is the (rest) mass density of the fluid, $r$ and $z$ are radial and vertical coordinates, and $U^r$ is the radial component of the 4-velocity of the flow. 

The second governing equation of our analysis will be the thermodynamic entropy evolution equation.  To derive this expression {one begins} with the stress energy tensor describing ideal GRMHD $T^{\mu\nu}$ (i.e., {a perfect fluid stress tensor in the presence of magnetic fields where there} are no radiative losses, resistive or viscous terms), 
and takes the contraction
\begin{equation}
    U_\nu \nabla_\mu T^{\mu\nu} = 0,
\end{equation}
where $\nabla_\mu$ is a covariant derivative, and finds \citep[see e.g.,][section 4]{MummeryBalbus2023}
\begin{equation}\label{entropy_cons}
    {{\rm d}e\over {\rm d}\tau } - \left(e + P\right) {{\rm d} \over {\rm d} \tau}\ln \rho  = 0, 
\end{equation}
where 
\begin{equation}
    {{\rm d} \over {\rm d}\tau } \equiv U^\mu {\partial \over \partial x^\mu} ,
\end{equation}
{and  $P$ and $e$ are the isotropic pressure and energy density of the flow respectively.}
This Newtonian-looking expression holds in full general relativity. In a non-ideal GRMHD flow, the right hand side of equation (\ref{entropy_cons}) would contain a series of terms describing the physical effects of (e.g.,) radiative losses, viscous heating, etc. If an equation of state (i.e., a constant $\Gamma$) is chosen such that 
\begin{equation}
    e = {P \over \Gamma - 1}, 
\end{equation}
then equation (\ref{entropy_cons}) simplifies to 
\begin{equation}\label{eos}
 K \equiv   P \rho^{-\Gamma} = {\rm constant},
\end{equation}
which is the second of our governing equations. 

We then make three further assumptions, which are traditionally associated with taking the thin disc limit (although we shall show in later sections that they describe thick discs remarkably well). The first is to replace integrals over the vertical extent of the disc with simple multiplications by a ``disc scale height'' $H$, i.e., 
\begin{equation}\label{mass}
    \dot M = 2\pi r \int \rho U^r \, {\rm d}z \approx 2\pi r \rho U^r H = {\rm constant}. 
\end{equation}
This disc scale height is typically (in thin disc theory) assumed to be small, in the sense that the disc aspect ratio satisfies $H/r\ll 1$, and all other quantities are now evaluated in the equatorial plane. 

If taken to be small (relative to the disc radius), the disc's scale height can be computed from the assumption that in the vertical direction the disc remains in hydrostatic equilibrium. \cite{Abramowicz97} derived the solution to these constraints, finding a relationship which holds down to the event horizon, with formal corrections at the level of ${\cal O}(H/r)^2$ 
\begin{equation}\label{height}
H = \sqrt{P r^4 \over \rho (U_\phi^2 + a^2 c^2 (1 - U_0^2))}  .
\end{equation}
In this expression $U_\phi$ and $U_0$ are the angular momentum and energy components of the disc fluids 4-momentum, and $a$ is the Kerr metric angular momentum parameter. 

Our final assumption regards the dynamics of the accretion flow in this region. We will make the assumption that within the ISCO gravity dominates the dynamics, and that fluid elements simply {\it advect} their angular momentum across the event horizon once they have been perturbed off a near circular orbit at the ISCO. Formally this assumption regards the relevant magnitude of gravitational, pressure and magnetic accelerations within the ISCO \citep[see section 3 of][for a more precise description of this assumption]{MummeryBalbus2023}. Under this assumption, we have 
\begin{align}
    U_\phi(r<r_I) &\approx U_\phi(r_I) , \\
    U_0(r<r_I) &\approx U_0(r_I) , 
\end{align}
where $r_I$ is the ISCO radius. This assumption suffices to close our set of equations, as the radial inflow velocity can be derived from the 4-velocity normalisation condition $g_{\mu\nu} U^\mu U^\nu = -c^2$, or explicitly \citep{Cunningham75, MummeryBalbus22PRL}
\begin{equation}\label{flow}
U^r \simeq - c \sqrt{2r_g \over 3 r_I} \left( {r_I \over r} - 1\right)^{3/2}  - u_I ,
\end{equation}
where $u_I > 0$ is a boundary condition, corresponding to the trans-ISCO radial velocity. Other notation is standard $r_g = GM_\bullet/c^2$ where $c$ is the speed of light and $G$ the gravitational constant.

The equations (\ref{eos}), (\ref{mass}), (\ref{height}) and (\ref{flow}) now form a closed set, and can be solved for the density and scale height of the disc. The final steps are spelled out in \cite{MummeryBalbus2023}, and the resulting expressions are 
\begin{align}
\left({\rho \over \rho_I}\right)^{1 + \Gamma} &= \left({r_I \over r} \right)^{6} \left[ \varepsilon^{-1} \left({r_I \over r} - 1\right)^{3/2} + 1\right]^{-2} , \label{rhrh}\\
\left({H \over H_I}\right)^{1 + \Gamma} &= \left({r_I \over r} \right)^{\Gamma - 5} \left[ \varepsilon^{-1} \left({r_I \over r} - 1\right)^{3/2} + 1\right]^{1 - \Gamma} \label{hh}, 
\end{align}
where $\varepsilon$ has the definition 
 \begin{equation}
 \varepsilon \equiv {u_I \over c} \sqrt{3 r_I \over 2 r_g} \ll 1 ,
 \end{equation}
and each variable with subscript $I$ corresponds to the value of that variable at the ISCO. 

Before we move on to a description of the azimuthal structure of the intra-ISCO disc density, it is important to recap the approximations made in this section, and examine some of the implications of these results. 

Firstly, we reiterate the ``thin disc'' assumptions made in this derivation. We have assumed that integration over the disc's vertical extent can be well approximated by a multiplication by a ``disc height'', and that this disc height can be well approximated by the leading order solution of vertical hydrostatic equilibrium. In addition, we assume that gravity dominates the dynamics of the intra-ISCO fluid -- an assumption which will be accurate when pressure gradients and magnetic fields are unable to hold up the rapid intra-ISCO gravitational acceleration. While  these expansions are formally invalid in thicker discs (particularly the equation of vertical hydrostatic equilibrium, which has corrections at the level of ${\cal O}[H^2/r^2]$), it is not necessarily obvious that these approximations will be {\it practically} invalid even in a thicker $H/r \sim 1$ disc, owing to the sheer magnitude of gravitational accelerations on near-horizon scales. The validity, or lack thereof, of these expressions can only be tested via numerical simulations. 

Secondly, we highlight some of the most novel implications of these expressions. The scale height $H$ of the disc drops substantially over the intra-ISCO region,  a result of the rapidly increasing vertical acceleration due to gravity of the near-horizon neighbourhood (which scales like $g_z \sim -GMz/r^3$ in a Newtonian theory), and the decrease in supporting pressure driven by adiabatic radial expansion. This prevents the disc density from dropping too rapidly over the intra-ISCO region, despite the rapid radial acceleration (and the constraint of mass conservation, eq. \ref{mass}). In fact, the density is only predicted to drop over the first half of the intra-ISCO inspiral, as can be verified by differentiating equation (\ref{rhrh}) and finding the location of the density minimum
\begin{equation}
     {\partial \rho \over \partial r} = 0 \Rightarrow - \left({r_I \over r} - 1\right) +{1\over 2} {r_I \over  r} \simeq 0 \Rightarrow r \simeq {r_I \over 2} ,
\end{equation}
where we have assumed that $\varepsilon \ll 1$. This is an important result, as it highlights how the zero intra-ISCO density assumption is likely a gross over-simplification. 

In addition, unlike conventional theories valid in the main body of the disc, there is no strong requirement that the disc fluid has settled into some sort of a global steady-state. The one steady-state requirement is that of a constant radial mass flux, which only need be valid over a small radial interval. Outside of this, the intra-ISCO solutions do not ``know'' anything about the global structure of the disc; each thermodynamic quantity is simply scaled by a normalising ISCO value set by the larger scale disc physics. This is an important result to test numerically, as these expressions should be accurate at both early and late times in a numerical simulation. 

Before we move onto numerical GRMHD simulations of black hole accretion flows, we derive a natural extension of the results presented here -- a description of the azimuthal structure of accretion flows within the ISCO.  

\section{ Fluid spirals: The azimuthal structure of the intra-ISCO disc density  }\label{spiralsec}
In the following section we derive properties describing the azimuthal structure of the intra-ISCO disc fluid. In particular, we demonstrate how turbulently induced over-densities in an accretion flow will be sheared out into spiral structures with a well-defined shape. This description is fundamentally time and azimuthally dependent, and therefore must be described with a coordinate system which is well behaved near to the horizon of black holes. We therefore begin by introducing a convenient coordinate system for studying fluid flows near to the event horizon of black holes (this also represents the coordinate system used in most modern GRMHD codes, such as {\tt AthenaK}). After introducing this Eddington-Finkelstein coordinate system, we derive both a simple test particle and more general fluid description of the $r-\phi$ structure of intra-ISCO accretion flows. 

\subsection{ The Schwarzschild metric in Eddington-Finkelstein coordinates  }
The metric describing a non-spinning black hole  in the usual Schwarzschild coordinates $(t, r, \theta, \phi)$ is given by 
\begin{multline}
     g_{\mu\nu}{\rm d}x^\mu {\rm d}x^\nu =  -\left( 1 - {2 / r}\right) {\rm d}t^2 + {{\rm d}r^2 \over \left(1 - {2 / r}\right)}   \\ + r^2\left({\rm d}\theta^2 + \sin^2\theta \, {\rm d}\phi^2\right) . 
\end{multline}
In this and all future sections we shall work in geometric units where both $GM_\bullet = 1 = c$, where $G$ is Newton's gravitational constant, $c$ is the speed of light and $M_\bullet$ is the black hole's mass. In this coordinate system both $r$ and $t$ are dimensionless (to return to physical units multiply $r$ by $GM_\bullet/c^2$ and $t$ by $GM_\bullet/c^3$). The event horizon of the Schwarzschild metric is at $r = r_+ = 2$, where it is clear that this coordinate system is poorly behaved. 

A more convenient coordinate choice for studying physics close to the event horizon ($r_+ = 2$) of the Schwarzschild black hole is the so-called Eddington-Finkelstein coordinate system.  The radial and angular Eddington-Finkelstein coordinates are identical to the more familiar Schwarzschild coordinates, while a new time coordinate $T$ is introduced, which is related to the Schwarzschild $t$ and $r$ coordinates by \citep[e.g.,][]{MTW}
\begin{equation}
    {\rm d}T = {\rm d}t + \left({2 \over r - 2} \right){\rm d}r . 
\end{equation} 
When written in these {ingoing} Eddington-Finkelstein coordinates\footnote{Note that some references define {``Eddington-Finkelstein''} coordinates differently, and use a time coordinate $v$ defined by ${\rm d}v = {\rm d}T + {\rm d}r$. The coordinate {systems used in this paper are the same as those used in {\tt AthenaK} GRMHD simulations; see Appendix \ref{KerrApp} for more details.} } $(T, r, \theta, \phi)$, the line element is given by 
\begin{multline}
    g_{\mu\nu}{\rm d}x^\mu {\rm d}x^\nu =  -\left( 1 - {2\over r}\right) {\rm d}T^2 + \left({4\over r}\right) {\rm d}T {\rm d}r \\ + \left(1 + {2\over r}\right){\rm d}r^2  + r^2\left({\rm d}\theta^2 + \sin^2\theta \, {\rm d}\phi^2\right) . 
\end{multline}
{This coordinate system corresponds to the $a\to0$ limit of the spherical Kerr-Schild coordinate system used in many modern GRMHD codes.} This mixing of Schwarzschild radial and temporal coordinates prevents the $T$ coordinate diverging as a particle crosses the event horizon $r_+ = 2$, an important property for studying horizon crossing fluid elements. As motion in the Schwarzschild spacetime is confined to a plane, we shall henceforth take $\theta = \pi/2$ without loss of generality.  Close to the equatorial plane we shall occasionally use the vertical coordinate $z$, defined by $z = r \cos\theta$. 

\subsection{ Test particle circular velocity profiles }
Fluid motion in an accretion disc about a black hole typically has a high angular momentum content, and is to leading order well-described by standard test-particle circular motion. Upon crossing the innermost stable circular orbit (ISCO), the point at which these orbits become unstable to inward perturbations, test particles transition onto a plunging orbit. 

The circular 4-velocity components of particles in the Eddington-Finkelstein coordinate system are described by the solutions of the following coupled equations 
\begin{align}
    &g_{\mu\nu} U^\mu U^\nu  = -1 \\ 
    &g'_{\mu\nu} U^\mu U^\nu  = 0, 
\end{align}
where a prime denotes a radial derivative, and the only non-zero velocity components are $U^\phi$ and $U^T$. As the Eddington-Finkelstein coordinate transformation only mixes up the $r$ and $T$ coordinates, these orbital profiles are unchanged from the usual values in the  Schwarzschild coordinate system 
\begin{align}
    U^T_g &= {1\over \sqrt{1 - 3/r}} \\ 
    U^\phi_g &= {r^{-3/2} \over \sqrt{1 - 3/r}}, \\
    \Omega_g &\equiv {U^\phi_g \over U^T_g} = r^{-3/2} .
\end{align}
We use the notation $X_g$ to denote that the quantity $X$ is evaluated for a geodesic orbit with constants of motion which describe a circular orbit.   The innermost stable circular orbit (ISCO) is at radial coordinate $r_I = 6$, and the conserved quantities $U_T$ and $U_\phi$ take the following values at this location 
\begin{equation}
    -U_{T, g}(r_I) \equiv {\cal E}_I = {2\sqrt{2}/3}, \quad U_{\phi, g}(r_I) \equiv {\cal J}_I = \sqrt{12}. 
\end{equation}
A particle perturbed infinitesimally inwards from the ISCO radius will spiral into the black hole while conserving its constants of motion ${\cal E}_I$ and ${\cal J}_I$. These conservation constraints lead to a radial velocity which grows as 
\begin{equation}
    U^r_g = - {1\over 3} \left({6 \over r} - 1\right)^{3/2} ,
\end{equation}
a result identical to that in the usual Schwarzschild coordinates \citep{MummeryBalbus22PRL}. As this particle spirals in towards the singularity it has an evolving $U^\phi$ given by 
\begin{equation}
    U^\phi_g = g^{\phi\phi} {\cal J}_I = {\sqrt{12} \over r^2}.
\end{equation}
The time component of the 4-velocity $U^T$ can then be calculated from $U_T = g_{T\mu} U^\mu$, and is equal to 
\begin{equation}\label{ut}
    U^T = {-U_T + 2U^r/r \over 1-2/r} = {2\sqrt{2}r  - 2\left({6 / r} - 1\right)^{3/2}  \over 3(r-2)}
\end{equation} 
Note that this converges to a finite value in the limit of the particle radius approaching the event horizon $r\to r_+ = 2$, and therefore fluid elements can cross the event horizon with no divergent four-velocity components (unlike in the standard Schwarzschild coordinate system). 

The key property of the intra-ISCO inspiral relevant for our calculations is the rapid radial acceleration of test particles. This property, we shall show, results in the pronounced shearing-out of initially localised perturbations in the disc flow into macroscopic spiral arms. 

We will ultimately show that this shearing-out is the leading order solution of the fluid equation describing mass conservation, but first we begin with an intuitive test particle picture, which highlights the key physics with a simpler calculation. 

\subsection{ An intuitive test-particle picture of spiral {arm} formation}
In this section we show how a series of test particles, released one after another from a rotating source of particles located at the ISCO, eventually trace out a spiral arm with properties which are simple to understand mathematically. A similar simple picture was discussed by \cite{Miller98} in the context of high frequency QPOs in Neutron star systems.

Given a radius-dependent four-velocity profile (like those derived above), it is simple to evolve a particle forward on an inspiral from the ISCO. One simply solves the integrals 
\begin{equation}\label{rinspiral}
    r(T) = r_I + \int_0^T {U^r\left(r(T')\right) \over U^T\left(r(T')\right)} \, {\rm d} T' , 
\end{equation}
and 
\begin{equation}\label{phinspiral}
    \phi(T) = \phi_0 + \int_0^T {U^\phi\left(r(T')\right) \over U^T\left(r(T')\right)} \, {\rm d} T' . 
\end{equation}
This individual trajectory will trace out a spiral structure described by a curve with cylindrical coordinates by $(r, \hat \phi(r))$, where 
\begin{equation}
    \hat \phi(r) = \phi_0 + \int_{r_I}^r {U^\phi\left(r'\right) \over U^r\left(r'\right)} \, {\rm d} r' . 
\end{equation}
If we take the pure geodesic values for $U^\phi$ and $U^r$ described above, then this spiral takes the form presented in \cite{MummeryBalbus22PRL}. 

The analytical forms for the 4-velocities described above  assume that a particle starts at rest and then is perturbed infinitesimally across the ISCO. In an evolving fluid this is a poor approximation, as an individual fluid element will cross the ISCO with finite radial velocity $|U^r(r_I)| \equiv u_I$. A fluid element with a finite radial velocity at the ISCO must be rotating at some sub-Keplerian rate, by which we mean that its angular velocity is given by 
\begin{equation}
    \Omega = f_K \Omega_g , \quad 0 \leq f_K \leq 1.
\end{equation}
This expression defines the ``Keplarity'' factor $f_K$ which shall be the one free parameter of our fluid spirals. For simplicity we shall assume that the reduction in the rotational frequency occurs because the $U^\phi$ component is reduced, i.e., $U^\phi = f_K U^\phi_g$ and the time component is unchanged $U^T = U^T_g$.  For a given choice of $f_K$ we can compute the trans-ISCO velocity from $g_{\mu\nu} U^\mu U^\nu = -1$, namely 
\begin{equation}
    {4\over 3} u_I^2 + {2\sqrt{2}\over 3} u_I - {1\over 3} + {f_K^2 \over 3} = 0 ,
\end{equation}
where we have substituted the ISCO values of the various parameters $r = r_I = 6$, $U^T_g(r_I) = \sqrt{2}$, and $U^\phi_g = 1/6\sqrt{3}$. Note that with $f_K=1$ we recover $u_I=0$, as required.  With $u_I$ determined from the choice of $f_K$, we approximate the particles radial 4-velocity by 
\begin{equation}\label{ur}
    U^r \simeq U^r_g - u_I = -{1\over 3} \left({6 \over r} - 1\right)^{3/2} - u_I, 
\end{equation}
and the angular 4-velocity component by 
\begin{equation}\label{uphi}
    U^\phi \simeq  f_K U^\phi_g = f_K {\sqrt{12} \over r^2}.  
\end{equation}

\begin{figure}
    \centering
    \includegraphics[width=1\linewidth]{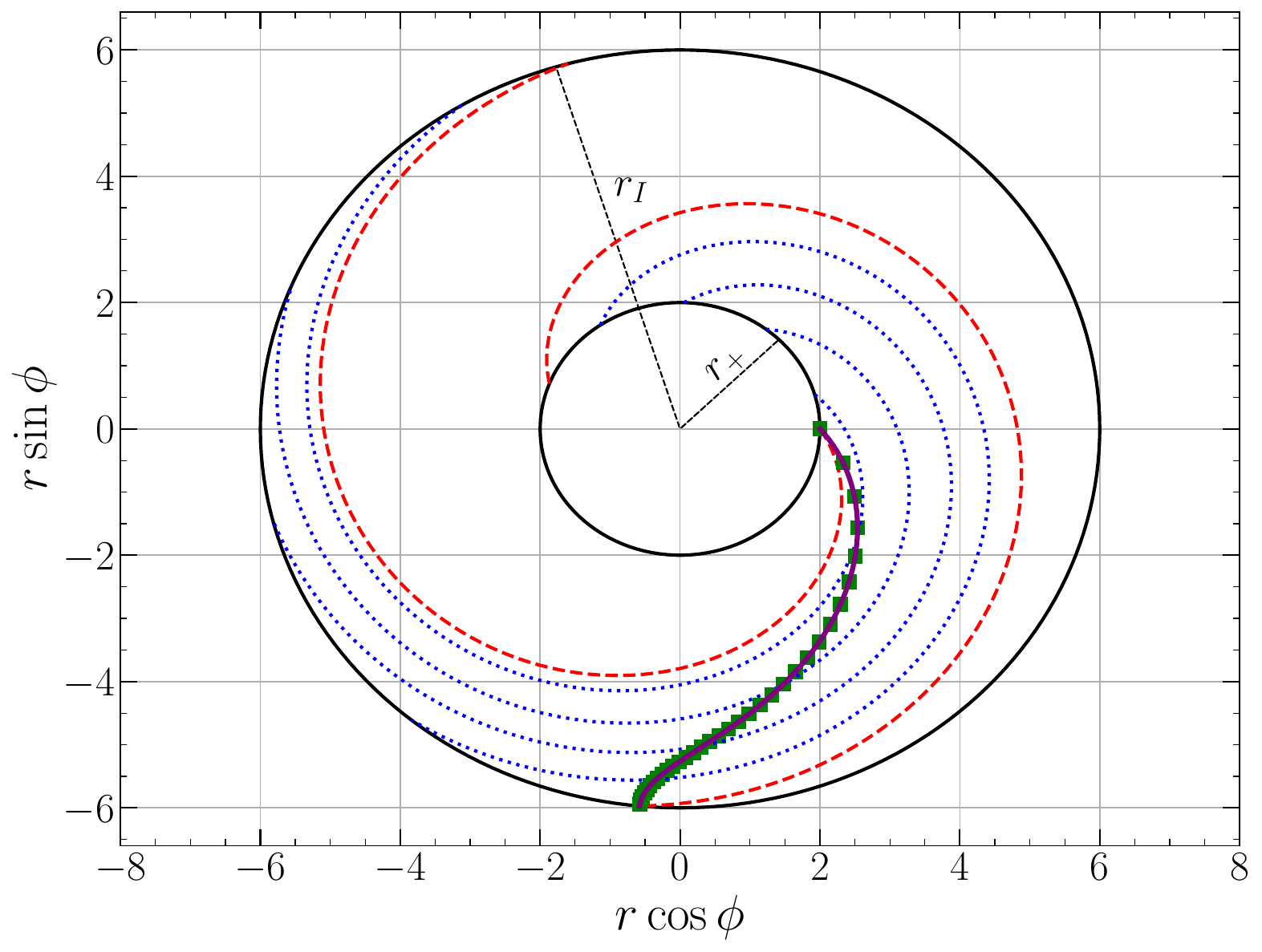}
    \caption{The formation of a spiral {arm} from a series of test particle inspirals. {A sequence of particles (denoted by green squares) are sent over the ISCO with a radial velocity $u_I$, one after another, from a source of particles which is rotating around the ISCO with angular velocity $\Omega_I = f_K \Omega_g(r_I)$ (in this figure $f_K = 0.9$). The trajectory of the particle which is just crossing the event horizon $r_+ = 2$ is shown by the (upper) red dashed curve. If a new particle is sent over the ISCO every $\Delta T$, then each particle will subsequently trace out an identical spiral (some examples are shown by blue dotted curves), but each spiral will have  an initial angular offset of $\Delta \phi = \Omega_I \Delta T$. As the angular velocity of each particle increases as $1/r^2$ during its infall, subsequent particles will appear to be lagging (azimuthally) their counterparts which crossed the ISCO earlier. After a time $\Delta T_{\rm fill}$ when the first particle crosses the event horizon, this differential rotation will result in the apparent formation of a spiral arm stretching from the event horizon out to the ISCO (displayed by a purple solid curve).  Thereafter, this spiral {arm} will maintain its shape, and will simply rotate around the black hole with angular velocity set by the ISCO rotation rate $f_K \Omega_g$. }} 
    \label{fig:spiral-demo}
\end{figure}

Using equations \ref{ut}, \ref{rinspiral}, \ref{phinspiral}, \ref{ur} \& \ref{uphi} we can trivially evolve a test particle from the ISCO to horizon for a given $f_K$. An example inspiral is shown by the red dashed curve in Figure \ref{fig:spiral-demo}. 

Consider now, rather than a single particle,  a series of particles which are sent over the ISCO with a radial velocity $u_I$, one after another, from a source of particles which is rotating around the ISCO with angular velocity $\Omega_I = f_K \Omega_g(r_I)$. If a new particle is sent over the ISCO every $\Delta T$, then each particle will subsequently trace out an identical spiral, but each spiral will have  an initial angular offset of $\Delta \phi = \Omega_I \Delta T$. As the angular velocity of each particle increases as $1/r^2$ during its infall, subsequent particles will appear to be lagging (azimuthally) their counterparts which crossed the ISCO earlier. After a time $\Delta T_{\rm fill}$ when the first particle crosses the event horizon, this differential rotation will result in the apparent formation of a spiral {arm} stretching from the event horizon out to the ISCO\footnote{In the language of fluid dynamics the quantity we are computing is the so-called streakline of the flow in the frame rotating with the fluid at the ISCO}.  Thereafter, this spiral {arm} will maintain its shape, and will simply rotate around the black hole with angular velocity set by the ISCO rotation rate $f_K \Omega_g$. 

An example of this spiral {arm} formation is shown in Figure \ref{fig:spiral-demo}. The outer source of particles is rotating with a Keplarity factor $f_K = 0.9$, and therefore each particle crosses the ISCO with radial velocity $u_I \simeq 0.06$. The trajectory of the innermost particle is shown by the (upper) red dashed curve, and it is just crossing the event horizon $r_+ = 2$ at the moment of plotting. Particles which crossed the ISCO at later times are shown by green squares, and are each following identical trajectories (some examples are shown by blue dotted curves).  The particle which is just crossing the ISCO will go on to follow the lower red dashed curve.  The locus of all the points undergoing these inspirals is displayed by the purple solid curve, and is the spiral {arm} to which we referred earlier. 

Mathematically this test-particle spiral {arm} is described by the curve $(\bar r, \bar \phi)$, where $\bar r = r(T)$ is given by equation \ref{rinspiral}, and 
\begin{equation}
\bar \phi(\bar r) = \phi(T) - f_K\Omega_g T,    
\end{equation} 
where $\phi(T)$ is the solution of the phi inspiral integral (equation \ref{phinspiral}). 

While this discussion involves a much simplified test-particle picture, the point is that if turbulent over densities near the ISCO have a relatively (azimuthally) narrow transition region across the ISCO, after which the fluid elements follow a near-geodesic plunge, then spiral arms should be ubiquitous in the inner regions of black hole accretion systems. Intuitively this seems like a reasonable description of fluid motion close to black holes, as turbulent eddies will likely be limited in size by the disc scale height $\sim H$ (and so will not be too azimuthally extended), and the dominant accelerations in these regions will be gravitational (unless magnetic fields reach extreme levels, perhaps in a so-called magnetically arrested disc state). 

This test-particle picture can be generalised to a more rigorous fluid dynamic model, as we now demonstrate. 

\subsection{ Mass conservation and characteristic curves }
\begin{figure}
    \centering
    \includegraphics[width=\linewidth]{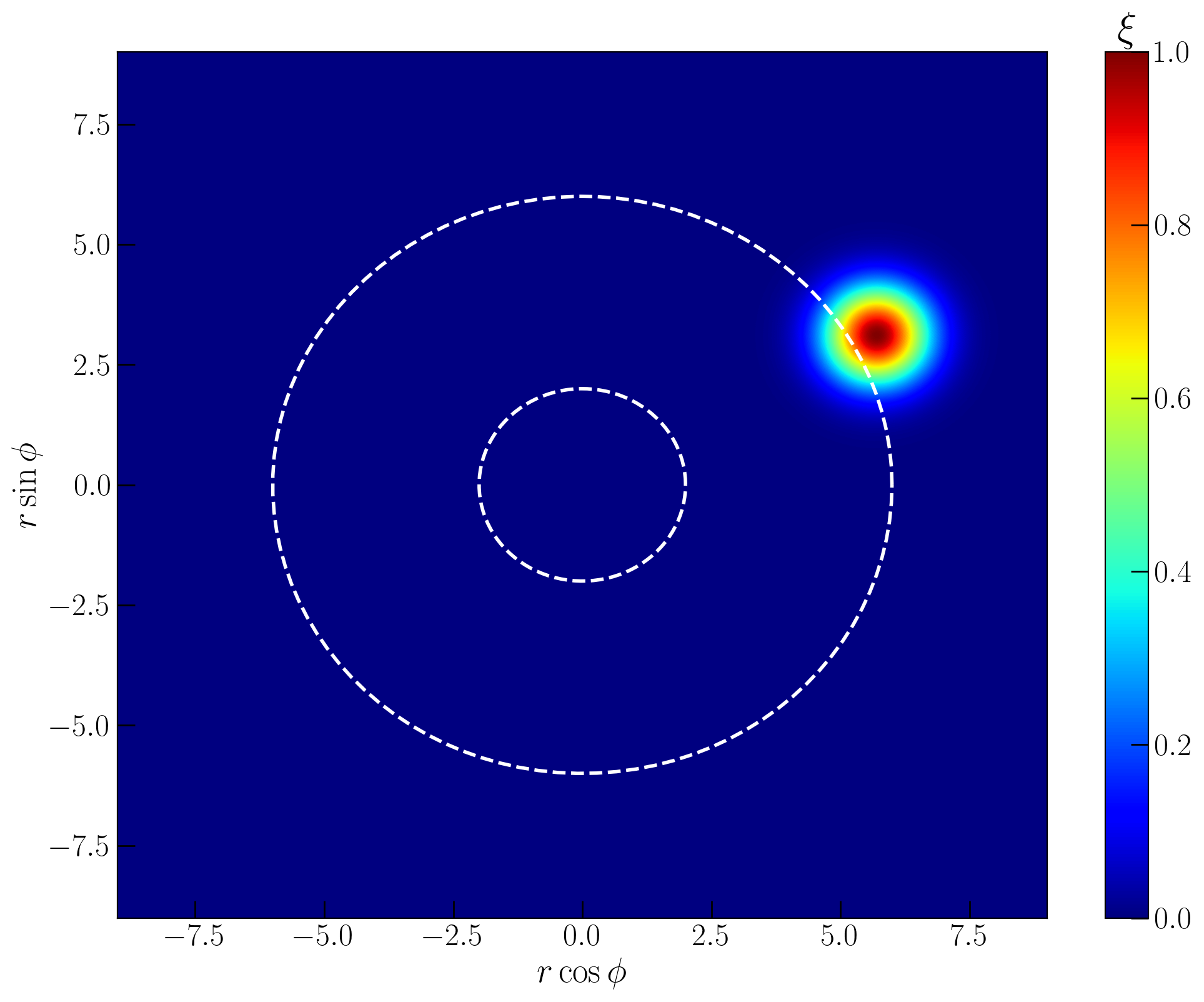}
    \includegraphics[width=\linewidth]{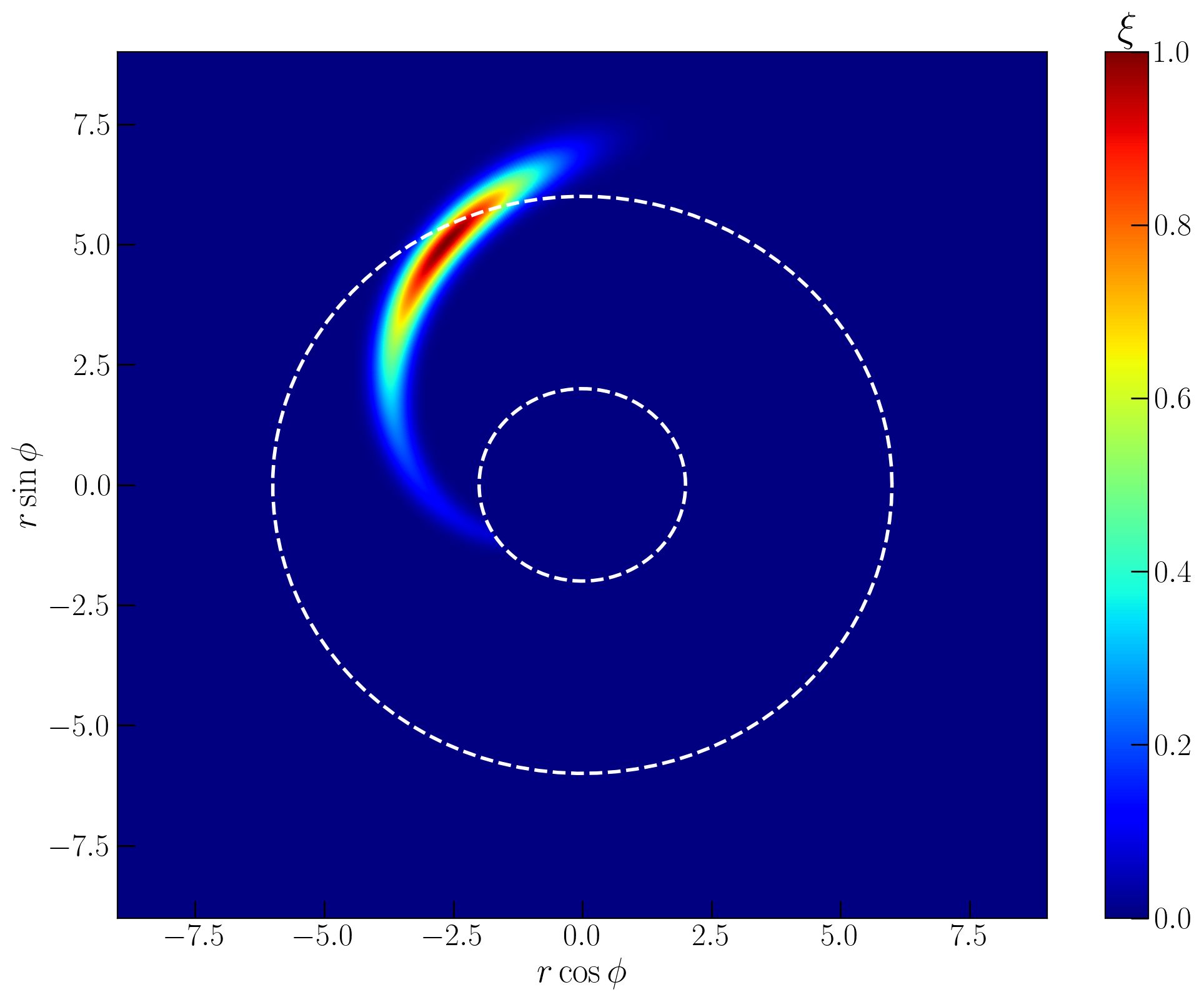}
    \includegraphics[width=\linewidth]{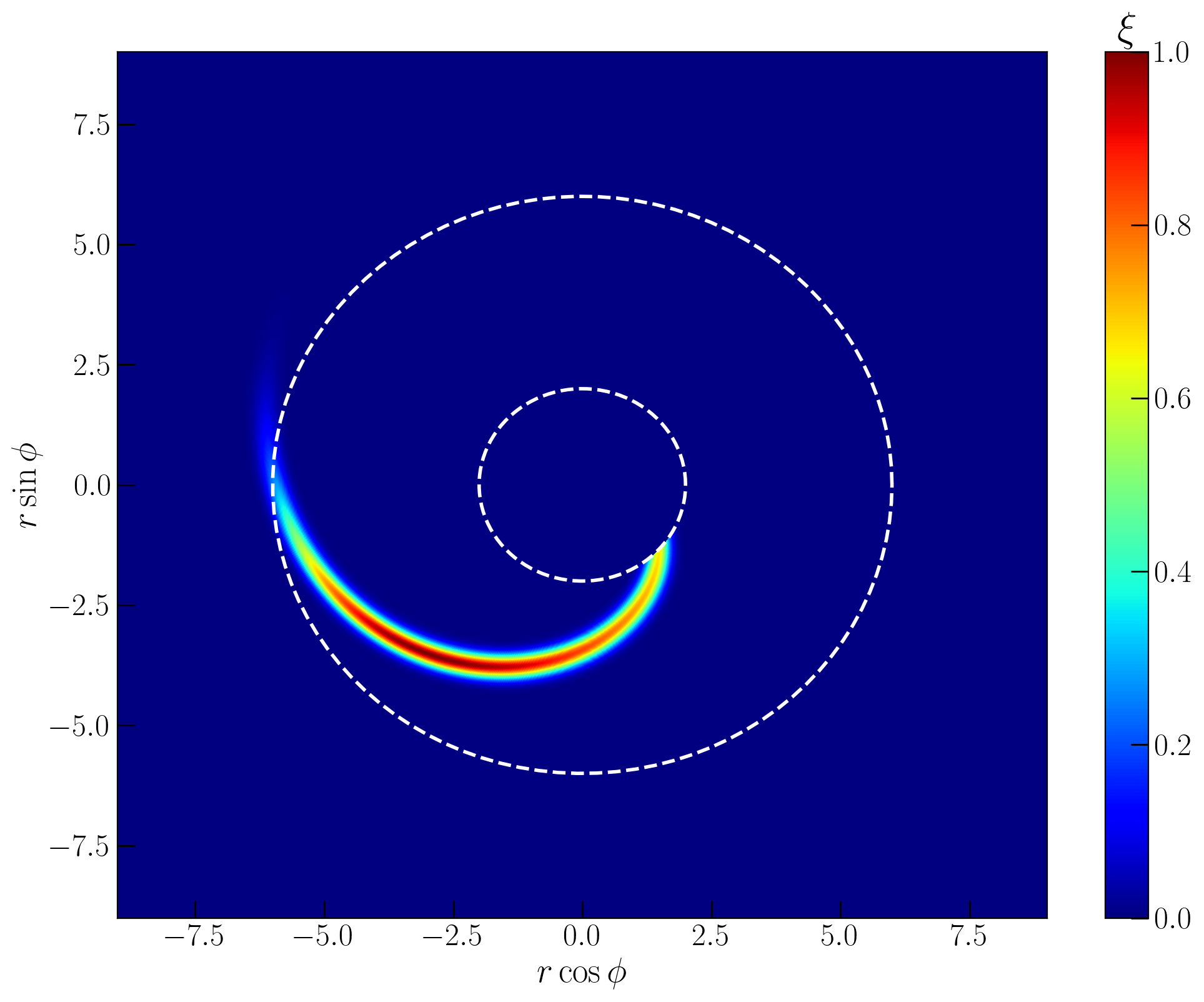}
    \caption{The shearing-out of an initially Gaussian $\xi_0$ into a spiral arm, caused by the rapid radial acceleration of the intra-ISCO region. The initial assumed perturbation $\xi$ is shown in the upper panel, and its subsequent evolution is shown in the lower two panels, with the middle panel produced at $\Delta T = 25$, and the lowest panel $\Delta T = 50$ after the initial condition. The orbital timescale at the ISCO for this system is $\Delta T_{\rm orb} \approx 100$, and so this spiral {arm} formation happens very quickly. The two dashed circles display the event horizon (inner) and ISCO (outer) locations.  }
    \label{fig:charac_spiral}
\end{figure}
Consider an accretion flow with density $\rho_0$, which is in a steady state. Turbulence within the accretion disc will result in fluctuations in the density $\rho \to \rho_0 + \delta \rho(r, \phi, z, T)$. Our first formal ``gravitationally dominated flow'' assumption is that this density perturbation results in only a small fluctuation in the {amplitude of the} flow's velocity field, or explicitly $|\delta U^\mu / U^\mu| \ll |\delta \rho  / \rho |$. In the inner regions of a black hole accretion flow this is reasonable, given the highly relativistic mean motion.   The conservation of rest mass density in the flow is then given by 
\begin{equation}
    \nabla_\mu (\rho U^\mu) = 0 \to \nabla_\mu (U^\mu \, \delta \rho ) = 0, 
\end{equation}
where $\nabla_\mu$ is a covariant derivative with respect to coordinate $x^\mu$, and the second equality follows as by definition $\rho_0$ is a steady state solution, and we neglect the (relatively) small velocity fluctuation $\delta U^\mu$. Expanding this expression in full, we find 
\begin{equation}
     {\partial \over \partial T} \left[ U^T\delta \rho \right]  + {1\over r}{\partial \over \partial r}\left[ r U^r \delta  \rho  \right] +  {\partial \over \partial \phi}\left[ U^\phi \delta \rho \right] + {\partial \over \partial z} \left[ U^z \delta \rho \right ] = 0. 
\end{equation}
Integrating this equation over height, assuming no vertical mass flux out of the disc $U^z(\pm H) = 0$, and defining $\delta \Sigma \equiv \int \delta \rho \, {\rm d}z$ leaves 
\begin{equation}
     {\partial \over \partial T} \left[ U^T \delta \Sigma \right] + {1\over r}{\partial \over \partial r}\left[ r U^r \delta  \Sigma  \right] +  {\partial \over \partial \phi}\left[ U^\phi \delta \Sigma \right]  = 0. 
\end{equation}
In gravitational dynamics $U^T, U^r$ and $U^\phi$ depend only on radius, and so this equation may be re-written 
\begin{equation}
    {\partial \over \partial T} \left( r U^r \delta \Sigma\right) + {U^r\over U^T}{\partial \over \partial r} \left( r U^r \delta \Sigma\right)+ {U^\phi \over U^T} {\partial \over \partial \phi}\left( r U^r \delta \Sigma\right)  = 0. 
\end{equation}
The statement $U^T \simeq U^T(r), U^\phi \simeq U^\phi(r)$ and $U^r \simeq U^r(r)$ is our second and final ``gravitationally dominated flow'' assumption. 

This equation has a clear mathematical and physical interpretation. Mathematically,  the quantity $\xi \equiv r U^r \delta \Sigma$ is a Riemann invariant along the characteristic curves $s(r, \phi, T)$, or explicitly 
\begin{equation}
    {{\rm d} \over {\rm d} s }\xi = 0, 
\end{equation}
where 
\begin{align}
    & {\partial T \over \partial s} = 1 , \label{char1}\\
    & {\partial r \over \partial s} = {U^r \over U^T} , \label{char2}\\
    & {\partial \phi \over \partial s} = {U^\phi \over U^T} \label{char3} ,
\end{align}
define the characteristic curves $s(r, \phi, T)$. 

Physically we can identify $\xi$ with a radial  accretion rate associated with the fluctuation $\delta \dot m \equiv r U^r \int \delta \rho \, {\rm d}z$, which is conserved along the characteristics. Interestingly, despite this being a fully time-dependent evolution (in which a global accretion rate $\dot M$ is not a constant), in the plunging region each perturbed density fluctuation gives rise to its own conserved quantity. 

This means that the invariant $\xi$ satisfies 
\begin{equation}
    \left.\xi(r, \phi, T)\right|_s = \xi(r_0, \phi_0, 0) \equiv \xi_0(r_0, \phi_0), 
\end{equation}
where the function $\xi_0$ encapsulates the shape of the original density perturbation, and the notation $\left.X\right|_s$ denotes an evaluation of $X$ along a characteristic curve $s$. The quantities $r_0, \phi_0$ represent an ``origin point'' for each fluid element within the density fluctuation.  Along the characteristics we have 
\begin{align}
    &r(T)\big|_s = \int_0^T {U^r \over U^T} \, {\rm d}T + r_0, \\
    &\phi(T)\big|_s = \int_0^T {U^\phi \over U^T} \, {\rm d}\phi + \phi_0 ,
\end{align}
a result which follows from the governing characteristic curve equations (\ref{char1}--\ref{char3}). 

In full therefore, we have an analytical solution for any initial perturbation  $\xi_0$ 
\begin{equation}
    \xi(r, \phi, T) = \xi_0\left( r - \int_0^T {U^r \over U^T} \, {\rm d}T,\, \phi - \int_0^T {U^\phi \over U^T} \, {\rm d}T \right) . 
\end{equation}
It is to be understood in these integrals that the fluid 4-velocity components depend implicitly on the time coordinate $T$ through their radial dependence $r(T)$ in a moving flow. 

In Figure \ref{fig:charac_spiral} we show the analytical solution of the mass conservation equations (computed using the method of characteristics) for an initial Gaussian perturbation of $\xi$. We assume that outside of the ISCO $U^r \simeq u_I$, and use a Keplarity factor $f_K = 0.95$. {Note that the relevant Keplarity factor for each spiral feature will be that of the {\it local} velocity structure of the turbulent over/under density. It is likely that this quantity will vary stochastically over the course of a simulation, and spiral features formed at different times in the disc will have subtly different structures.  }

What is clear from this analytical solution is that initially localised perturbations (top panel of Fig. \ref{fig:charac_spiral}) are rapidly sheared into macroscopic (covering many gravitational radii) spiral arms. This happens on timescales shorter than the orbital period of the disc fluid -- the upper and lower panels are separated in coordinate time by a factor $\Delta T / T_{\rm orb} \simeq 1/2$. As turbulent eddies in the disc are expected to be excited on timescales $\gtrsim T_{\rm orb}$, this means that the turbulence will not have sufficient time to churn up the density 
fluctuations before they have been sheared into these spiral features. 

Unlike in the simplified test-particle case, in the full fluid case the properties of the resulting spiral {arm} show a (slight) dependence on the functional form of the initial perturbation $\xi_0$. They also, unlike in the test-particle case, show no ``kink'' at the ISCO (contrast Figures \ref{fig:charac_spiral} and \ref{fig:spiral-demo}), which is entirely a result of the simplifications used in computing the test-particle spiral. 

\subsection{ Properties of the spiral {arm} }
The properties of the spiral {arm} which forms from the collective test-particle behaviour (e.g., Figure \ref{fig:spiral-demo}) is entirely determined by the Keplarity factor $f_K$, and is therefore simplest to analyse. These test particle spiral {arm}s are an excellent approximation to the full fluid calculations computed above, except for right at the ISCO. 

\begin{figure}
    \centering
    \includegraphics[width=1\linewidth]{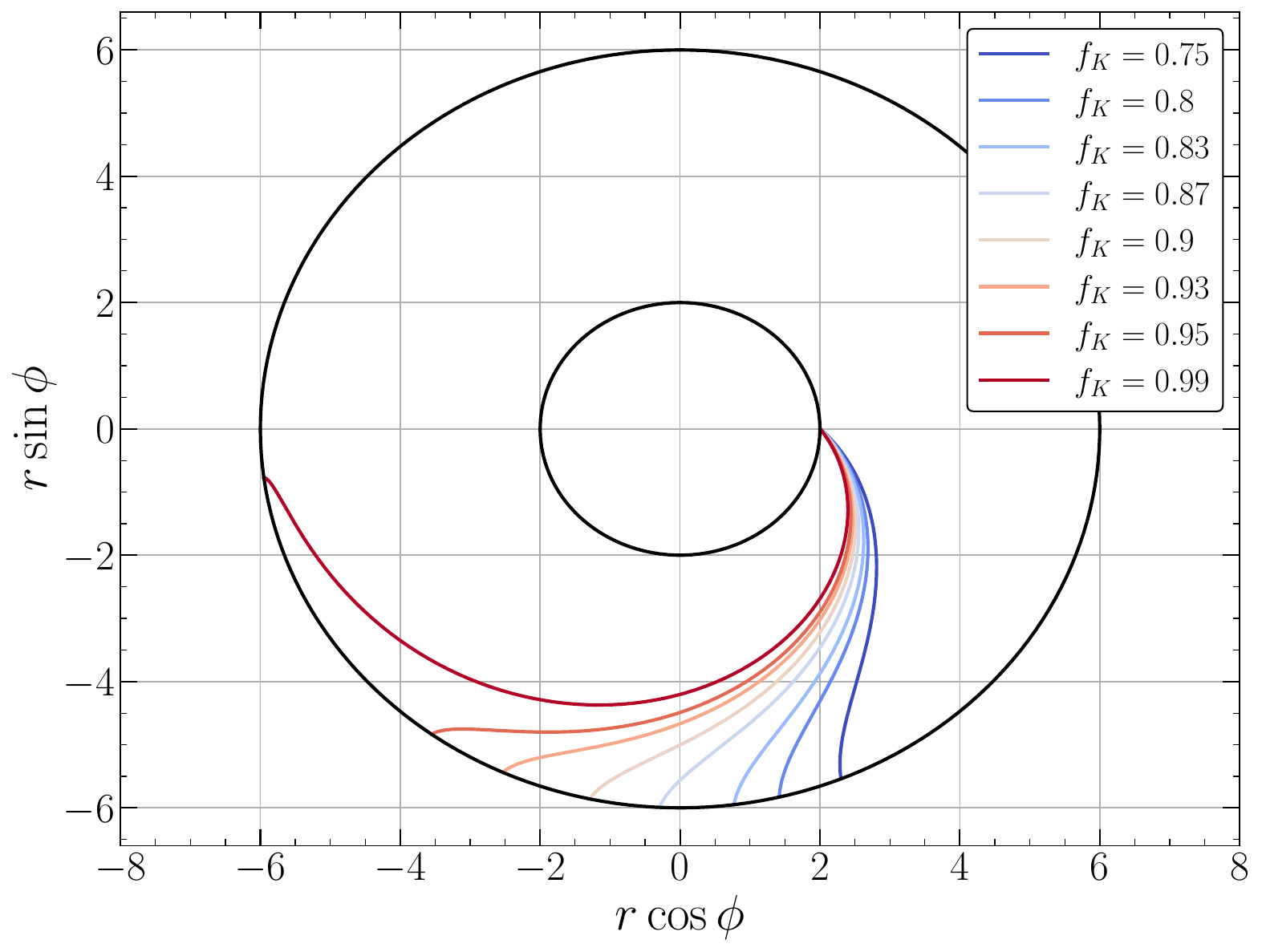}
    \includegraphics[width=1\linewidth]{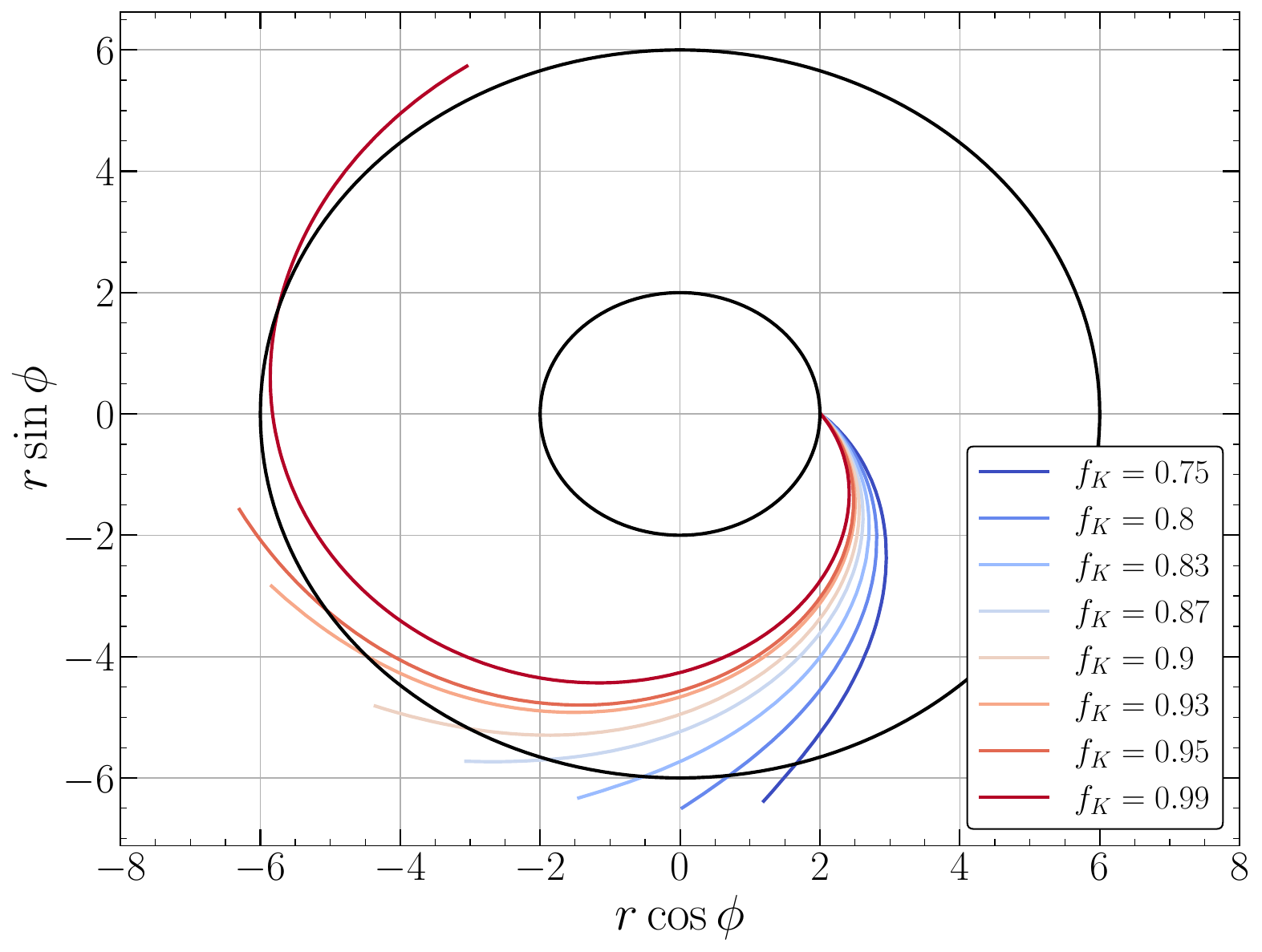}
    \caption{spiral {arm}s for different Keplarity factors $f_K$ (denoted in figure legend). Flows with lower Keplarity form straighter bars, which fill in quicker, and will then rotate around the black hole more slowly. The upper panel shows the test particle approximation, while the lower panel shows the method of characteristics solution for an initial Gaussian over density. The spirals are very similar in both cases, except right at the ISCO.   }
    \label{fig:spiral-keplarity}
\end{figure}

\begin{figure}
    \centering
    \includegraphics[width=\linewidth]{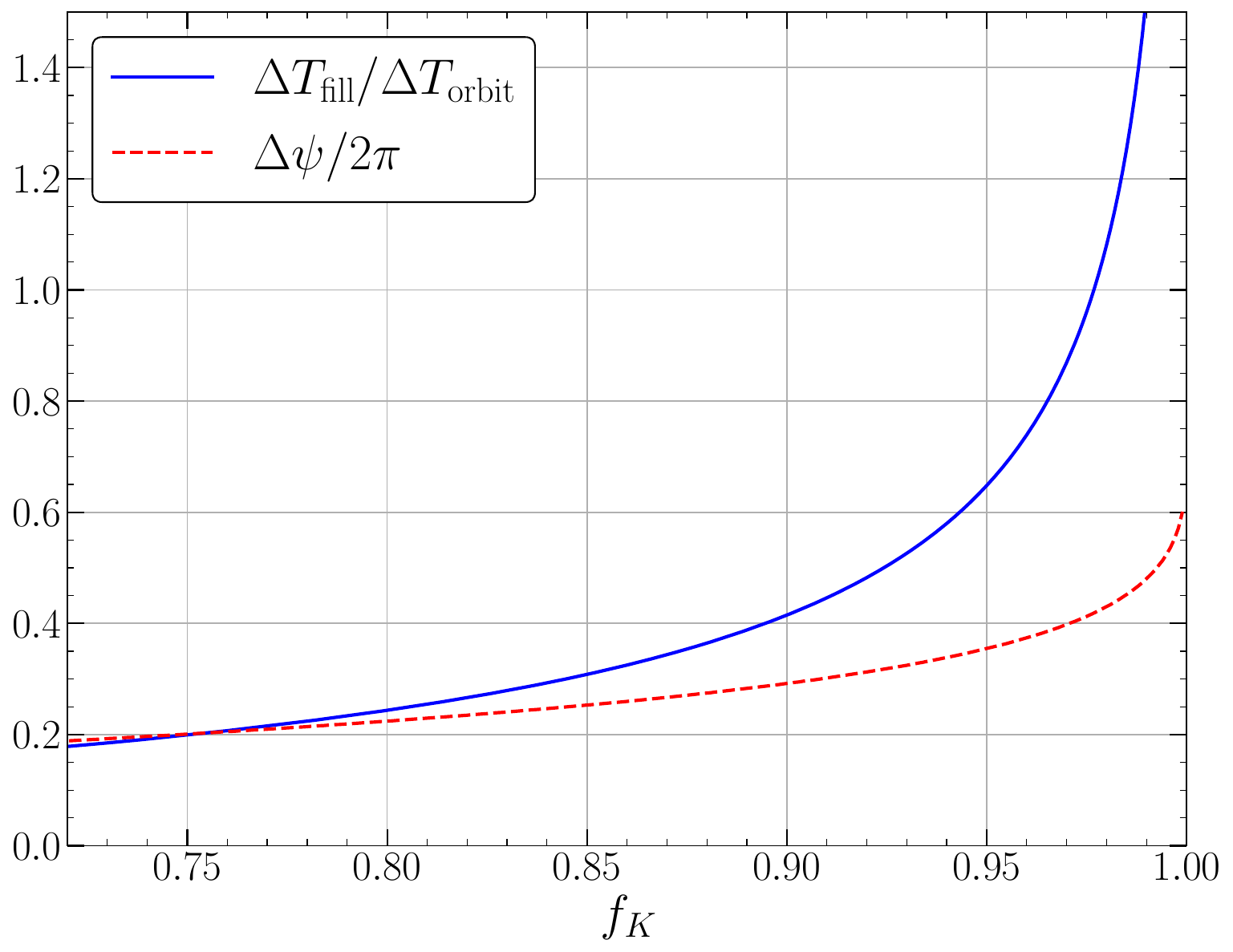}
    \caption{The filling time (blue solid curve) for the spiral {arm}s, in units of the orbital timescale at the ISCO, plotted against Keplarity factor $f_K$. In red (dashed) we plot  as a function of Keplarity the global spiral {arm} pitch angle (defined in text) as a fraction of $2\pi$.  }
    \label{fig:time-keplarity}
\end{figure}

Example {spiral} structures for other values of the Keplarity factor $f_K$ are shown in Figure \ref{fig:spiral-keplarity} (in which we also display the structure of the Gaussian over density model, cf. Fig. \ref{fig:charac_spiral}, with individual curves formed from the contour of maximum density).  Flows with lower Keplarity (and therefore larger trans-ISCO velocities $u_I$) form straighter bars, which fill in quicker, and will then rotate around the black hole more slowly.  Note that the global pitch angle of the {spiral} $\Delta \psi$, which we define as 
\begin{equation}
    \Delta \psi \equiv \bar \phi(r_+) - \bar \phi(r_I), 
\end{equation}
is a relatively weak function of Keplarity across $0.75 \lesssim f_K \lesssim 0.95$ (Fig. \ref{fig:time-keplarity}). This means that the predictions of this theory are falsifiable, only certain pitch angles should be seen in GRMHD simulations of black hole accretion.   It is interesting to note that despite it taking a formally infinite time for a particle with $f_K=1$ to asymptote off the ISCO radius, this particle only accumulates a finite phase difference from the perfectly circular orbit at $r=r_I$ over its inspiral (and therefore the resulting {spiral} has finite global pitch).

The time it takes, from the beginning of the first particles orbit, for the spiral {arm} to be filled out is given by the time it takes for a single particle to cross the distance from the ISCO to the event horizon 
\begin{equation}
    \Delta T_{\rm fill} = \int_{r_I}^{r_+} {U^T(r') \over U^r(r')} \, {\rm d}r' ,
\end{equation}
and is a function only of the Keplarity factor (through the trans-ISCO velocity $u_I$). We plot the filling time as a function of Keplarity in Fig. \ref{fig:time-keplarity}. The relevant timescale with which to compare $\Delta T_{\rm fill}$ is the orbital timescale of a particle on a circular orbit at the ISCO, namely $\Delta T_{\rm orb} = 2\pi / \Omega$. As we stressed in the previous section, for Keplarity factor $f_K \lesssim 0.95$ this is substantially shorter than the orbital period at the ISCO, and it seems unlikely therefore that turbulent fluctuations will have time to wash out these spiral structures. 

The geometric properties of spiral {arm}s are typically examined mathematically through two local properties, the local pitch angle $\alpha$ and local curvature $\kappa$. The local pitch angle of a spiral is defined as the angle between the tangent of the spiral at a point $(r, \phi)$ and the tangent of a circle at that point with the same radius $r$. This definition is equivalent to 
\begin{equation}
    \tan \alpha = {r'\over r} , \quad r' \equiv  {{\rm d} r \over {\rm d}\phi} . 
\end{equation}
The curvature at a point on a curve is defined as the curvature of its osculating circle (the circle that best approximates the curve near this point). In polar coordinates $r, \phi$ the local curvature is given by 
\begin{equation}
    \kappa = {r^2 + 2(r')^2 - r r'' \over (r^2 + (r')^2)^{3/2}} ,
\end{equation}
and can be interpreted as the inverse of the radius of its osculating circle $\kappa = 1/R$. 

\begin{figure}
    \centering
    \includegraphics[width=\linewidth]{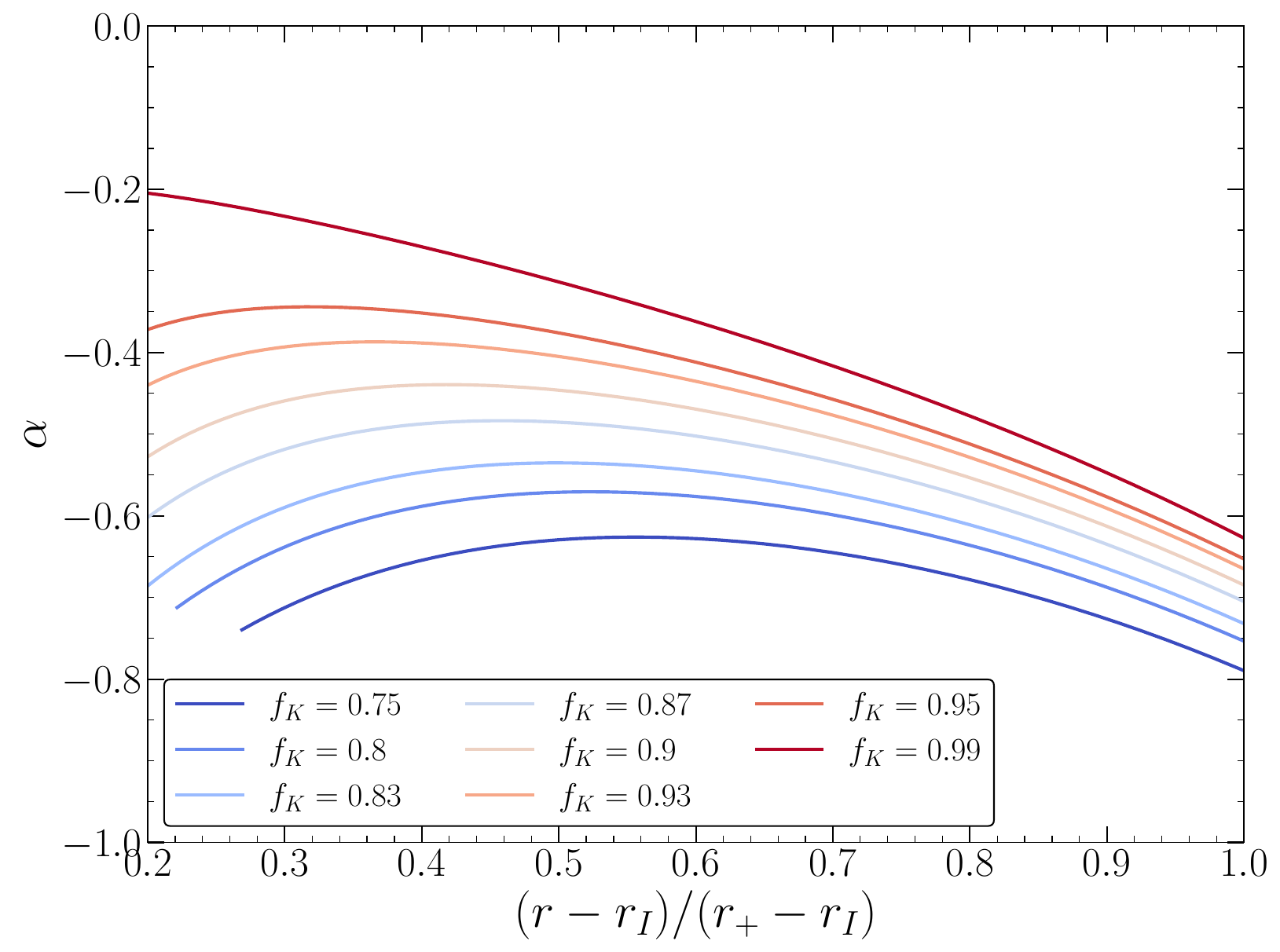}
    \includegraphics[width=\linewidth]{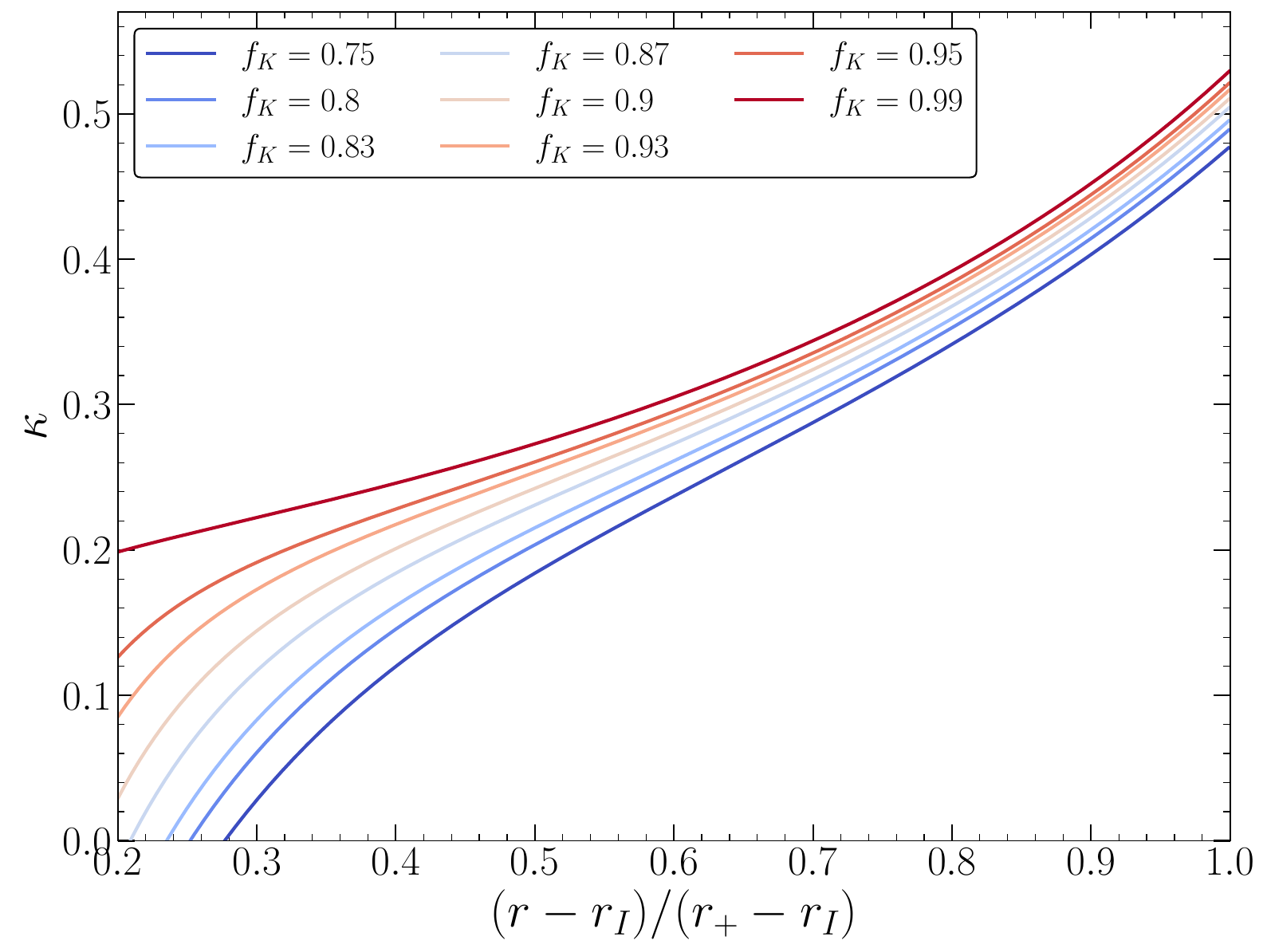}
    \caption{ The local pitch angle (upper) and curvature (lower) of the spiral {arm}s, as a function of the normalised distance traversed over the inspiral from the ISCO, for a number of different Keplarity factors.  Higher Keplarity factors generally lead to higher curvature and lower pitch angles, but the curvature of the geodesic spiral {arm}s is in general a weak function of $f_K$. The pitch angle $\alpha(r)$ is, for most of the inspiral, roughly constant, a characteristic property of logarithmic spirals.   }
    \label{fig:spiral-props}
\end{figure}

We plot both the local pitch angle $\alpha$ and curvature $\kappa$ of the test-particle spiral {arm}s in Figure \ref{fig:spiral-props}. In this figure we note that these characteristic spirals have a roughly constant local pitch angle $\alpha$. This is a property of so-called logarithmic spirals, which have the functional form $r \propto \exp(\phi \tan \alpha)$. We stress that the geodesic spirals constructed here are not precisely logarithmic spirals, but are simply relatively closely approximated by this class of spiral, which will prove useful in later sections. 

Both the local pitch angle $\alpha$ and curvature $\kappa$ of the characteristic spirals are functions of the Keplarity factor of the flow. Flows which are better approximated by circular motion (those with Keplarity closer to $1$) result in intra-ISCO spirals which have larger curvature and smaller pitch angles. It is likely that thinner discs (i.e., those with lower random energy contents) will have $f_K$ closer to 1, and therefore will likely have larger intra-ISCO spiral curvature. 

\subsection{Kerr generalisation}
The relevant metric for the analogous calculation to that carried out above, but for the spacetime around a rotating black hole, is the spherical Kerr-Schild coordinate system, which has the general form 
\begin{multline}
    {\rm d}s^2  =  g_{TT} \, {\rm d}T^2 + 2 g_{Tr} \, {\rm d}T \, {\rm d}r  + g_{rr} \, {\rm d}r^2  + 2 g_{T\phi} \, {\rm d}T \, {\rm d}\phi \\ + 2 g_{r\phi} \, {\rm d}r \, {\rm d}\phi  + g_{\theta\theta} \, {\rm d}\theta^2 + g_{\phi\phi}  {\rm d}\phi^2. 
\end{multline}
We write out in full each of the $g_{\mu \nu}$ components in terms of the physical parameters of the system in Appendix \ref{KerrApp}, although in this section we use a compact $g_{\mu\nu}$ notation. Note that the quantities $T$ and $\phi$ differ from their more familiar Boyer-Lindquist $t$ and $\varphi$ by the relationships 
\begin{align}\label{SKS2BL}
    {\rm d}T &= {\rm d}t + {2r \over r^2 - 2r + a^2} \, {\rm d}r, \\
    {\rm d}\phi &= {\rm d}\varphi + {a \over r^2 - 2r + a^2} \, {\rm d}r . \label{blp}
\end{align}
Here $a$ is the usual (dimensionless) Kerr metric spin parameter ($|a|\leq 1$ for a black hole spacetime), and $r$ is the usual Boyer-Lindquist radial coordinate, which is identical to the Kerr-Schild radial coordinate. The $\theta$ coordinate is similarly identical in both systems. This coordinate system, unlike the Boyer-Lindquist system, has well behaved $T$ and $\phi$ coordinates on the event horizon of the Kerr black hole.  This behaviour is essential for our purposes of studying trans-event horizon flows. 

As the radial coordinate is identical in both Boyer-Lindquist and Kerr-Schild coordinates, the  radial velocity over an inspiral from the ISCO is given by \citep{Cunningham75, MummeryBalbus22PRL}
\begin{equation}
    U^r_g = -\sqrt{2\over 3r_I} \left({r_I \over r} - 1\right)^{3/2} . 
\end{equation}
As the coordinate transformations (equations \ref{SKS2BL}) only mix up the Boyer-Lindquist temporal and azimuthal coordinates with the radial coordinate, and circular motion has by definition fixed $r$, the circular motion 4-velocity components in Kerr-Schild coordinates are unchanged from their Boyer-Lindquist form. For the inspiral problem we are primarily interested in the conserved ISCO energy and angular momenta, which are given by 
\begin{align}
    -U_{T, g}(r_I) = \left(1 - {2\over 3r_I}\right)^{1/2}, \\
    U_{\phi, g}(r_I) = 2\sqrt{3} \left(1 -{2a \over 3 \sqrt{r_I}} \right) . 
\end{align}
Where we remind the reader that subscript $g$ denotes the results relevant for a circular orbit. Assuming an inspiral from the ISCO which is completely dominated by gravity, the ISCO values of $U_T$ and $U_\phi$ will be conserved. The two equations $U_T = g_{T\mu } U^\mu$ and $U_\phi = g_{\phi \mu} U^\mu$ can be rearranged for the azimuthal and temporal 4-velocity components, which are ultimately given by 
\begin{align}
    U^T &= { g_{\phi \phi} U_T - {g_{\phi T}} U_\phi + \left[{g_{\phi r}g_{\phi T}} - g_{\phi \phi}g_{Tr} \right] U^r \over g_{TT}g_{\phi \phi} - g_{\phi T}^2 } , \\
    U^\phi &= { g_{TT} U_\phi - {g_{\phi T}} U_T + \left[{g_{T r}g_{\phi T}} - g_{TT}g_{\phi r} \right] U^r \over g_{TT}g_{\phi \phi} - g_{\phi T}^2 } .
\end{align}
The ISCO angular frequency is given by 
\begin{equation}
    \Omega_g \equiv \left.{U^\phi \over U^T}\right|_{r_I} = {r_I^{-3/2} \over 1 + a r_I^{-3/2} } .
\end{equation}
Just as in the Schwarzschild case considered above, by specifying a single Keplarity parameter $f_K$, and assuming that at and within the ISCO $\Omega = f_K \Omega_g$ and $U^\phi = f_K U^\phi_g$, one can compute a trans-ISCO velocity $u_I(f_K)$ from solving the quadratic $\left. g_{\mu\nu}U^\mu U^\nu \right|_{r_I} = -1$, which provides a modified inflow velocity $U^r = U^r_g - u_I$. From these results the shape of an intra-ISCO spiral {arm} can be computed from the two integrals (eq. \ref{rinspiral}) and (eq. \ref{phinspiral}). The actual structure of these spirals looks qualitatively identical to the Schwarzschild case, the principle difference being the trivial change in radial extent of the spiral itself, which simply tracks the change in size of the plunging region (which is a function only of the Kerr spin parameter $a$). As such we postpone a discussion of Kerr spirals until the GRMHD analysis portion of this paper. 

\section{GRMHD simulations  }\label{numsec}
We now have a complete $(r, \phi, z)$ analytical model for how the density of accretion flows should behave within the ISCO.  It is the purpose of this section to compare the model of \cite{MummeryBalbus2023} with the results of numerical GRMHD simulations run using {\tt AthenaK}, a new performance portable version of the {\tt Athena++} code \cite{Stone20} written using the Kokkos library \cite{Trott21}. In this paper we focus on thick disc simulations (i.e., those with an aspect ratio $H/r \sim 1$). 

The reason for this is two fold; firstly, thicker discs are most simple to simulate with a resolution which is sufficient to accurately capture the physics of the  magnetorotational instability (MRI) which drives  angular momentum transport within the disc \citep{BalbusHawley91}. Secondly, many astrophysical black hole accretion flows are in fact thick, owing to inefficient radiative cooling. Most notably of course are the supermassive black holes located in the centre of our own Galaxy (Sagittarius A$^*$) and in the centre of the giant elliptical galaxy M87. Both of these sources have been the subject of Event Horizon Telescope (EHT) analysis \citep[e.g.,][]{EHT19, EHT22}, and analytical descriptions of the plunging region disc density may well be of use in future studies of these systems. 

\subsection{Simulation setup}
We initialise a \cite{Fishbone76} torus with inner edge at $r_{\rm in} = 20$, and a density maximum at $r_{\rm peak} = 41$. We use an ideal equation of state with adiabatic index $\Gamma = 13/9$. We initialise the magnetic field with a SANE (Standard And Normal Evolution) setup, consisting of poloidal field loops parallel to density contours with alternating sign. Comparison with numerical models
initialised with a MAD (magnetically arrested disk) configuration is left for future work.

Our simulation domain is defined by Cartesian coordinates $x, y, z$, with an outer edge at $x_{\rm max}, y_{\rm max}, z_{\rm max} = \pm 1024, \pm 1024, \pm 1024$. Static mesh refinement is used to resolve the innermost disc regions. Each refinement region is a cube, inside of which the resolution doubles, with the edges of each cube located at $x, y, z$ coordinates which are half the size of the parent grid, i.e. $\pm 512$, down to $\pm 8$. This leads to a total of eight levels. The flow is resolved with $256^3$ cells on each level.  We have tested that the flow geometry on grids with both twice and half this resolution are qualitatively similar, indicating for our purposes the flow is resolved at our fiducial resolution. 

We solve the GRMHD equations using {\tt AthenaK}. Each model required about 100 GPU hours to run to a final time $T_{\rm max} = 8500$ $[GM_\bullet/c^3]$. We save snapshots of the $z=0$ and $y=0$ planes every $\Delta T_{2d} = 10$$[GM_\bullet/c^3]$, and save the full state of the simulation every $\Delta T_{3d} = 100$$[GM_\bullet/c^3]$.  

It will be convenient for our analysis to convert from the simulation's Cartesian coordinates back into spherical Eddington-Finkelstein coordinates. We do this using the standard {\tt AthenaK} data analysis scripts\footnote{Which can be found here: \url{https://github.com/c-white/athenak_scripts}.}. {See Appendix \ref{KerrApp} for a discussion of the different coordinate systems.}

\subsection{Schwarzschild analysis}
Our first simulation involves a Schwarzschild $a_\bullet  = 0$ black hole. 

We will be comparing analytical models to various averaged simulation quantities, denoted by $\Xbar_{y}$, where $X$ is the quantity of interest and $y$  denotes the particular average taken. Of particular interest are the following 
\begin{align}
    \Xbar_{\phi} &\equiv {1 \over 2\pi} \int_{0}^{2\pi} X(r, \theta, \phi, T) \, \sin \theta \, {\rm d}\phi , \\
    \Xbar_{\theta} &\equiv {1 \over 2} \int_{0}^{\pi} X(r, \theta, \phi, T) \,   {\rm d}\theta , \\
    \Xbar_{T} &\equiv {1 \over {\Delta T}} \int_{T}^{T+\Delta T} X(r, \theta, \phi, T') \,  {\rm d}T', 
\end{align}
which all have particularly simple (and Newtonian-esque) functional forms in Eddington-Finkelstein coordinates. Double averages will also be computed, which are rather trivially given by 
\begin{align}
    \Xbar_{\theta , \phi} &\equiv {1 \over 4\pi} \int_{0}^{2\pi} \int_{0}^{\pi}  X(r, \theta, \phi, T)  \, \sin \theta  \, {\rm d}\theta \, {\rm d}\phi \\
    \Xbar_{\theta, T} &\equiv {1 \over 2\Delta T} \int_{T}^{T+\Delta T} \int_{0}^{\pi} X(r, \theta, \phi, T') \, {\rm d}\theta  \,  {\rm d}T', \\
    \Xbar_{\phi, T} &\equiv {1 \over 2\pi {\Delta T}}  \int_{T}^{T+\Delta T} \int_{0}^{2\pi} X(r, \theta, \phi, T') \,  \sin \theta \,  {\rm d}\phi \,  {\rm d}T'. 
\end{align}
The variable $\Xbar$ remains a function of all quantities which have not been averaged over.

\subsubsection{Vertical structure}

We begin by analysing the vertical structure of the accretion disc density. The prediction of the  theory developed in \cite{MummeryBalbus2023} is that the vertical contours of the accretion disc density should be described by the functional form 
\begin{equation}\label{hp}
    {H(r) \over H_I} = \left({6\over r}\right)^{-16/11} \left[ {c \over 3 u_I} \left({6 \over r} - 1\right)^{3/2} + 1\right]^{-2/11} , 
\end{equation}
where we have substituted $\Gamma = 13/9$ into equation (\ref{hh}), along with the values of the ISCO radius $r_I = 6$ and $\varepsilon =  \sqrt{3 r_I u_I^2/2 r_g c^2}$. 

\begin{figure}
    \centering
    \includegraphics[width=\linewidth]{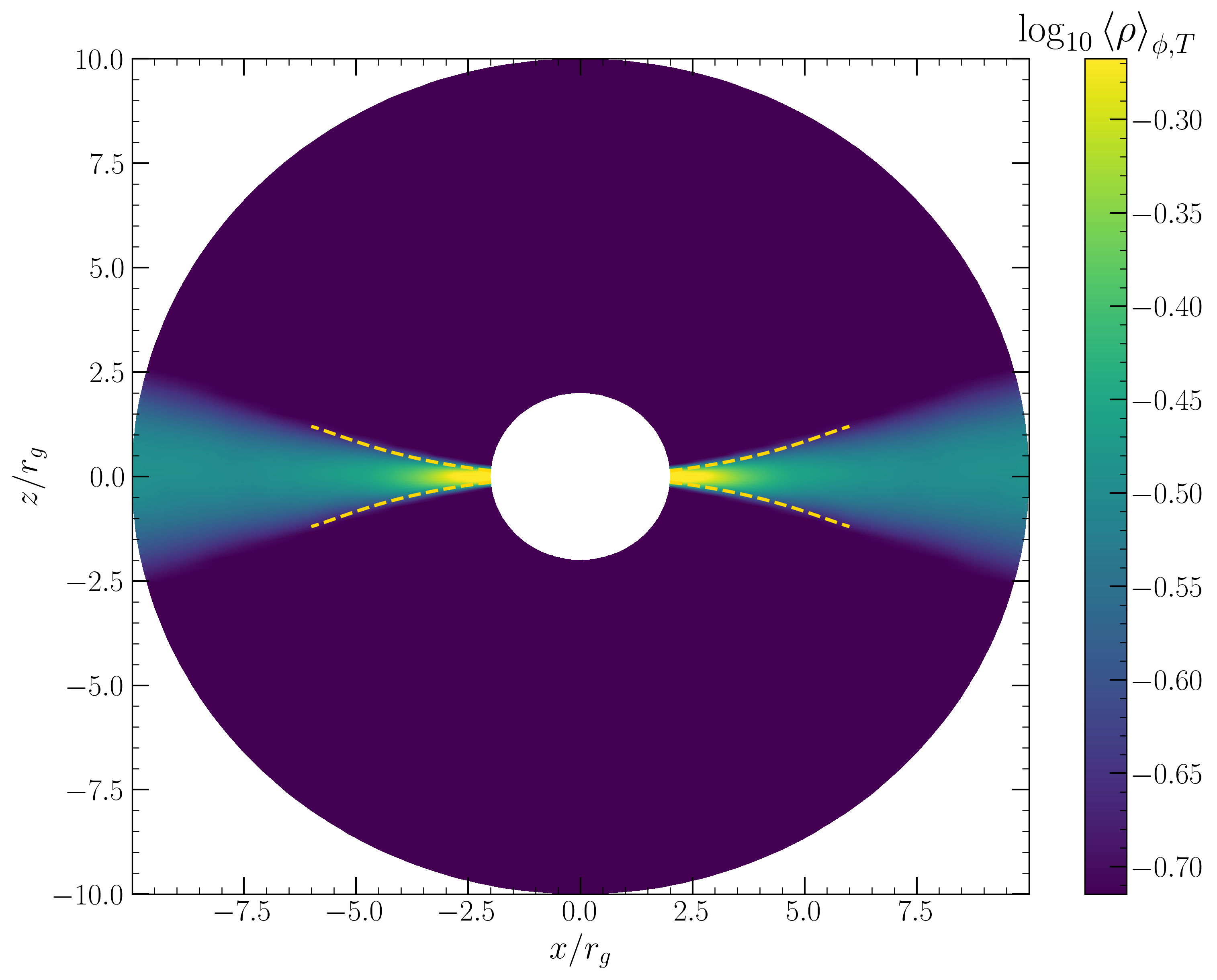}
    \includegraphics[width=\linewidth]{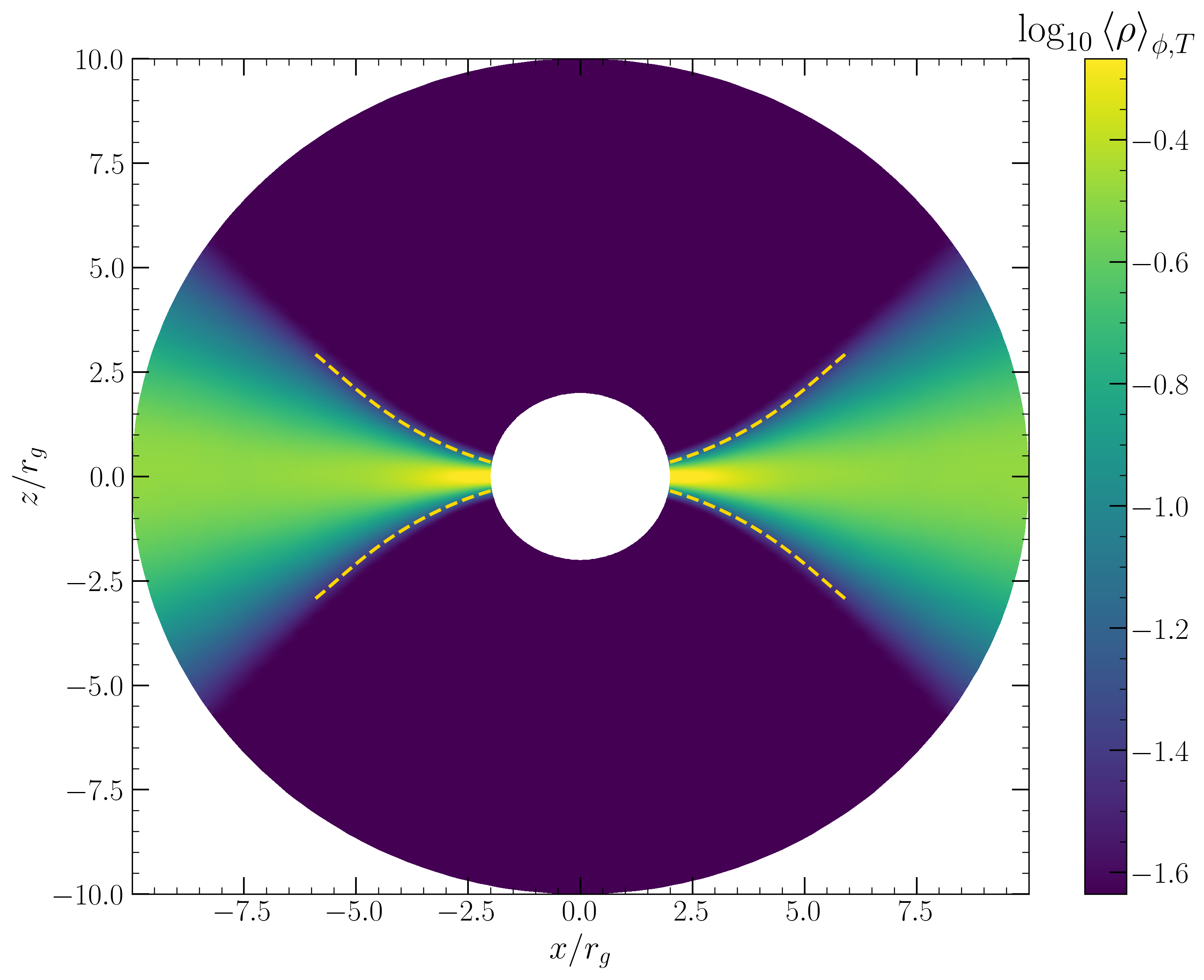}
    \includegraphics[width=\linewidth]{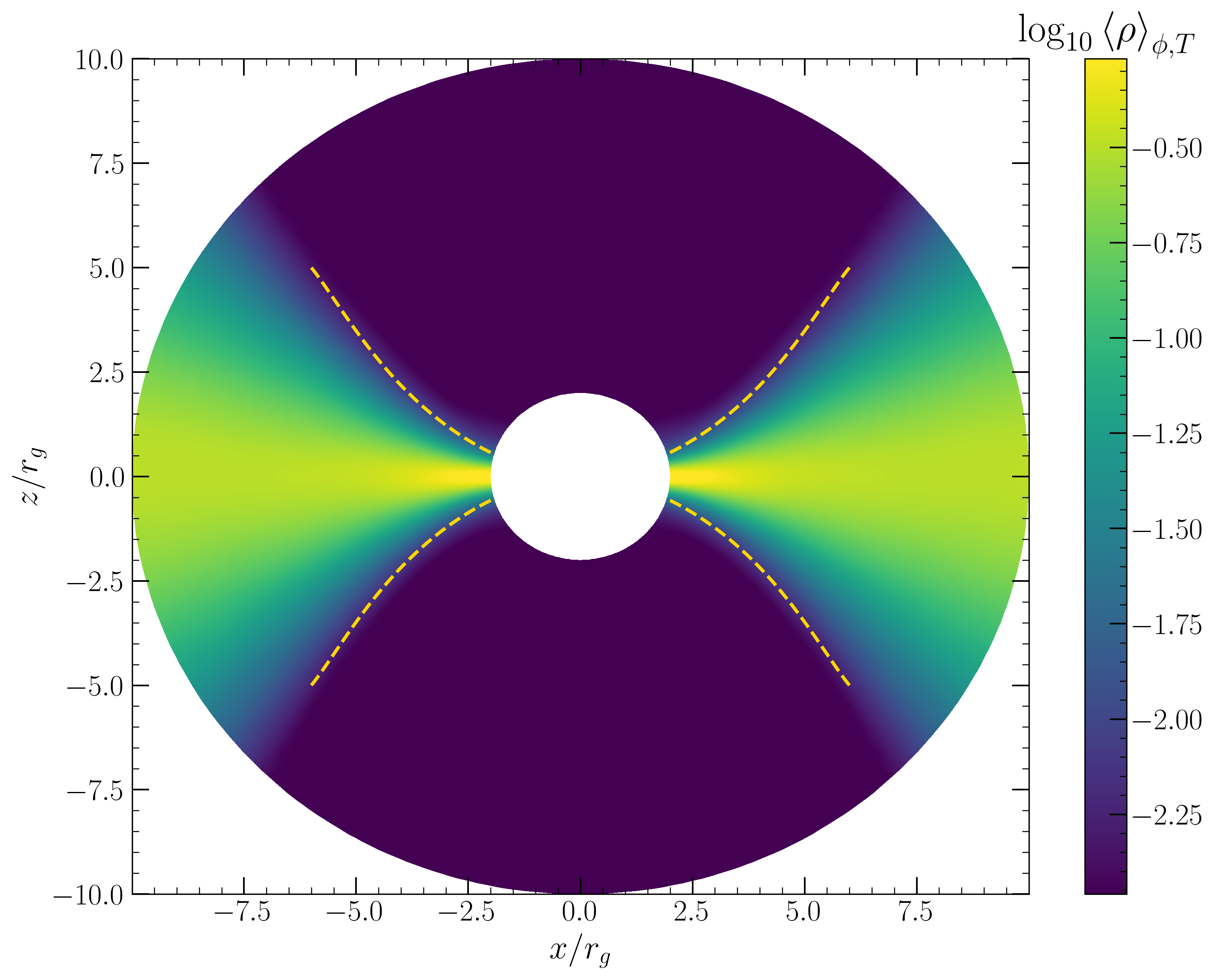}
    \caption{The azimuthally and temporally averaged disc density $\rhobar_{\phi, T}$ (code units), plotted on a logarithmic scale against the Eddington-Finkelstein $x = r \sin \theta$ and $z = r \cos \theta$ coordinates. The three plots are identical except for the lower limit of the colourbar, which is cropped at different levels to highlight different vertical density contours in the flow. The dashed curves show the analytical disc height function derived in \citealt{MummeryBalbus2023}, with different normalisation's chosen to match each density contour. Despite being derived in the thin-disc limit, this analytical result accurately describes the vertical structure of this thick disc GRMHD simulation.   }
    \label{fig:averaged_vertical}
\end{figure}

As a first test, we compute $\rhobar_{\phi, T}$ by averaging the disc density over its full $2\pi$ azimuth and the final 5000$GM_\bullet/c^3$ of the simulation run. The results of this averaging are shown in Figure \ref{fig:averaged_vertical}. In Figure \ref{fig:averaged_vertical} we display the logarithm of this averaged density (in code units). We reproduce this three times, and each panel is identical except for the lower limit of the colourbar, which is cropped at different levels to highlight different vertical density contours in the flow. The three dashed curves {show the function $z = H(r)$ and} are plotted from the ISCO $r_I = 6$, to the event horizon $r_+ = 2$, {each having} an identical functional form (equation \ref{hp}{, with $r$ the Eddington-Finkelstein radial coordinate}), and are distinguished only in their amplitude $H_I$. We see that each density contour we pick can be described by a function of the same form, merely differentiated by amplitude. As can be seen from equation (\ref{hp}), the dependence of the density scale height on the fitting parameter $\varepsilon$ is weak ($\sim \varepsilon^{2/11}$), and a wide range of $\varepsilon$ accurately reproduce the observed density scaling. For these plots we take a trans-ISCO velocity of $u_I = 0.05c$, as this well described the radial and azimuthal structure (to be discussed in later sections) which are more sensitive to $u_I$. 

\begin{figure*}
    \centering
    \includegraphics[width=.42\linewidth]{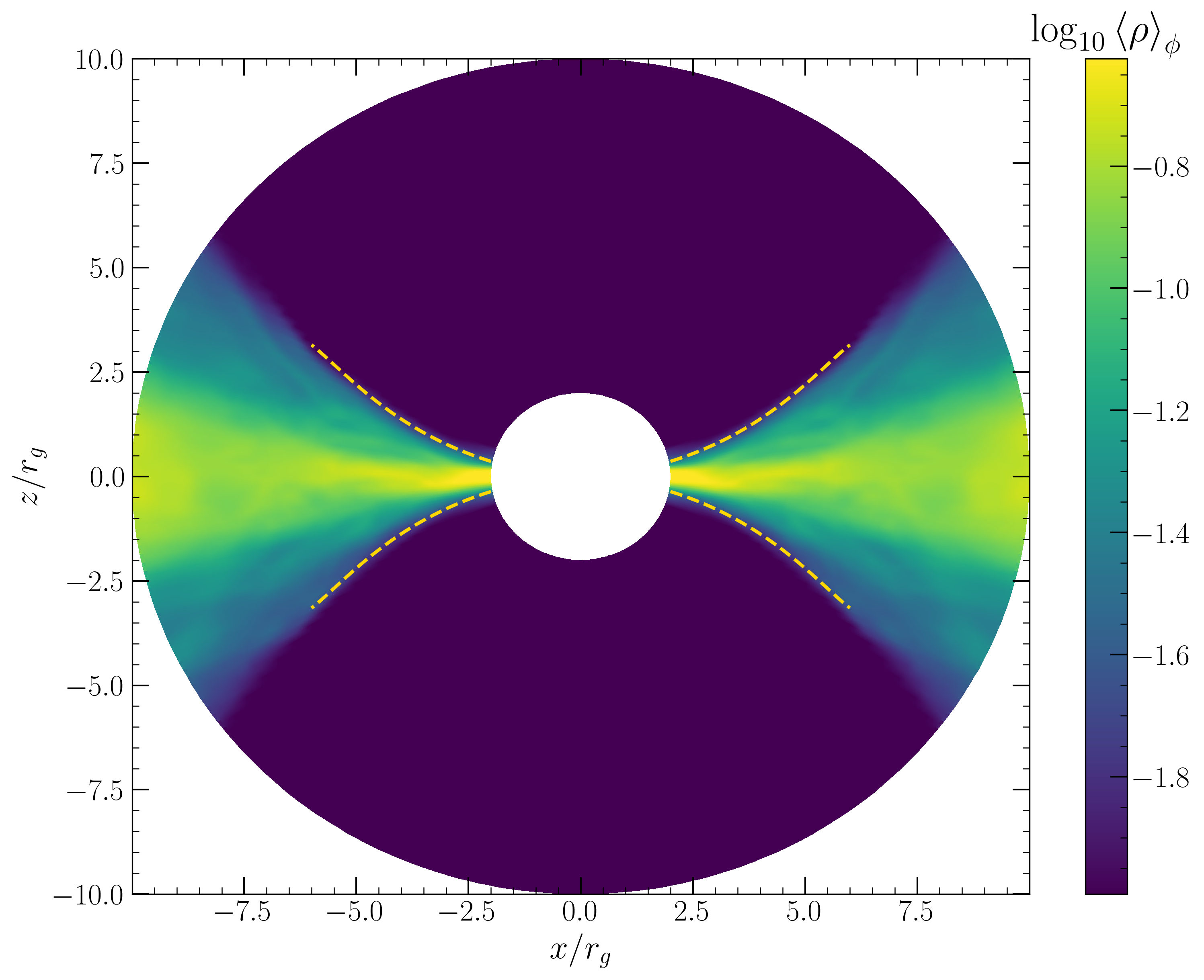}
    \includegraphics[width=.42\linewidth]{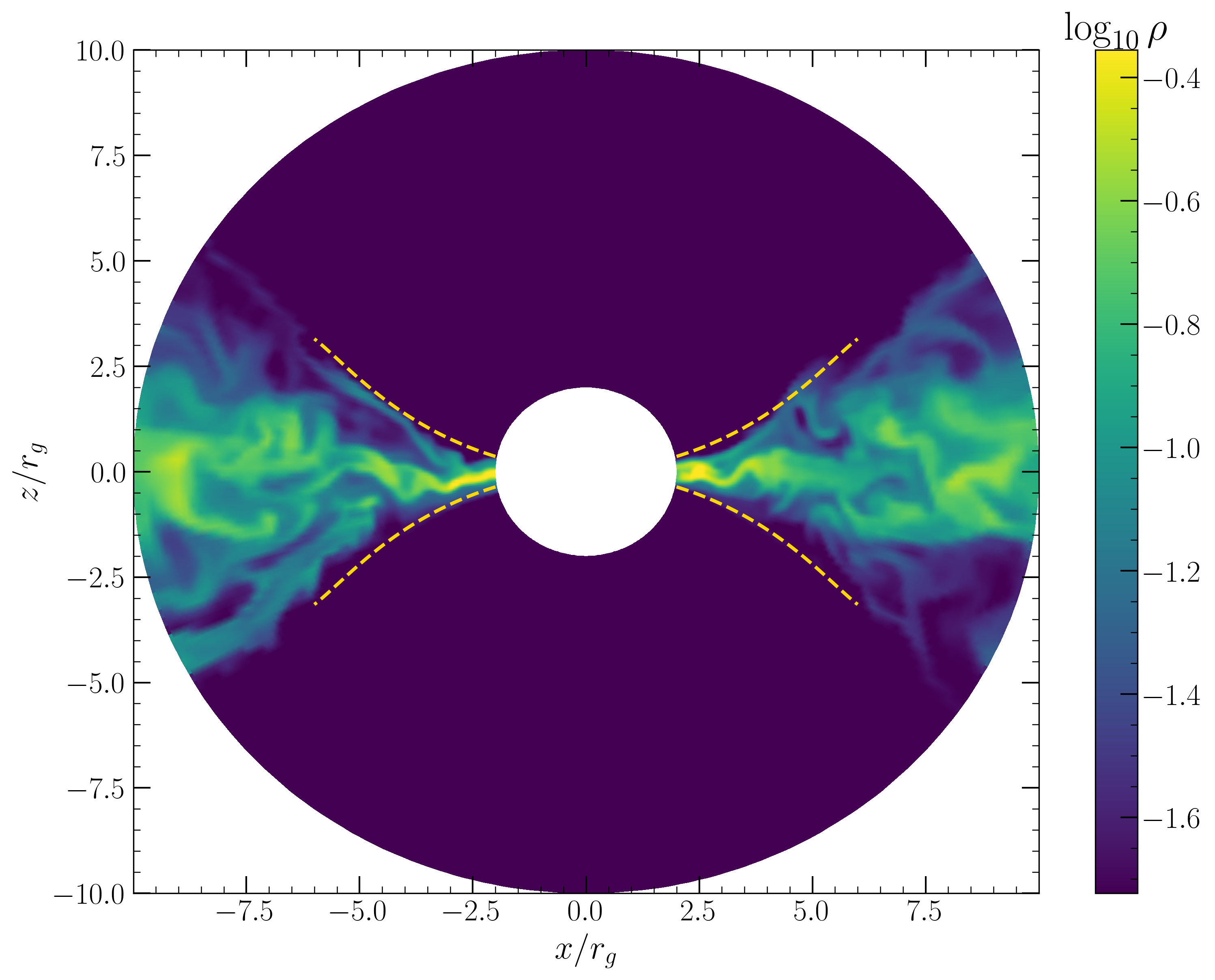}
    \includegraphics[width=.42\linewidth]{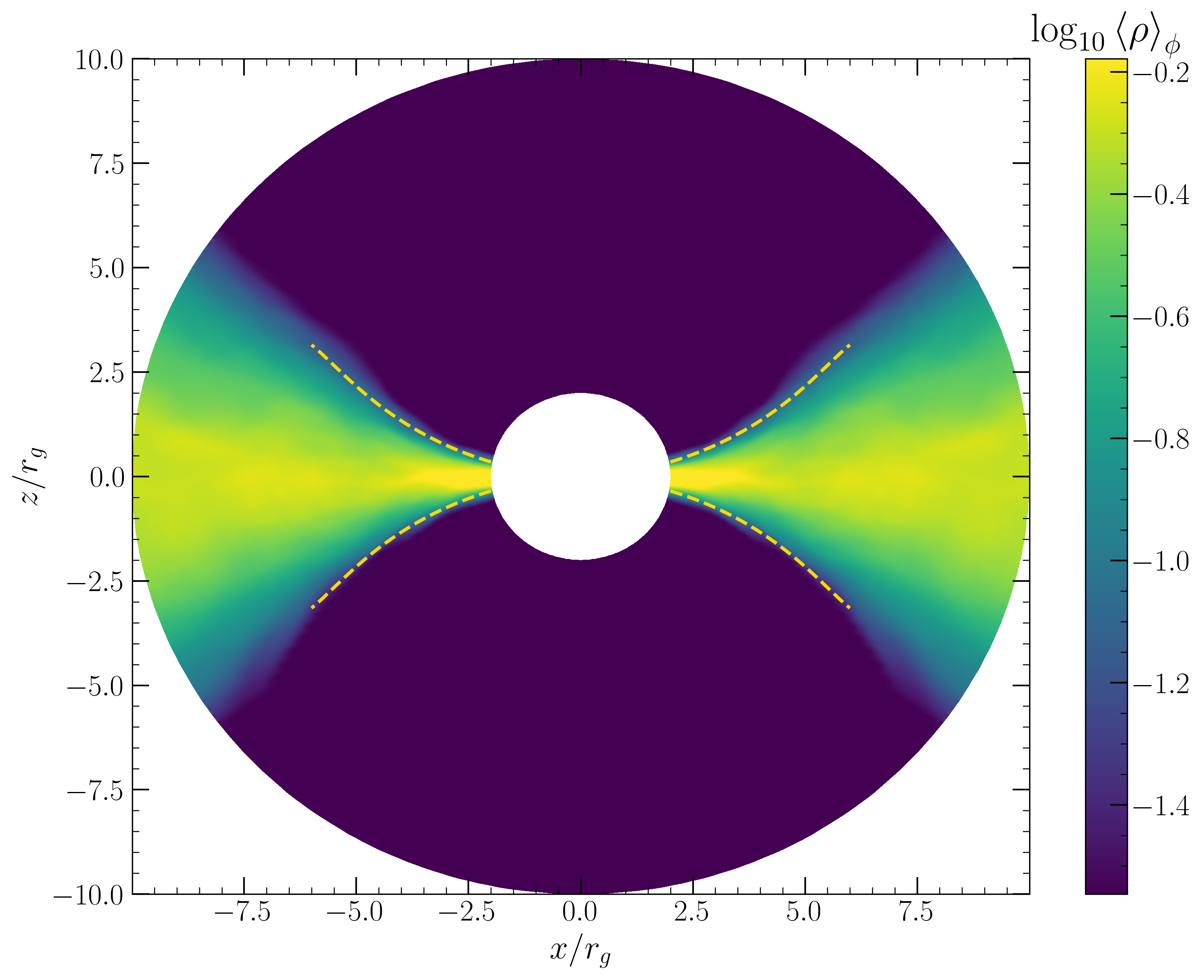}
    \includegraphics[width=.42\linewidth]{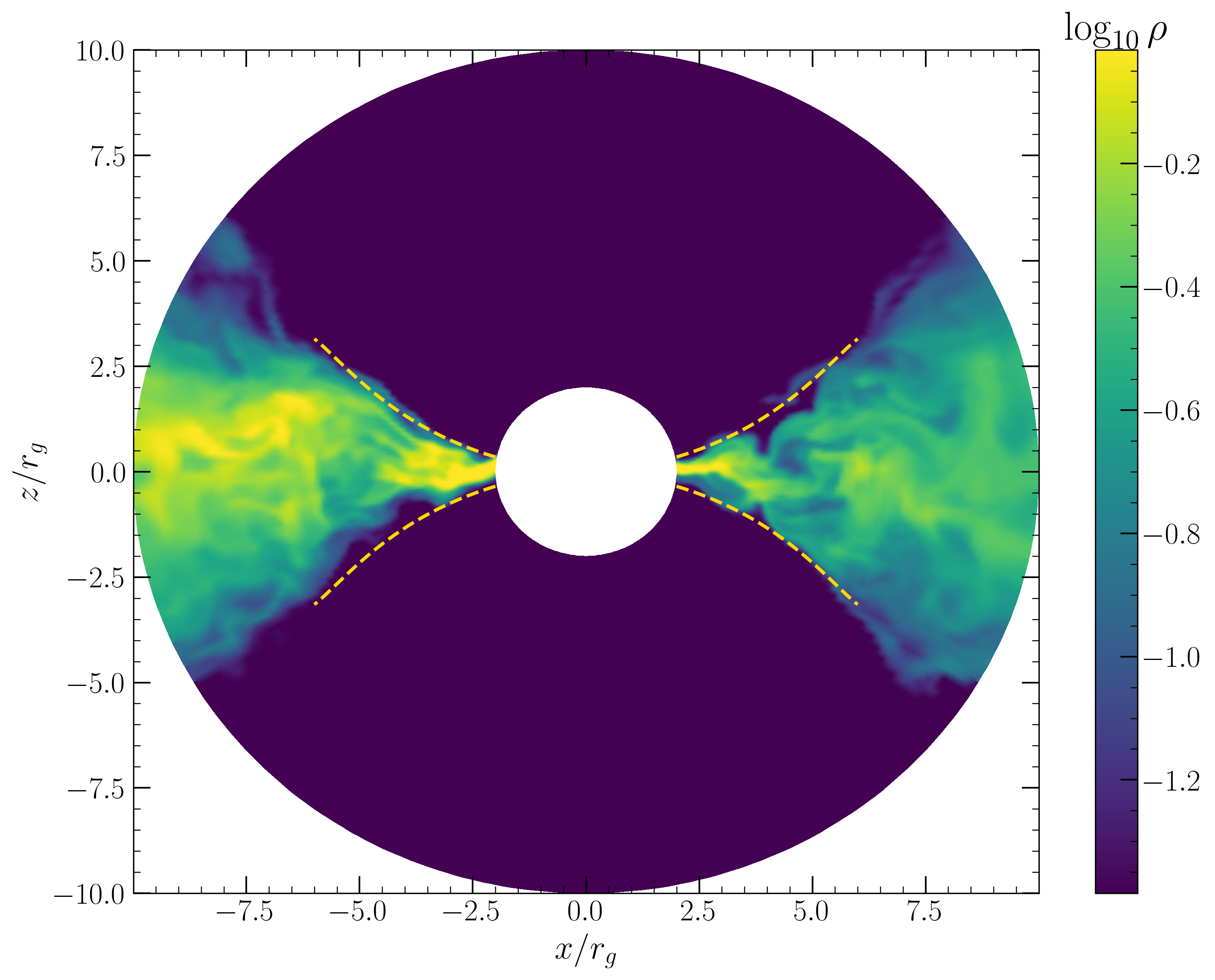}
    \caption{The vertical structure of the {\tt AthenaK} simulation disc density, at two snapshots in time. One snapshot was taken in the early phase $T = 3700$, and one towards the end of the simulation $T = 7300$. The colourmap shows the logarithm of the disc density (in code units), and has been cropped at $3\%$ of the maximum to highlight a particular density contour. In the left hand column we display density profiles averaged over the $2\pi$ azimuth at each snapshot. These angle-averaged snapshots have more structure than the temporally averaged profiles of Figure \ref{fig:averaged_vertical}, but clearly remain well described by the theory (equation \ref{hp}). Even more remarkable are individual slices through the disc plane, with no averaging whatsoever. These $\phi = 0$ slices are shown in the right hand column. Individual constant $\phi$ slices clearly display the turbulent nature of the accretion flow, and yet the structure of equation (\ref{hp}) is already apparent.  }
    \label{fig:vertical_snapshot}
\end{figure*}
\begin{figure}
    \centering
    \includegraphics[width=\linewidth]{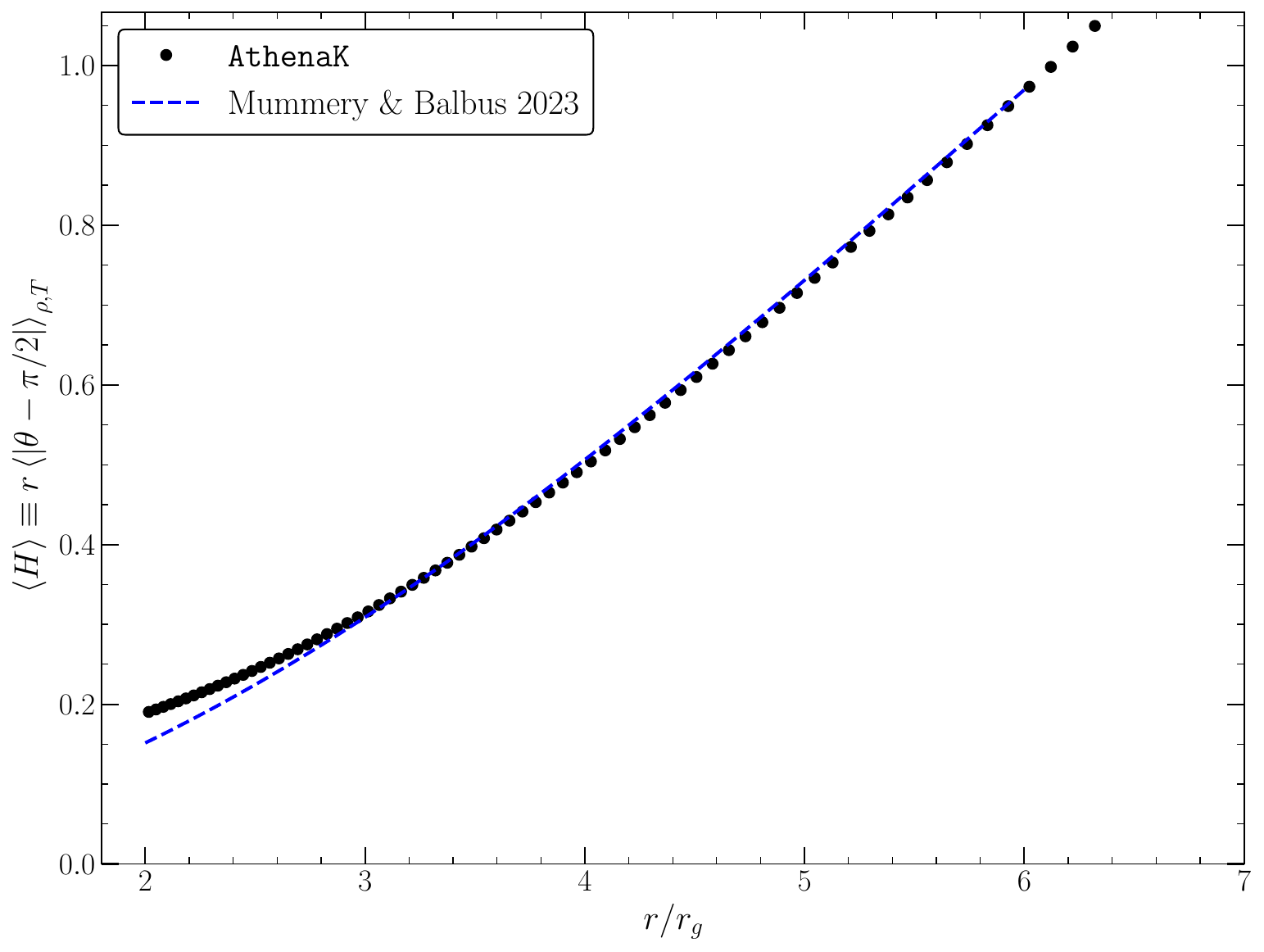}
    \caption{{The time and angle averaged disc density scale height, extracted from the {\tt AthenaK} simulation (solid points, see text)  compared to the theoretical prediction (blue dashed curve). }   }
    \label{fig:scale_height_schwarz}
\end{figure}
As we stressed in section \ref{rzsec} however, the intra-ISCO density solutions do not require that the disc fluid is in a global steady state, merely that the intra-ISCO radial accretion rate is constant. This means that the functional form described above should be accurate at describing individual snapshots of the disc density, in addition to temporal averages. We show that this is the case in Figure \ref{fig:vertical_snapshot}.

\begin{figure*}
    \centering
    \includegraphics[width=.45\linewidth]{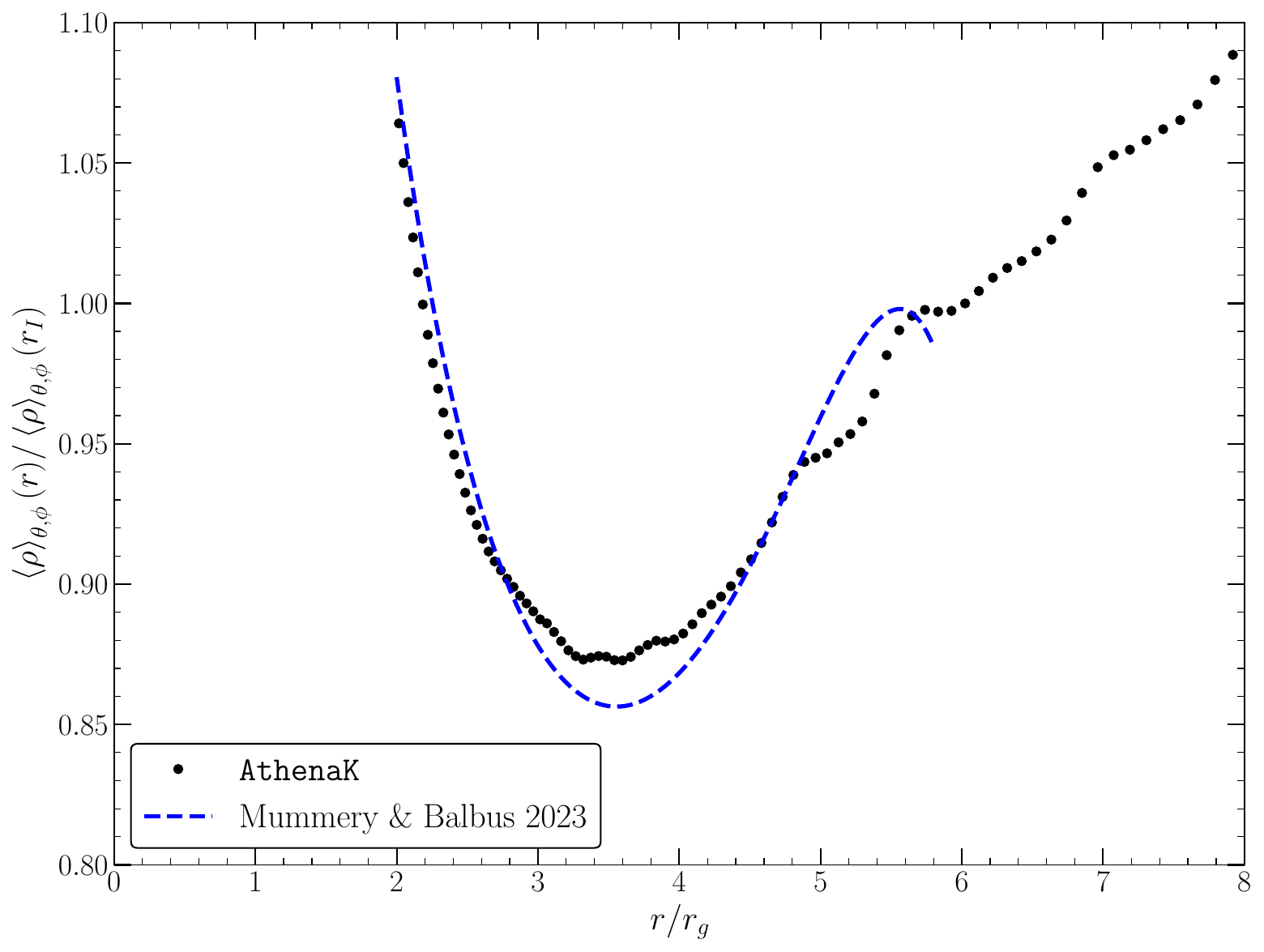}
    \includegraphics[width=.45\linewidth]{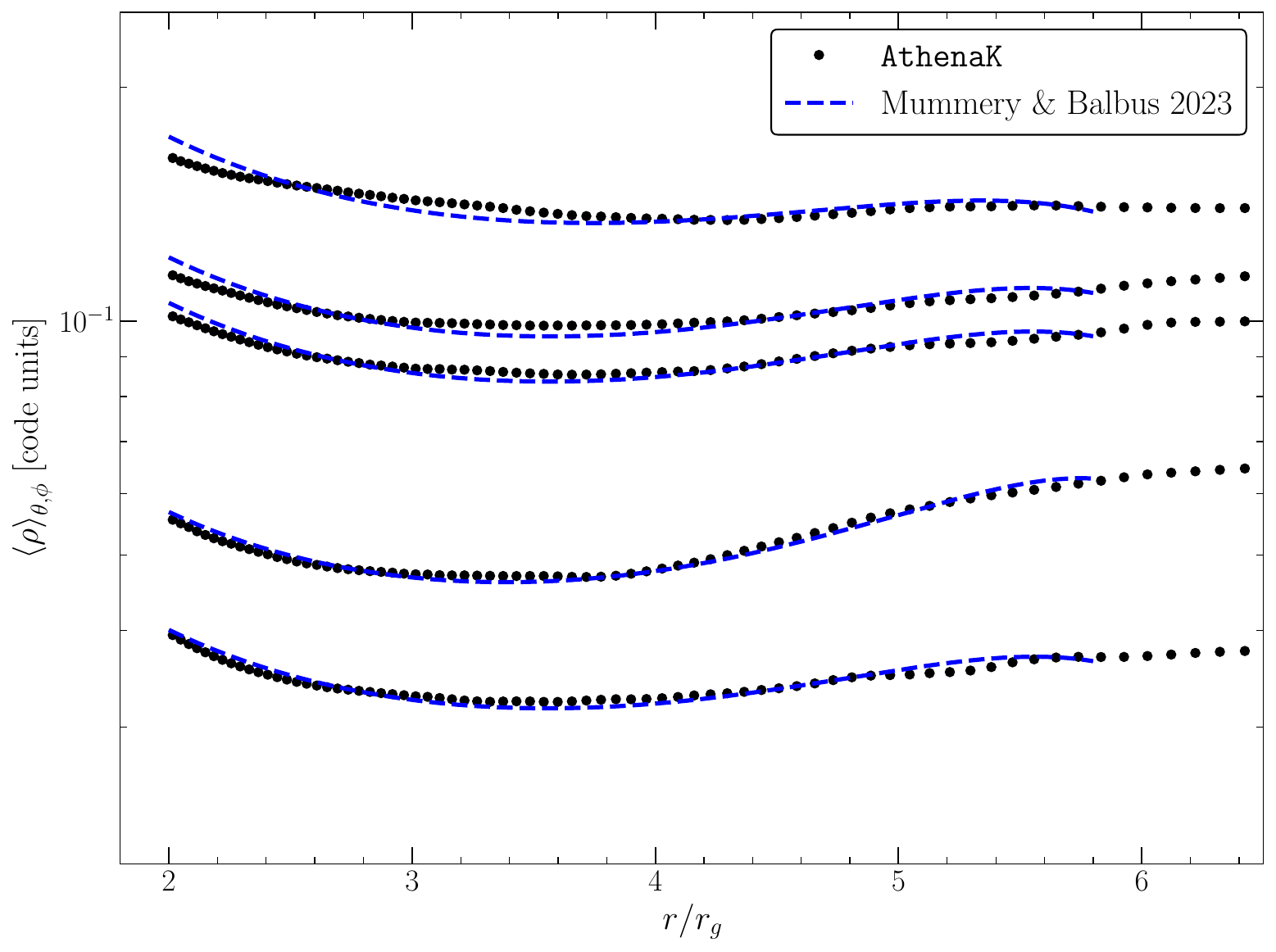}
    \centering
    \includegraphics[width=.45\linewidth]{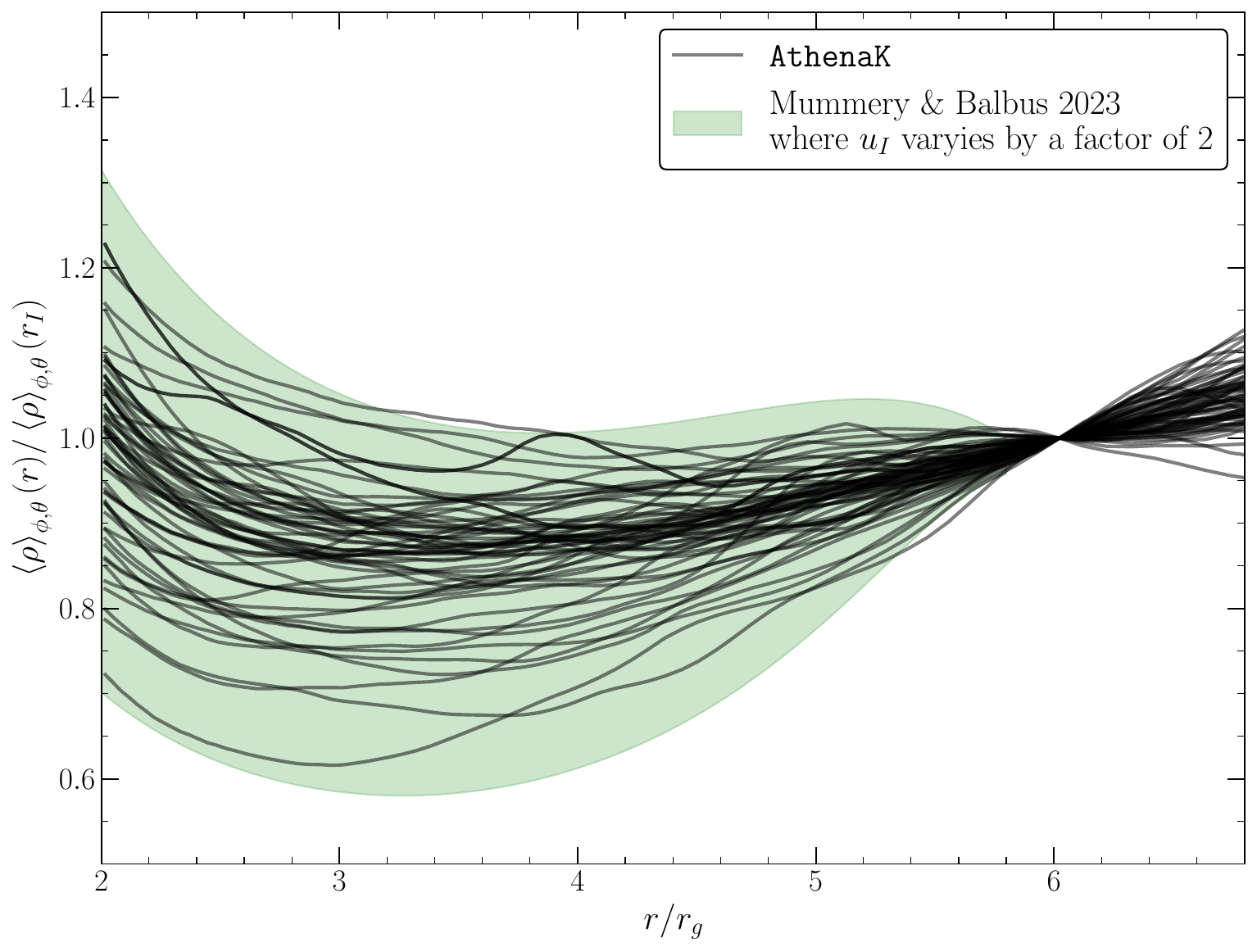}
    \includegraphics[width=.45\linewidth]{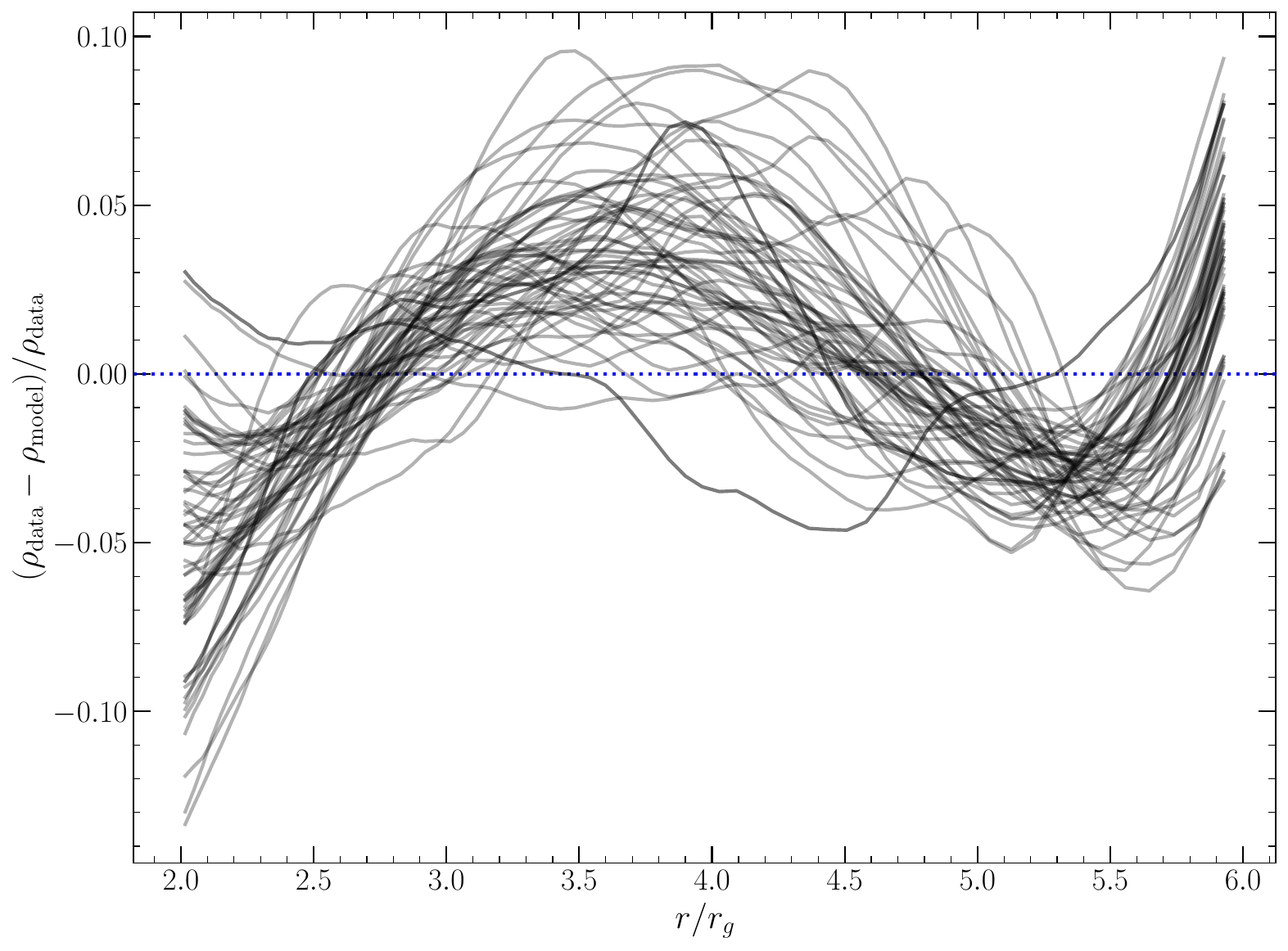}
       \caption{The radial structure of the accretion disc density extracted from the {\tt AthenaK} simulations. On the vertical axis of each plot we display the angle-averaged density $\rhobar_{\theta, \phi}$, and on the horizontal axis we display the Eddington-Finkelstein radial coordinate. Each numerical density curve is extracted from an individual snapshot of the simulation. In the upper left panel we show the density from a particular snapshot ($T = 3700$), normalised by its value at the ISCO, contrasted with the prediction of equation (\ref{rhp}). We see that the qualitative structure of the theory (equation \ref{rhp}) is imprinted on the numerical simulation, which does a remarkably good job at describing the numerical data. This good agreement between theory and simulation is independent of the magnitude of the ISCO density (or more generally the global structure of the flow), as is highlighted in the upper right panel, which shows 5 different snapshots covering early (lowest amplitude) to late (highest amplitude) times.  In the lower left panel we show the density of each snapshot of the final $5000 GM_\bullet/c^3$ of the simulation, normalised by its value at the ISCO. {We overplot as a shaded region curves with the functional form of equation \ref{rhp} with  trans-ISCO velocities $u_I$ which vary by a factor 2.} Each snapshot is qualitatively well described by the theory, and the variance can be explained by turbulent variability in $u_I$.  {In the lower right panel we display the fractional error of a best fitting model fit to each snapshot $\Delta = (\rho_{\rm data}- \rho_{\rm model})/\rho_{\rm data}$, showing that the theory is accurate at the $\sim 10\%$ level at all times in the simulation evolution.}  }
    \label{fig:radial_density}
\end{figure*}

In the left hand column of Figure \ref{fig:vertical_snapshot} we display density profiles averaged over the $2\pi$ azimuth at individual snapshots in time $\log_{10} \rhobar_\phi$. These angle-averaged snapshots have more structure than the temporally averaged profiles of Figure \ref{fig:averaged_vertical}, but clearly remain well described by the theory. Even more remarkable are individual slices through the disc plane, with no averaging whatsoever. These $\phi = 0$ slices are shown in the right hand column of Figure \ref{fig:vertical_snapshot}. Individual constant $\phi$ slices clearly display the turbulent nature of the accretion flow, and yet the structure of equation (\ref{hp}) is already apparent. 

{As an alternative to plotting individual density contours as above, an estimate of the density scale height can be computed directly from the {\tt AthenaK} simulation data itself. An estimate of the disc density scale height is given by  }
\begin{multline}\label{ha}
{\left \langle H \right \rangle \over r} = \left\langle |\theta - \pi/2| \right\rangle_{\rho, T} \\  \equiv  {\int_T^{T+\Delta T}\int_0 ^{2\pi} \int_{0}^\pi \rho \left| \theta - \pi/2 \right| \sin \theta \, {\rm d}\theta \, {\rm d}\phi \, {\rm d}T \over \int_T^{T+\Delta T}\int_0 ^{2\pi} \int_{0}^\pi\rho \, \sin \theta \, {\rm d}\theta \, {\rm d}\phi \, {\rm d}T} ,
\end{multline}
{i.e., the scale height is the density weighted opening angle of the disc, multiplied by the radius. This scale height is plotted in Figure \ref{fig:scale_height_schwarz} (back solid points), and is contrasted with the analytical result (equation \ref{hp}; blue dashed curve). As expected from the above analysis, the analytical theory accurately reproduces the numerical data.  }

\subsubsection{Radial structure}
To compare the radial structure of the intra-ISCO accretion flow extracted from the {\tt AthenaK} simulations with the theory of \cite{MummeryBalbus2023} we construct the angle-averaged quantity $\rhobar_{\theta, \phi}$ at each snapshot. The theoretical prediction for the radial dependence of this quantity is given by 
\begin{equation}
    {\rho \over \rho_I} = \left({6 \over r} \right)^{27/11} \left[ {c \over 3 u_I} \left({6 \over r} - 1\right)^{3/2} + 1\right]^{-9/11} , \label{rhp}
\end{equation}
where we have once again specialised the general expression (equation \ref{rhrh}) to the $\Gamma = 13/9$ and Schwarzschild limits. 

In Figure \ref{fig:radial_density} we plot a number of probes of the radial structure of the accretion disc density extracted from the {\tt AthenaK} simulations. On the vertical axis of each plot we display the angle-averaged density $\rhobar_{\theta, \phi}$, and on the horizontal axis we display the Eddington-Finkelstein radial coordinate. Each numerical density curve is extracted from an individual snapshot of the simulation. In the upper left panel we show the density from a particular snapshot ($T = 3700$), normalised by its value at the ISCO, contrasted with the prediction of equation (\ref{rhp}). We see that the qualitative structure of the theory (equation \ref{rhp}) is imprinted on the numerical simulation, which does a remarkably good quantitative job at describing the numerical data. In particular we note the radial density turning point predicted by \cite{MummeryBalbus2023} at $r \simeq r_I/2$ is reproduced in the simulation.  The best-fitting free parameter $u_I$ was found to be $u_I \simeq 0.043c$. 

This good agreement between theory and simulation is independent of the magnitude of the ISCO density (or more generally the global structure of the flow), as is highlighted in the upper right panel.  In this plot we display fits of equation (\ref{rhp}) to 5 different snapshots ranging from early to late times in the simulation. While the magnitude of the disc density varies by almost an order of magnitude at these different times, the functional form across the plunging region is unchanged. 

To further demonstrate this behaviour, in the lower {left} panel we show the density of each snapshot of the final $5000 GM_\bullet/c^3$ of the simulation, normalised by its value at the ISCO. We overplot as {a shaded region}  the functional form of equation (\ref{rhp}), with  trans-ISCO velocities $u_I$ {which vary by a factor of two}. Each snapshot is qualitatively well described by the theory, and the variance can be explained by turbulent variability in the trans-ISCO velocity $u_I$. {This can be further seen in the lower right panel, where we display the fractional error of a best fitting model fit to each snapshot $\Delta = (\rho_{\rm data}- \rho_{\rm model})/\rho_{\rm data}$, showing that the theory is accurate at the $\sim 10\%$ level at all times in the simulation evolution, although there do appear the be slight systematic deviations in the model fits, suggesting the theory can be improved upon.} Turbulent variability is more pronounced in the radial density structure, when compared to the vertical structure of the previous sub-section, as a result of the stronger functional dependence on $u_I$ of equation (\ref{rhp}) when contrasted with equation (\ref{hp}).  We see that each snapshot of the simulation has a growing density as the flow crosses the horizon ($r_+ = 2$), as predicted theoretically.

\subsubsection{ Azimuthal structure}

\begin{figure}
    \centering
    \includegraphics[width=\linewidth]{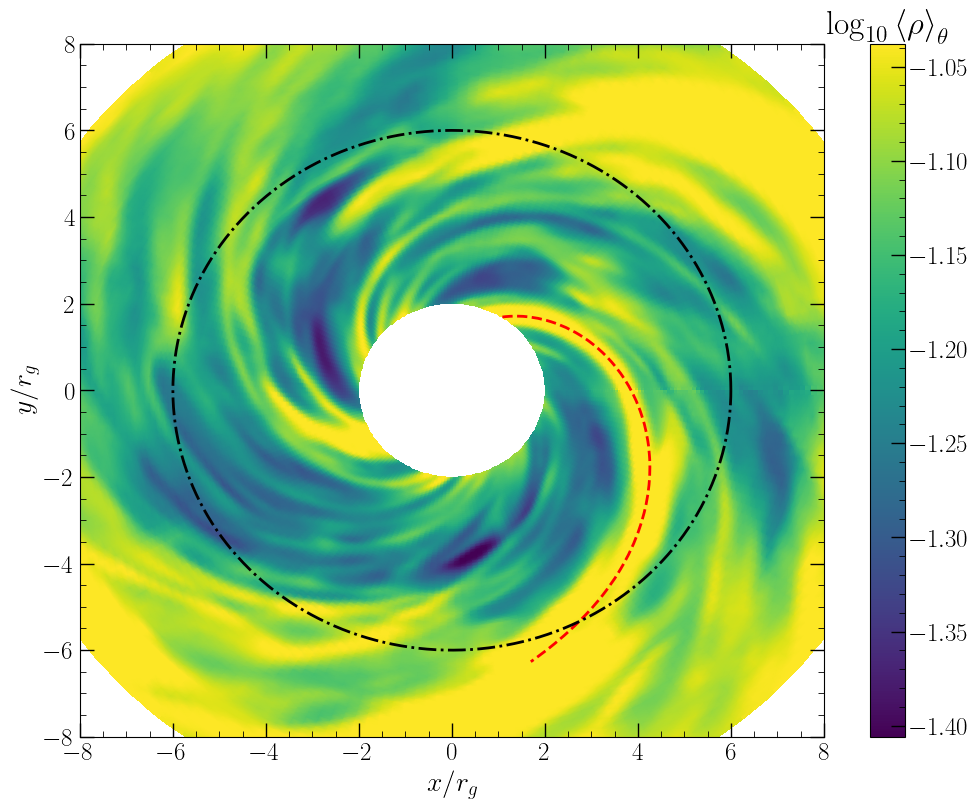}
    \includegraphics[width=\linewidth]{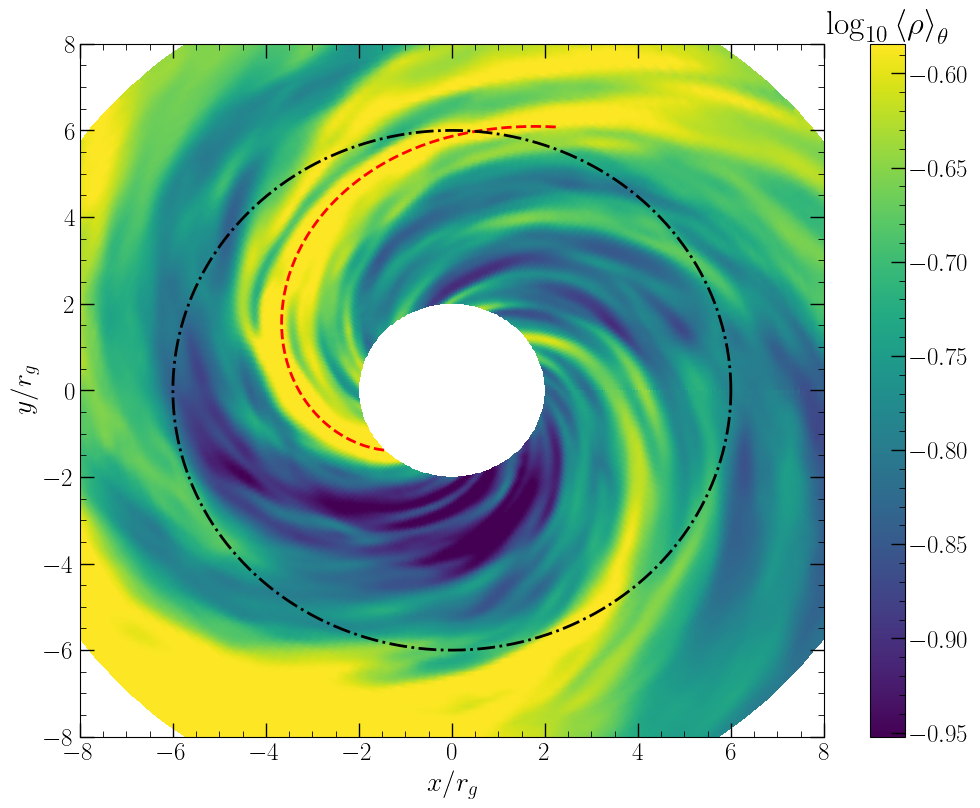}
    \includegraphics[width=\linewidth]{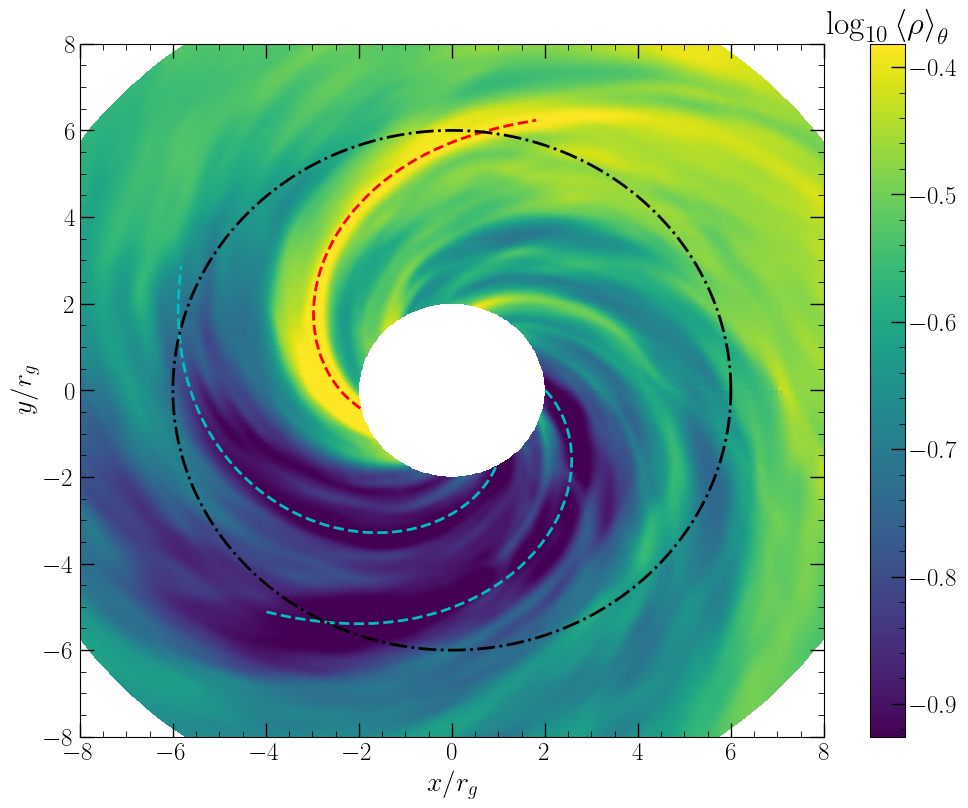}
    \caption{Two dimensional projections of the disc density (averaged over $\theta$ and presented in code units) plotted in the $r-\phi$ plane. Plotted as red dashed curves are spiral arms computed using the geodesic characteristic method of section \ref{spiralsec}, and by a black dashed circle we denote the ISCO radius. In blue we demonstrate how under densities also follow the same characteristic curves.    The Keplarity factors of the spirals are the following: upper panel $f_K = 0.9$, lower panel $f_K = 0.87$, and middle panel $f_K = 0.93$. }
    \label{fig:spirals}
\end{figure}

The prediction of the analytic theory developed in section \ref{spiralsec} is that accretion within the plunging region will be mediated along spiral structures, with a particular shape described by the geodesic characteristic curves (Figure \ref{fig:charac_spiral}).  This prediction is simple to test on a surface level, we simply plot the logarithm of the vertically averaged disc density $\rhobar_\theta$ in the inner disc regions and  inspect visually for spiral structures. Spiral structures are visually clear at all epochs, and a set of examples are shown in Figure \ref{fig:spirals}. By a black dot-dashed circle we display the location of the ISCO, and by red dashed curves we show the spiral characteristics (computed following the techniques developed in section \ref{spiralsec}, with an initial Gaussian perturbation). The colour-bar denotes $\log_{10} \rhobar_\theta$ in code units.  It is clear that the over-densities in the flow follow the characteristic curves developed in this paper. 

{For each epoch in Figure \ref{fig:spirals} we compute analytical models of spiral arm formation with different Keplarity factors $f_K$. The reason for this is that we found that the value of $f_K$ was highly stochastic in the {\tt AthenaK} simulation, varying randomly between $f_K \sim 0.87-0.97$ over the last $\Delta T = 5000 \, [GM/c^3]$  of the run. As such, we display the analytical models which best describe the observed shapes in each epoch, limiting the range of $f_K$ values to that seen in the course of the {\tt AthenaK} simulation.    }

\begin{figure}
    \centering
    \includegraphics[width=\linewidth]{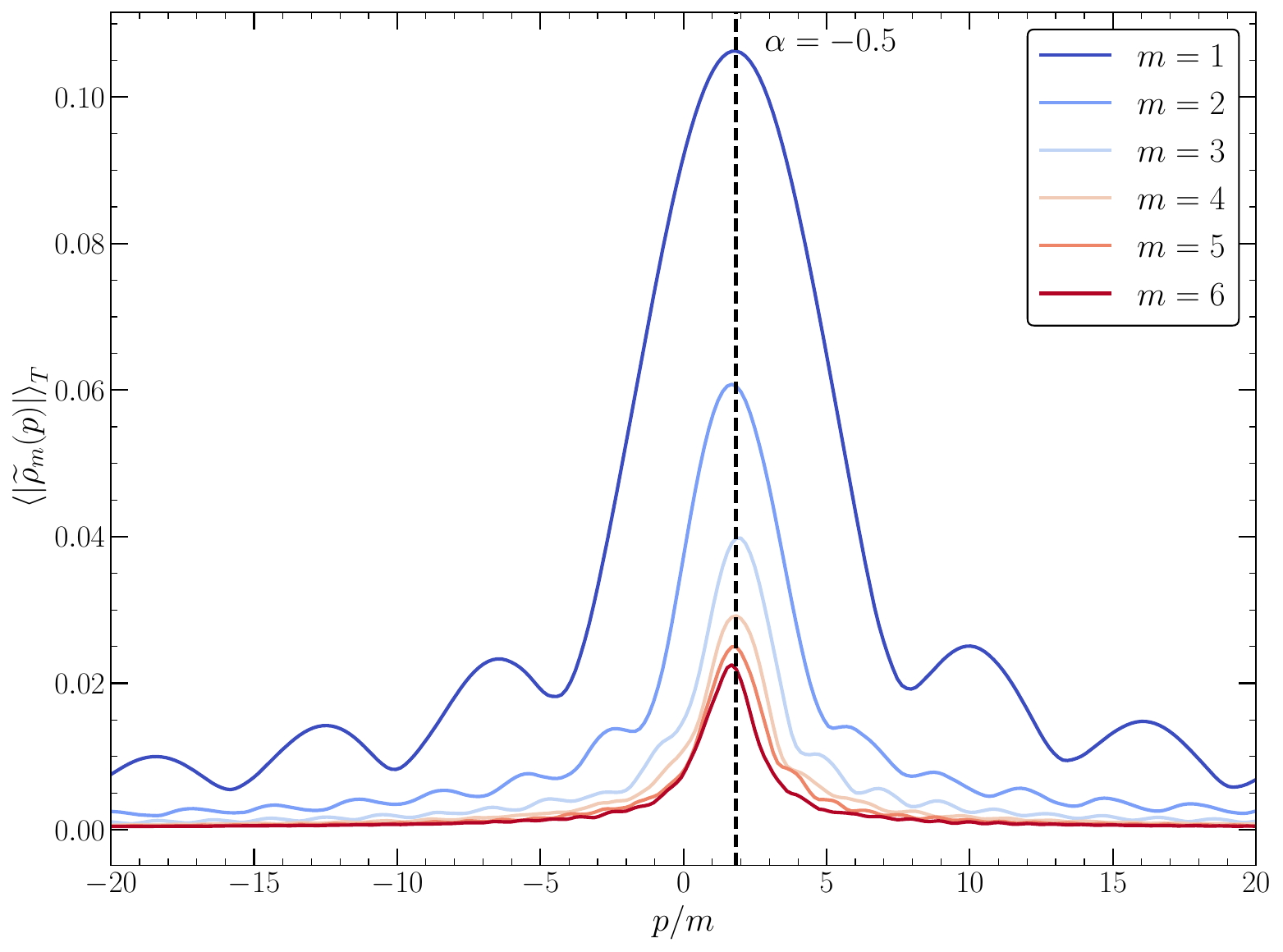}
    \caption{Time averaged Fourier spiral amplitudes of the final 5000 $GM_\bullet/c^3$ of the {\tt AthenaK} simulation, for different numbers of spiral arms ($m$).  For each number of spiral arms the peak of the Fourier amplitudes are clustered around a pitch angle $\alpha = -1/2$, a value predicted from theory (Fig. \ref{fig:spiral-props}). The largest power is in $m=1$ spiral structures, but there is significant power in  spiral structures with more arms. {As this is a time averaged quantity this suggest that for the majority of the evolution the plunging region is dominated by $m = 1$ spiral features, but less frequently $m = 2$ features dominate (in keeping with visual inspection, e.g. Figure \ref{fig:spirals}).}   }
    \label{fig:fourier}
\end{figure}

Interestingly, in the lowest panel we also highlight that under-densities in the flow also follow the same characteristic curves as the over densities. This is simple to understand analytically, as at no point in the derivation of the characteristic curves was the requirement that the density perturbation $\delta \rho$ be positive, and the same shape describes negative perturbations $\delta \rho < 0$. 

It is possible to be more precise and not simply rely on visual analysis to confirm the presence of spiral features in an image. To do this, we borrow analysis techniques used in the galactic observation community \citep[see e.g.,][]{Puerari92}. In observational astronomy it is often relevant to ask whether an image of a galaxy contains spiral features, and as such a set of Fourier techniques have been developed to answer this question. Given an intensity map $I(x, y)$ (in our case this will be the disc density $\rho(x, y)$) of a two dimensional image, one can project this image onto a basis of logarithmic spirals\footnote{Much as standard Fourier techniques decompose signals into a superposition of sinusoidal functions of different frequencies, but do not assume that the underlying signal is a pure sinusoid, this method does not assume that observed spiral structures are logarithmic. It simply decomposes the observed distribution into a superposition of logarithmic spirals of different pitch angles and numbers of arms. } $r = r_0 \exp(-m\phi / p)$ by computing 
\begin{equation}
    \widetilde I_m(p) = {1\over N} \int_{u_{\rm min}}^{u_{\rm max}}\int_{-\pi}^{+\pi} I(u, \phi)\, \exp\left(-i[m\phi + pu]\right) \, {\rm d}\phi \, {\rm d}u , 
\end{equation}
where $u \equiv \ln r$. To interpret this integral note that $m \in \mathbb{Z}^+$ represents the number of spiral arms of the projection, and $p \in \mathbb{R}$ is related to the pitch angle $\alpha$ of the spiral by $\tan \alpha = -m/p$. The factor $N$ is a normalization given by
\begin{equation}
    N =  \int_{u_{\rm min}}^{u_{\rm max}}\int_{-\pi}^{+\pi} I(u, \phi) \, {\rm d}\phi \, {\rm d}u . 
\end{equation}
Formally $u_{\rm max}/u_{\rm min} \to \pm \infty$, but a restricted interval is typically used in practice. In this work we use $r_{\rm min/max} = r_+/r_I$.    Just as in standard Fourier techniques, a ``detection'' of a spiral in an image corresponds to a large amplitude of this Fourier transform $|\widetilde \rho_m(p)|$, with the pitch angle of the most prominent spiral feature corresponding to $\alpha = -\arctan(m/p_{\rm max})$, where $p_{\rm max}$ is the $p$ value of maximum amplitude. 

To examine the properties of the spiral structures observed in our simulations, we compute the time average of the absolute value of this Fourier transformed density  
\begin{equation}
    \left\langle  | \widetilde \rho_m(p) | \right\rangle_T = {1 \over \Delta T}\int_{T}^{T+\Delta T}  |\widetilde \rho_m(p, T') | \,  {\rm d}T', 
\end{equation} 
which we plot in Figure \ref{fig:fourier}. It is clear that there is, throughout the final 5000 $GM_\bullet/c^3$ of the {\tt AthenaK} simulation, a clear signature of spiral features with pitch angle $\alpha \simeq -1/2$. This value of the pitch is predicted from theory (Fig. \ref{fig:spiral-props}). The largest power is in $m=1$ spiral structures, but there is significant power in more complex spiral structures. This suggests that for the majority of the time individual spiral features dominate (see e.g., Figure \ref{fig:spirals}), but spirals with more arms do still have significant non-zero power {and occur a non-zero fraction of the time (cf. Figure \ref{fig:spirals})}.

\begin{figure*}
    \centering
    \includegraphics[width=.49\linewidth]{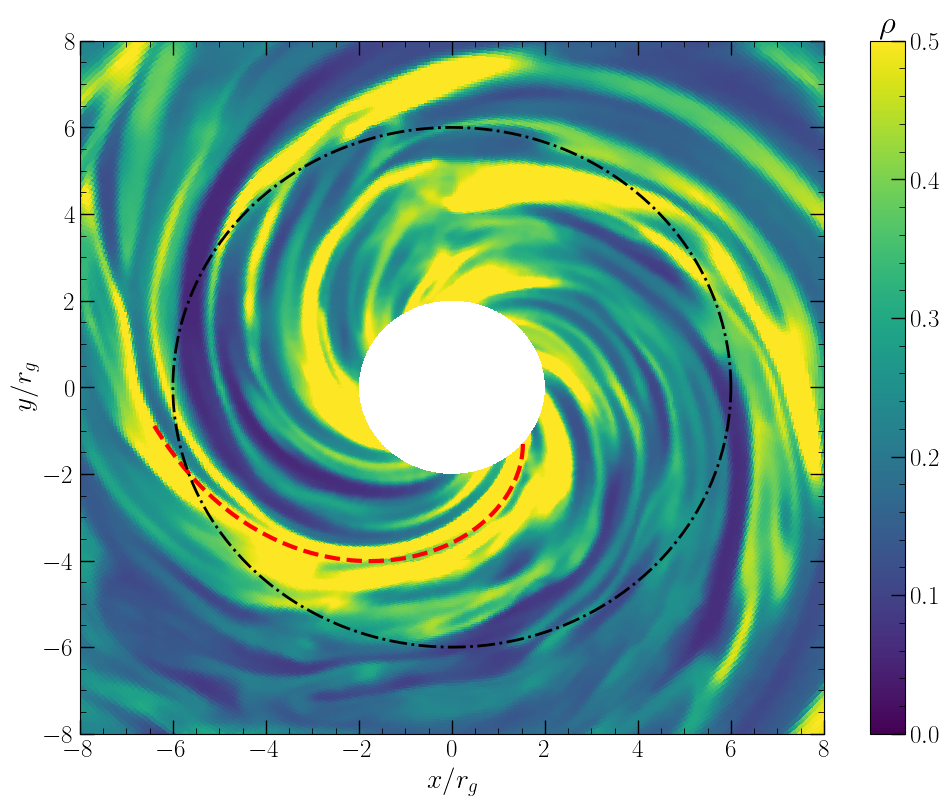}
    \includegraphics[width=.49\linewidth]{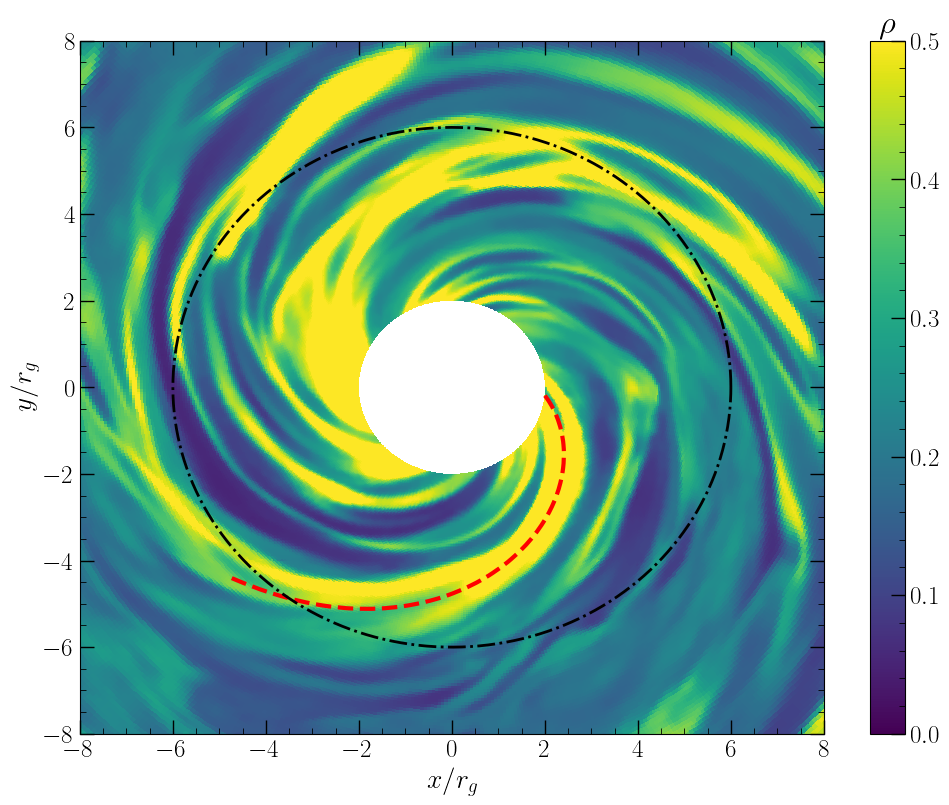}
    \includegraphics[width=.49\linewidth]{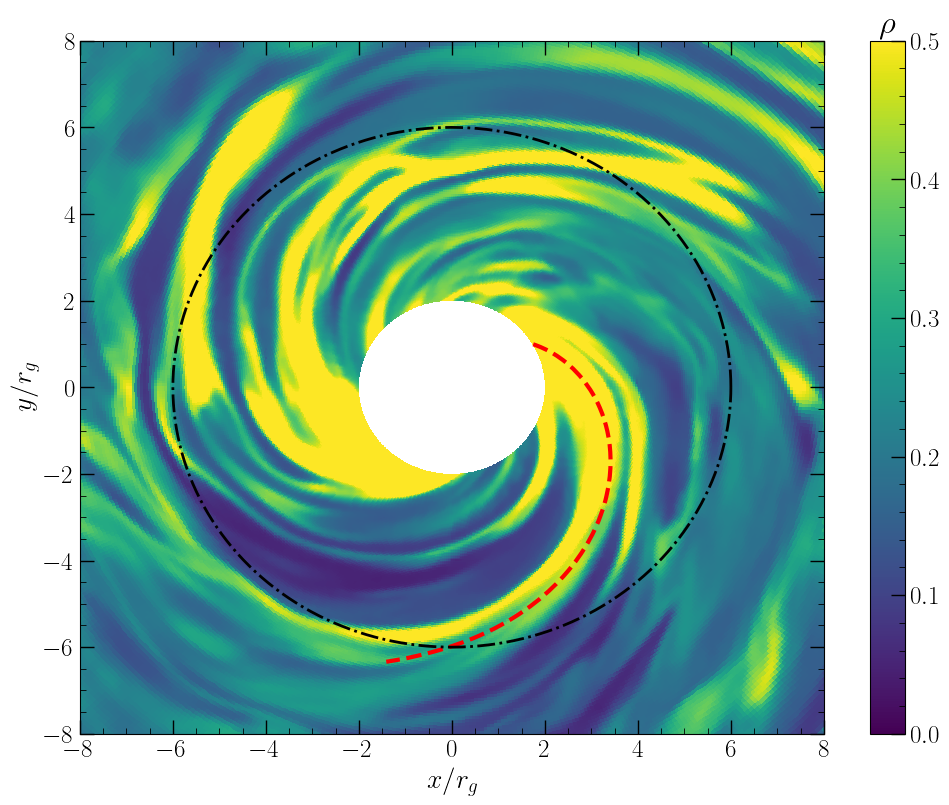}
    \includegraphics[width=.49\linewidth]{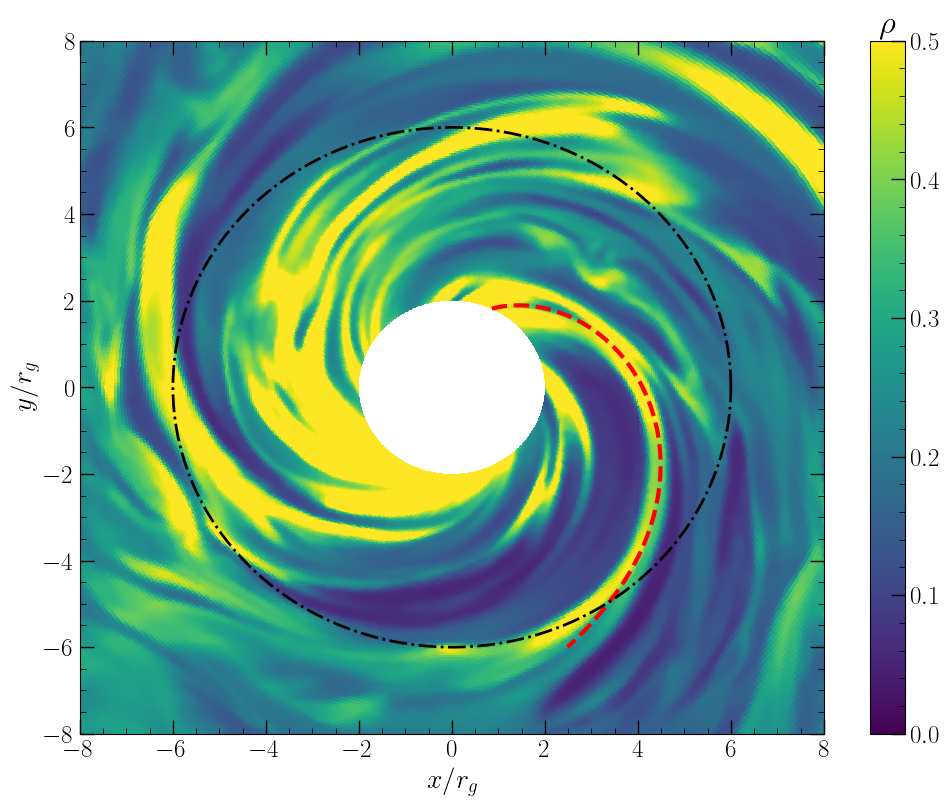}
    \includegraphics[width=.49\linewidth]{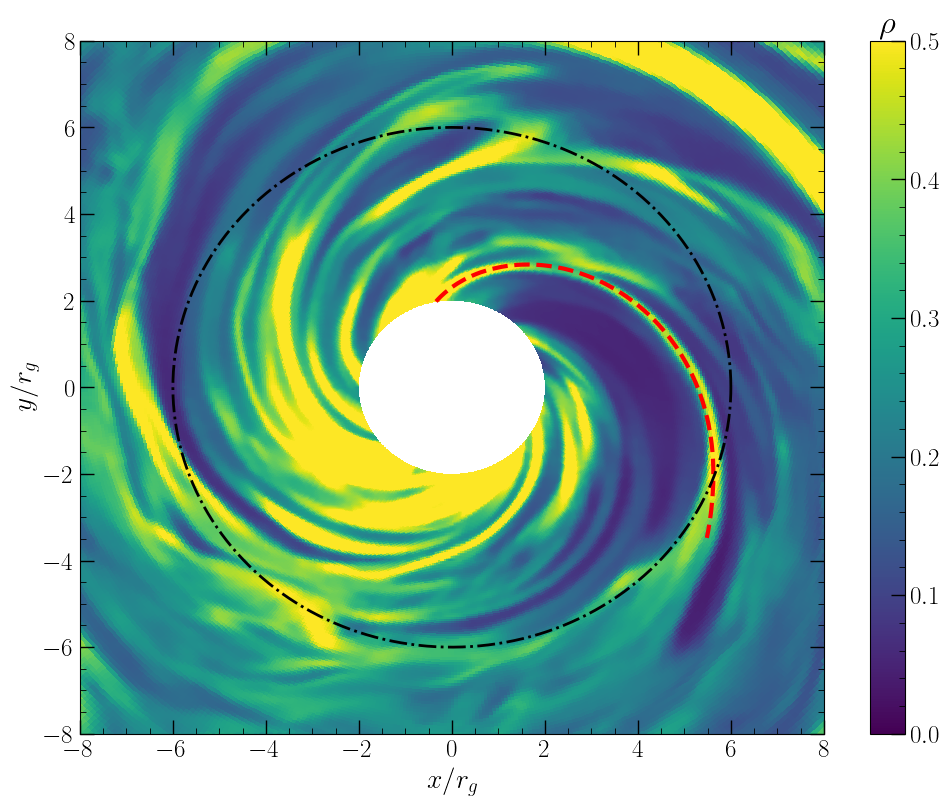}
    \includegraphics[width=.49\linewidth]{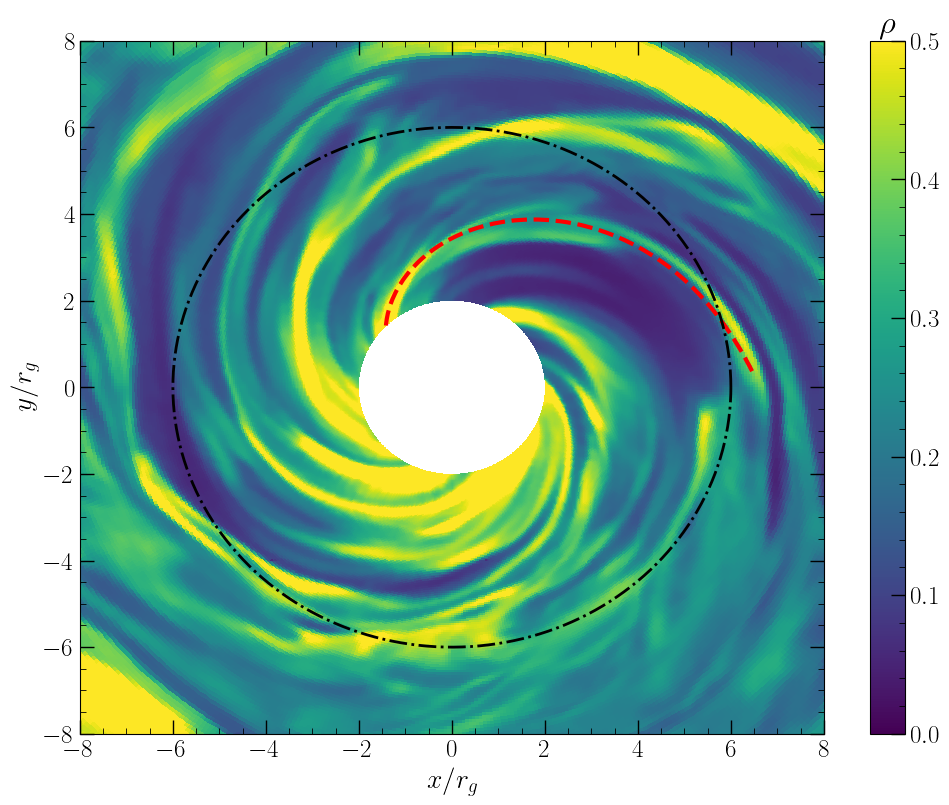}
    \caption{The time evolution of a particular intra-ISCO spiral feature over a timespan of $\Delta T = 50 GM_\bullet/c^3$. Each figure (starting in the upper left panel) is produced $\Delta T_{\rm 2d} = 10 GM_\bullet/c^3$ apart, starting at an initial time of $T_0 = 2950 GM_\bullet/c^3$. The colour-bar denotes the density on a linear scale recorded in a $z=0$ slice, while the red dashed curve shows a characteristic spiral produced with $f_K = 0.9$. This analytical spiral is then rotated at an angular velocity $f_K \Omega_g(r_I)$, which clearly describes well  the propagation of the observed spiral feature in the {\tt AthenaK} simulation.   }
    \label{fig:evolving_spiral}
\end{figure*}

As a final test of the model put forward in this paper, we examine the time evolution of these spiral structures using the two dimensional $z=0$ slice data, which is saved at a higher cadence. While this two-dimensional data is less clean than the height integrated data (owing to the turbulent structure of the disc flow over the scale height, see Fig. \ref{fig:vertical_snapshot}), the spiral structures are still clearly visible. In Figure \ref{fig:evolving_spiral} we display the time evolution of a particular intra-ISCO spiral feature over a timespan of $\Delta T = 50 GM_\bullet/c^3$. The colour-bar denotes the density on a linear scale, while the red dashed curve shows a characteristic spiral produced with $f_K = 0.9$. This spiral feature is then rotated at an angular velocity $f_K \Omega_I$, which clearly well describes the propagation of the observed spiral feature in the {\tt AthenaK} simulation.

\subsection{ Kerr analysis, $a = \pm 1/2$ }  
To verify that there is nothing peculiar about the Schwarzschild $a=0$ spacetime, in this section we perform an identical analysis as before but now for two Kerr spacetimes, with spin values $a = \pm 1/2$. One benefit of considering these two spacetimes is that their ISCO radii differ by a large factor (specifically $r_I(+1/2) \simeq 4.233$, while $r_I(-1/2) \simeq 7.555$), while they share a common event horizon location. Differences in the small radii density profiles can therefore be traced to their differing ISCO radii. 
\begin{figure*}
\centering{}
    \includegraphics[width=.49\linewidth]{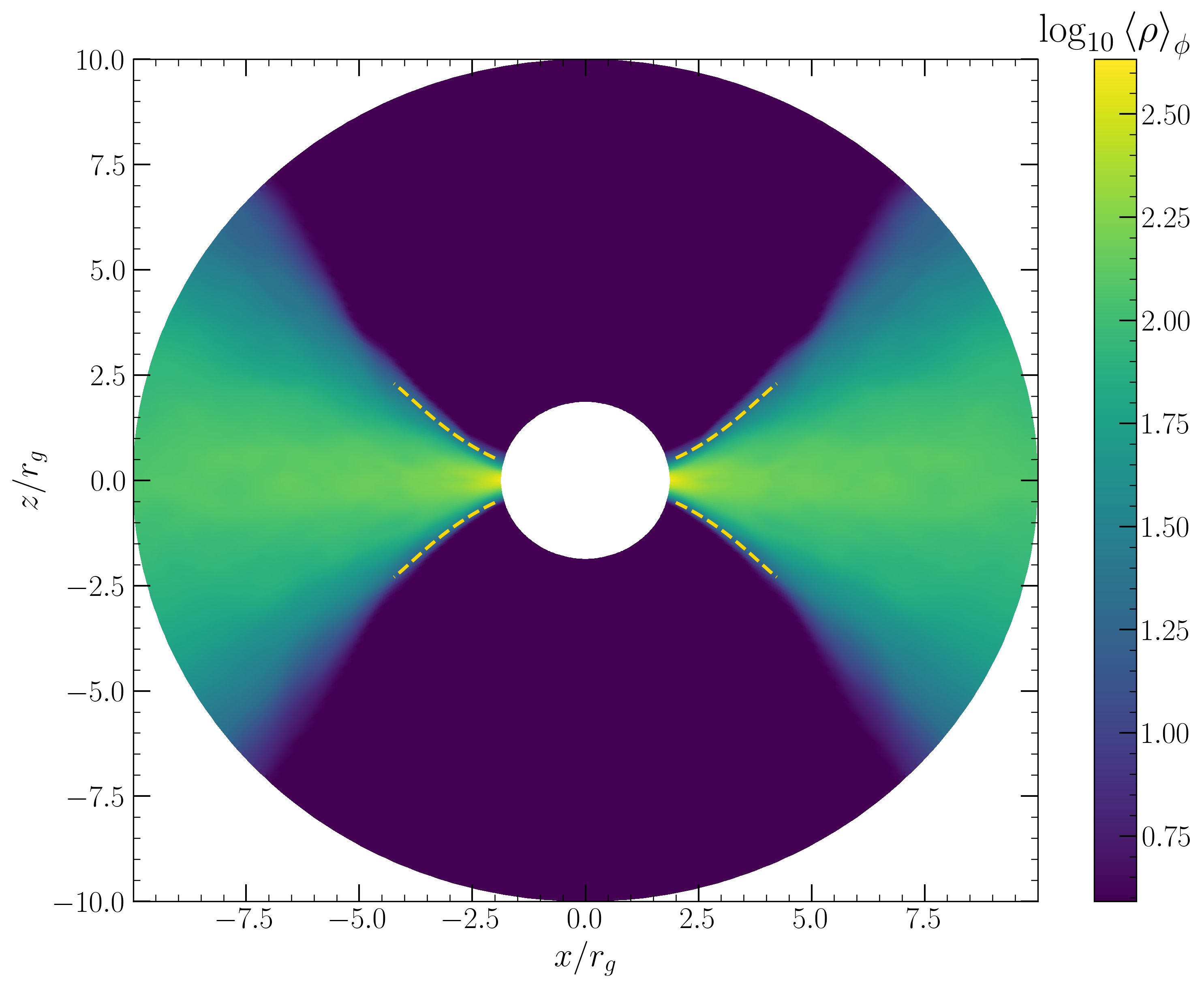}
    \includegraphics[width=.49\linewidth]{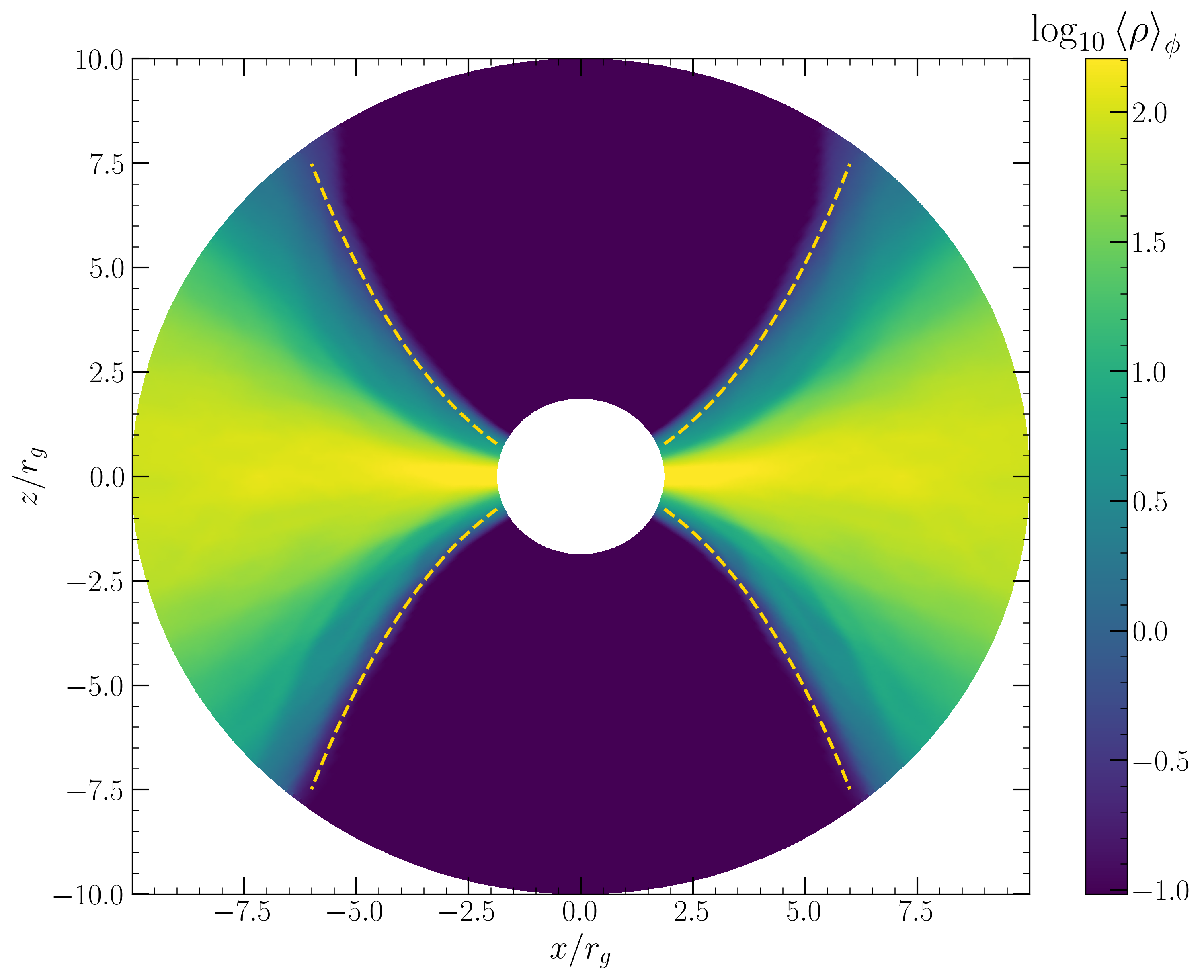}
    \includegraphics[width=.49\linewidth]{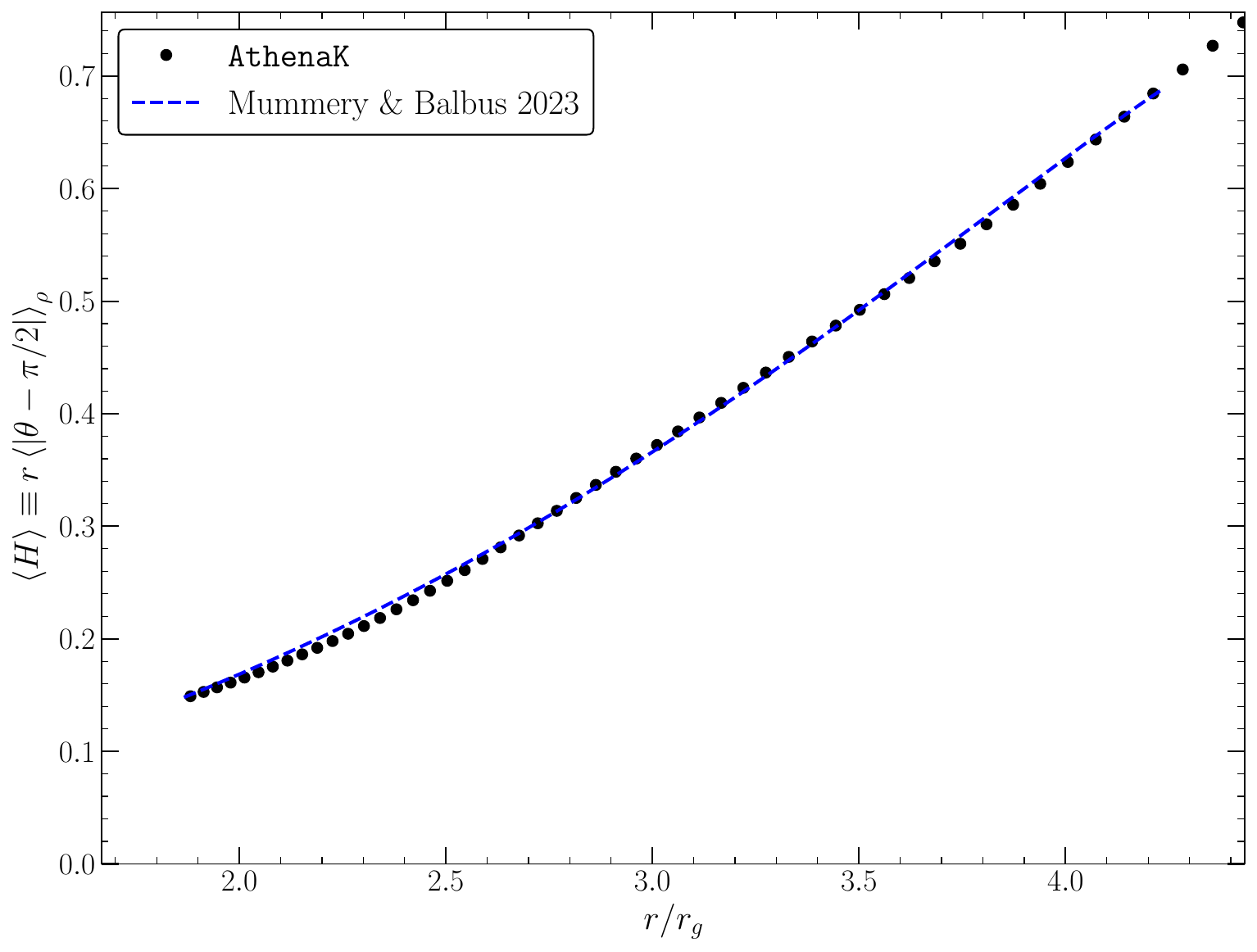}
    \includegraphics[width=.49\linewidth]{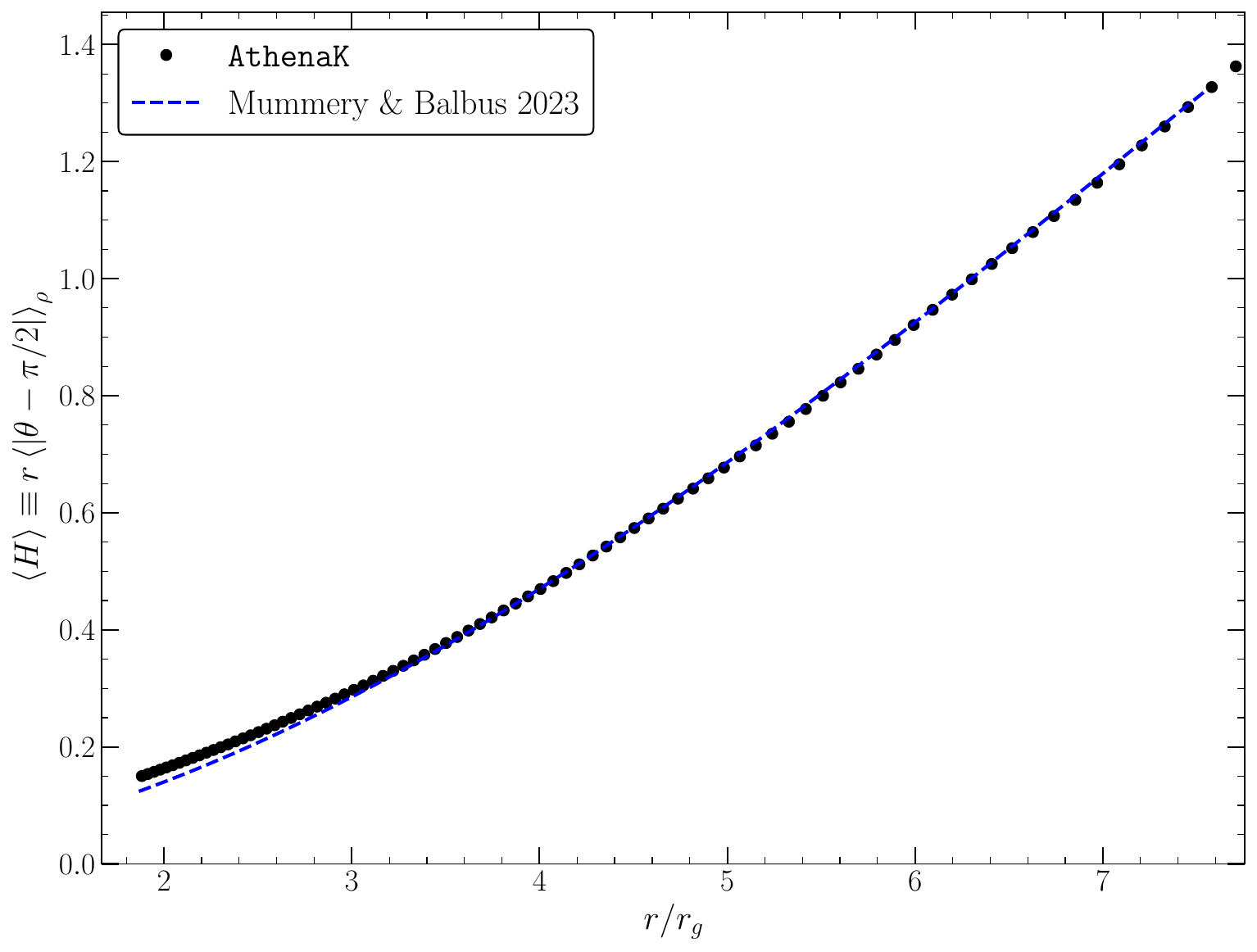}
        \caption{The vertical structure of the $\phi$-averaged density $\rhobar_{\phi}$ extracted from the Kerr black hole {\tt AthenaK} simulations, plotted against Cartesian Kerr-Schild $x-z$ coordinates. {In the upper row we individual contours in the disc, while in the lower panel we show the density scale height computed as in equation \ref{hak}.} On the left we show $a = +1/2$, and on the right $a = -1/2$.  By dashed curves we show how the scale height of the intra-ISCO flow should evolve according to the theory of \citealt{MummeryBalbus2023}, which describes the simulation results extremely well.   }
    \label{fig:kerr_vertical}
\end{figure*}
\subsubsection{Averages}
The Kerr metric in spherical Kerr-Schild coordinates has a series of non-trivial off diagonal components, which means some care must be taken when computing averages (see Appendix \ref{KerrApp} for an explicit list of the metric coefficients).  Analogously to in the Schwarzschild sub-section we define the following averages which are of particular interest 
\begin{align}
    \Xbar_{\phi} &\equiv  { \int_{0}^{2\pi} \sqrt{g_{\phi\phi}} \, X \, {\rm d}\phi \over \int_{0}^{2\pi} \sqrt{g_{\phi\phi}}  \, {\rm d}\phi }, \\
    \Xbar_{\theta} &\equiv   { \int_{0}^{\pi} \sqrt{g_{\theta\theta}} \, X \, {\rm d}\theta \over \int_{0}^{\pi} \sqrt{g_{\theta\theta}}  \, {\rm d}\theta }, \\
    \Xbar_{T} &\equiv  {\int_{T}^{T+\Delta T} \sqrt{g_{TT}}\, X \,  {\rm d}T' \over\int_{T}^{T+\Delta T} \sqrt{g_{TT}} \,  {\rm d}T'} .  
\end{align}
 Double averages will also be computed, which are given by 
\begin{align}
    \Xbar_{\theta , \phi} &\equiv { \int_{0}^{2\pi} \int_{0}^{\pi} \sqrt{g_{\phi\phi}g_{\theta\theta}} \, X \, {\rm d}\theta \, {\rm d}\phi \over \int_{0}^{2\pi}  \int_{0}^{\pi}  \sqrt{g_{\phi\phi}g_{\theta\theta}}  \, {\rm d}\theta\, {\rm d}\phi }, \\
        \Xbar_{\theta ,T } &\equiv { \int_{T}^{T+\Delta T} \int_{0}^{\pi} \sqrt{g_{TT}g_{\theta\theta}} \, X \, {\rm d}\theta \, {\rm d}T' \over \int_{T}^{T+\Delta T} \int_{0}^{\pi} \sqrt{g_{TT}g_{\theta\theta}} \, {\rm d}\theta \, {\rm d}T' }, \\
    \Xbar_{\phi, T} &\equiv { \int_{T}^{T+\Delta T} \int_{0}^{2\pi} \sqrt{g_{TT}g_{\phi\phi} - g_{T\phi}^2} \, X \, {\rm d}\phi \, {\rm d}T' \over \int_{T}^{T+\Delta T} \int_{0}^{2\pi} \sqrt{g_{TT}g_{\phi\phi} - g_{T\phi}^2} \, {\rm d}\phi \, {\rm d}T' } ,
\end{align}
where the final integral is non-trivial owing to the mixing of $\phi$ and $T$ coordinates in the Kerr metric.  As in the Schwarzschild metric the variable $\Xbar$ remains a function of all quantities which have not been averaged over.

\subsubsection{ Vertical structure }
\begin{figure*}
\centering{}
    \includegraphics[width=.49\linewidth]{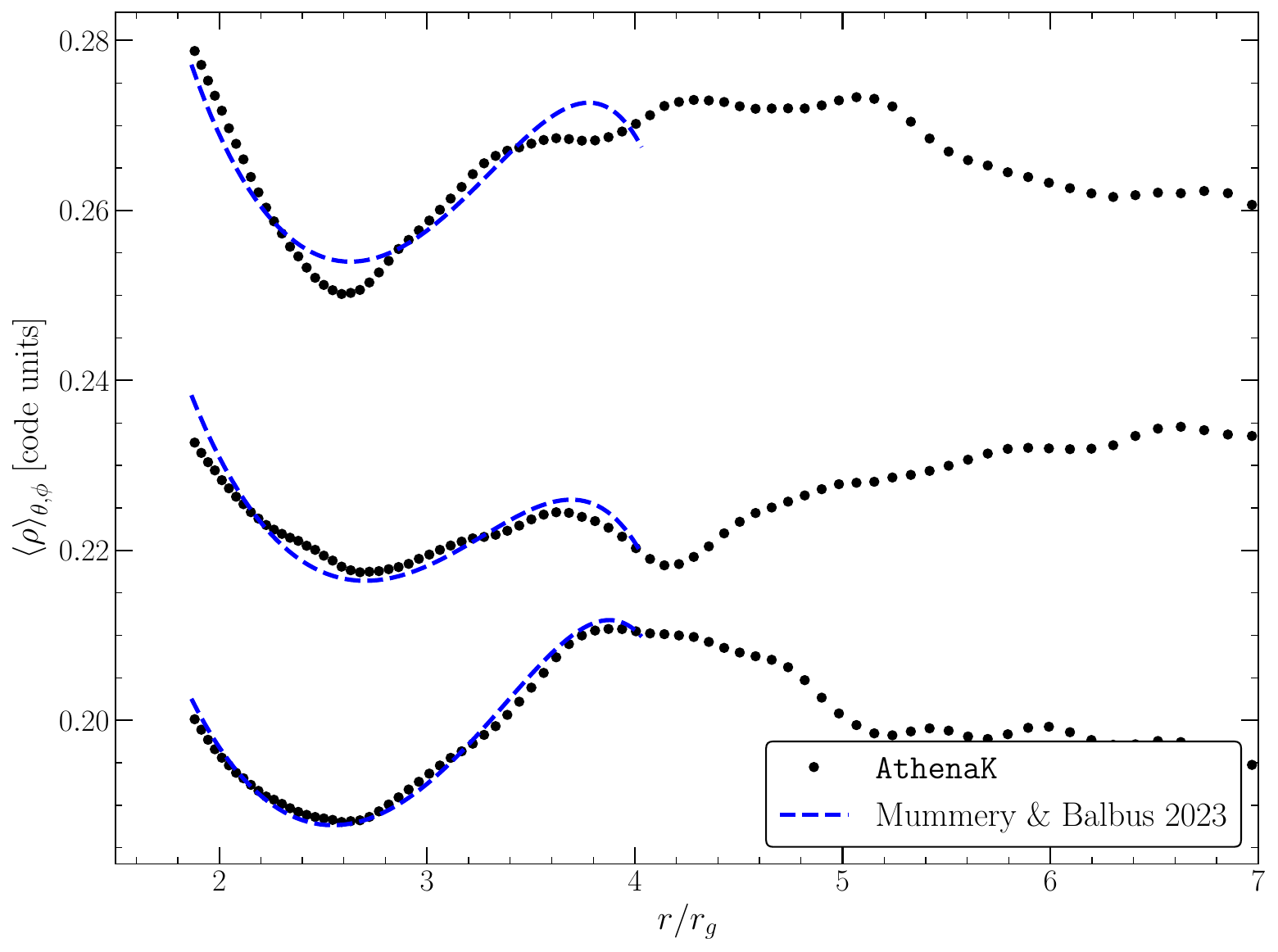}
    \includegraphics[width=.49\linewidth]{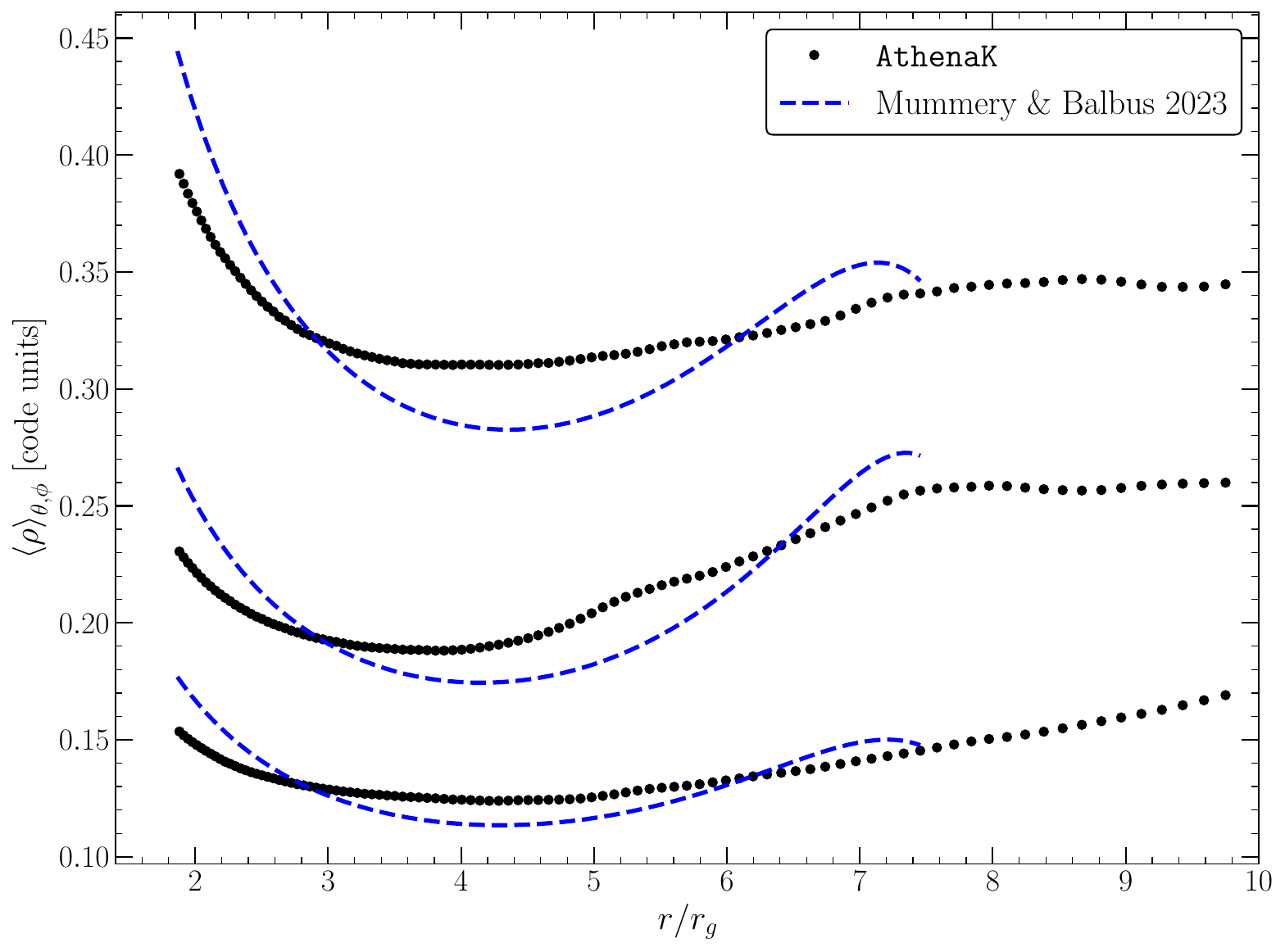}
        \caption{The radial structure of the accretion disc density extracted from the Kerr black hole {\tt AthenaK} simulations. On the vertical axis of each plot we display the angle-averaged density $\rhobar_{\theta, \phi}$, and on the horizontal axis we display the Spherical Kerr-Schild radial coordinate. The different epochs have been offset by a multiplicative constant to show more clearly the structure. Each numerical density curve is extracted from an individual snapshot of the simulation.  The left panel shows three snapshots from the prograde $a = +1/2$ Kerr metric, while the right panel shows three  snapshots from the retrograde $a = -1/2$ Kerr metric. The  disc theory works best for higher spins, with smaller plunging regions, but is {in} qualitative agreement with simulations of larger negative spins.  }
    \label{fig:kerr_radial}
\end{figure*}

We being with an examination of the vertical structure of the flow. As shown in the Schwarzschild spacetime, it is sufficient to consider the azimuthally averaged disc density $\rhobar_\phi$ of individual snapshots, as time-averaging barely changes the results.

The analytical theory of \cite{MummeryBalbus2023} predicts that the vertical density contours should follow (for $\Gamma = 13/9$)
\begin{equation}\label{hpk}
    {H \over H_I} = \left({r_I\over r}\right)^{-16/11} \left[ \sqrt{2c^2 \over 3r_I u_I^2} \left({r_I \over r} - 1\right)^{3/2} + 1\right]^{-2/11} .
\end{equation}
The results of the {\tt AthenaK} simulation are shown in Figure \ref{fig:kerr_vertical}, overplotted with the analytical theory (dashed curves). The prograde disc ($a = +1/2$) is shown on the left, and on the right we display $a = -1/2$. The analytical theory describes the data extremely well, even for the  discs which are extremely thick at the ISCO, like the retrograde $a = -1/2$ spacetime. As in the Schwarzschild spacetime, these profiles are relatively insensitive to $u_I$, which we fixed to those values found for fits to the radial and azimuthal structure (presented later). Similarly good fits to the vertical structure were found at all epochs, the particular epochs displayed are simply those of the final timestep of the evolution. {In the lower panels we display the density scale height, computed directly from the {\tt AthenaK} simulations }
\begin{equation}\label{hak}
{\left \langle H \right \rangle \over r} = {\int_0 ^{2\pi} \int_{0}^\pi \rho \left| \theta - \pi/2 \right| \sqrt{g_{\phi\phi}g_{\theta\theta}} \, {\rm d}\theta \, {\rm d}\phi \over \int_0 ^{2\pi} \int_{0}^\pi\rho \,\sqrt{g_{\phi\phi}g_{\theta\theta}} \,  {\rm d}\theta \, {\rm d}\phi } ,
\end{equation}
{which is the Kerr-generalisation of the Schwarzschild analysis presented earlier. }
\subsubsection{ Radial structure }
By averaging over both Kerr-Schild angles $\theta$ and $\phi$, the radial structure of the disc density can be examined. The analytical prediction of \cite{MummeryBalbus2023} is 
\begin{equation}
    {\rho \over \rho_I} =      \left({r_I\over r}\right)^{27/11} \left[ \sqrt{2c^2 \over 3r_I u_I^2} \left({r_I \over r} - 1\right)^{3/2} + 1\right]^{-9/11} , \label{rhpk}
\end{equation}
for $\Gamma = 13/9$.

In Figure \ref{fig:kerr_radial} we show snapshots of the disc density as a function of Kerr-Schild radial coordinate. These epochs represent the last three of each simulation, each offset by a multiplicative constant to show more clearly the structure.  While it is clear that the simulations are in qualitative agreement with the theory for both black hole spins (in particular the turning point in the density at roughly $r \sim r_I/2$), the  disc theory works best for higher spins, with smaller plunging regions. 

Once again we emphasise that there is nothing special about the epochs we have chosen to display here, and that all epochs are in qualitative agreement with the theory. The variation in profiles within the plunging region can be explained by turbulent variability in the trans-ISCO velocity $u_I$ (cf. the lower panel of Figure \ref{fig:radial_density}).

\subsubsection{ Azimuthal structure }
\begin{figure*}
\centering{}
    \includegraphics[width=.49\linewidth]{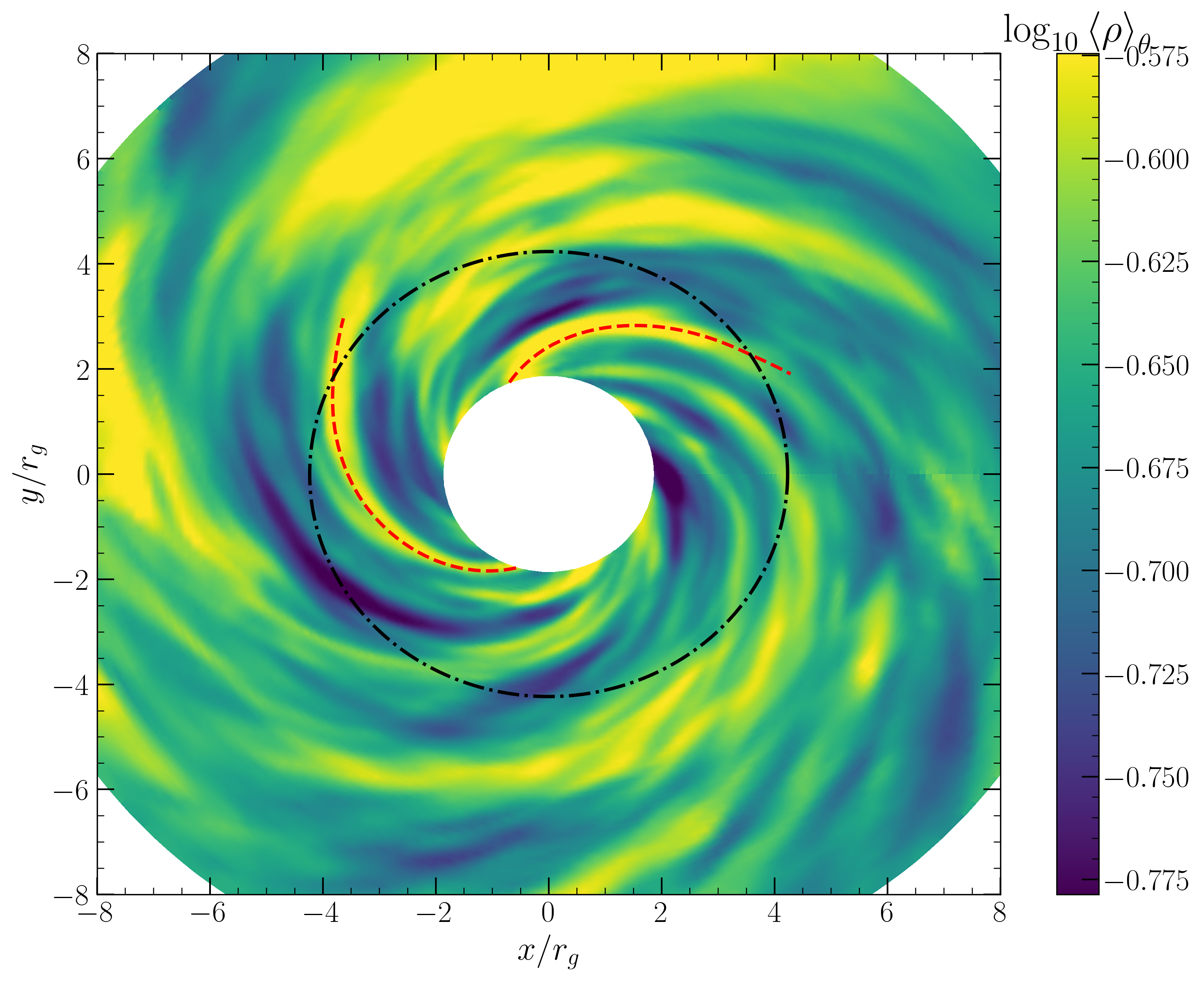}
    \includegraphics[width=.49\linewidth]{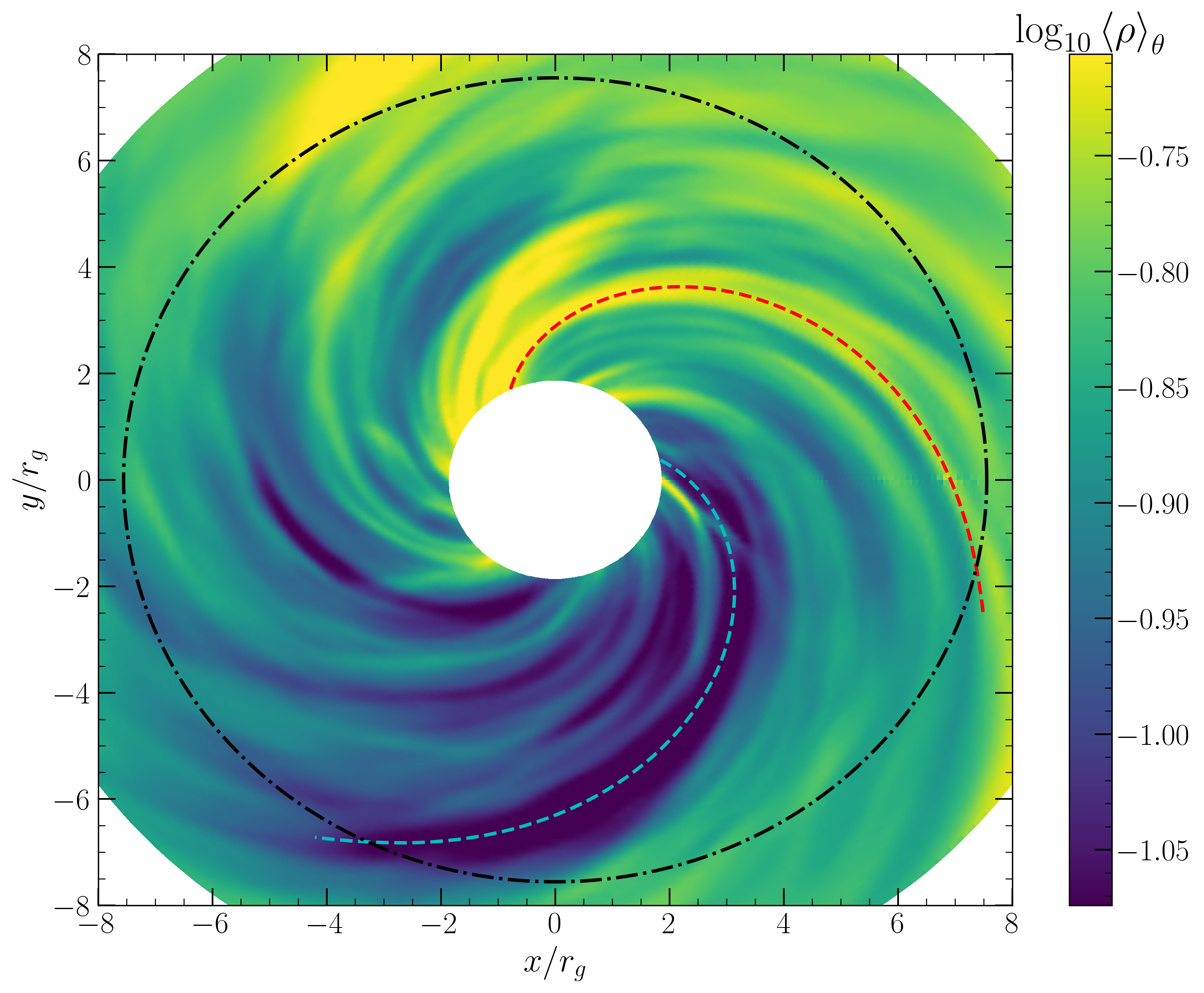}
        \caption{The azimuthal structure of the $\theta$-averaged density $\rhobar_{\theta}$ extracted from the Kerr black hole {\tt AthenaK} simulations, plotted against Cartesian Kerr-Schild $x-y$ coordinates. On the left we show $a = +1/2$, and on the right $a = -1/2$. By dashed curves we highlight how the over (red) and under (blue) densities in the inner flow follow the characteristic spirals derived in this paper.  }
    \label{fig:kerr_azimuthal}
\end{figure*}

Finally, we turn to the azimuthal structure of the Kerr accretion flows. This angular structure in particular highlights the differences induced in the flows  behaviour  by a change in ISCO radius. In Figure \ref{fig:kerr_azimuthal} we show the vertically averaged disc density $\rhobar_{\theta}$ extracted from the Kerr black hole {\tt AthenaK} simulations, plotted against Cartesian Kerr-Schild $x-y$ coordinates. On the left we show $a = +1/2$, and on the right $a = -1/2$. By dashed curves we highlight how the over (red) and under (blue) densities in the inner flow follow the characteristic spirals derived in this paper.  {We again find that the Keplarity factor $f_K$ underwent a highly stochastic evolution over the course of the simulation.} 

Perhaps the most interesting result highlighted by this pair of figures is the difference in the flow behaviour at $r \sim 5 r_g$. For $a = +1/2$ this radius is outside of the plunging region, and it is clear to see visually that the flow remains very much turbulent at this point. On the contrary, for $a = -1/2$ the radius $5r_g$ is well within the plunging region, and the fluid behaviour is clearly well described by simple, smooth, spiral structures. 

\begin{figure}
\centering{}
    \includegraphics[width=.99\linewidth]{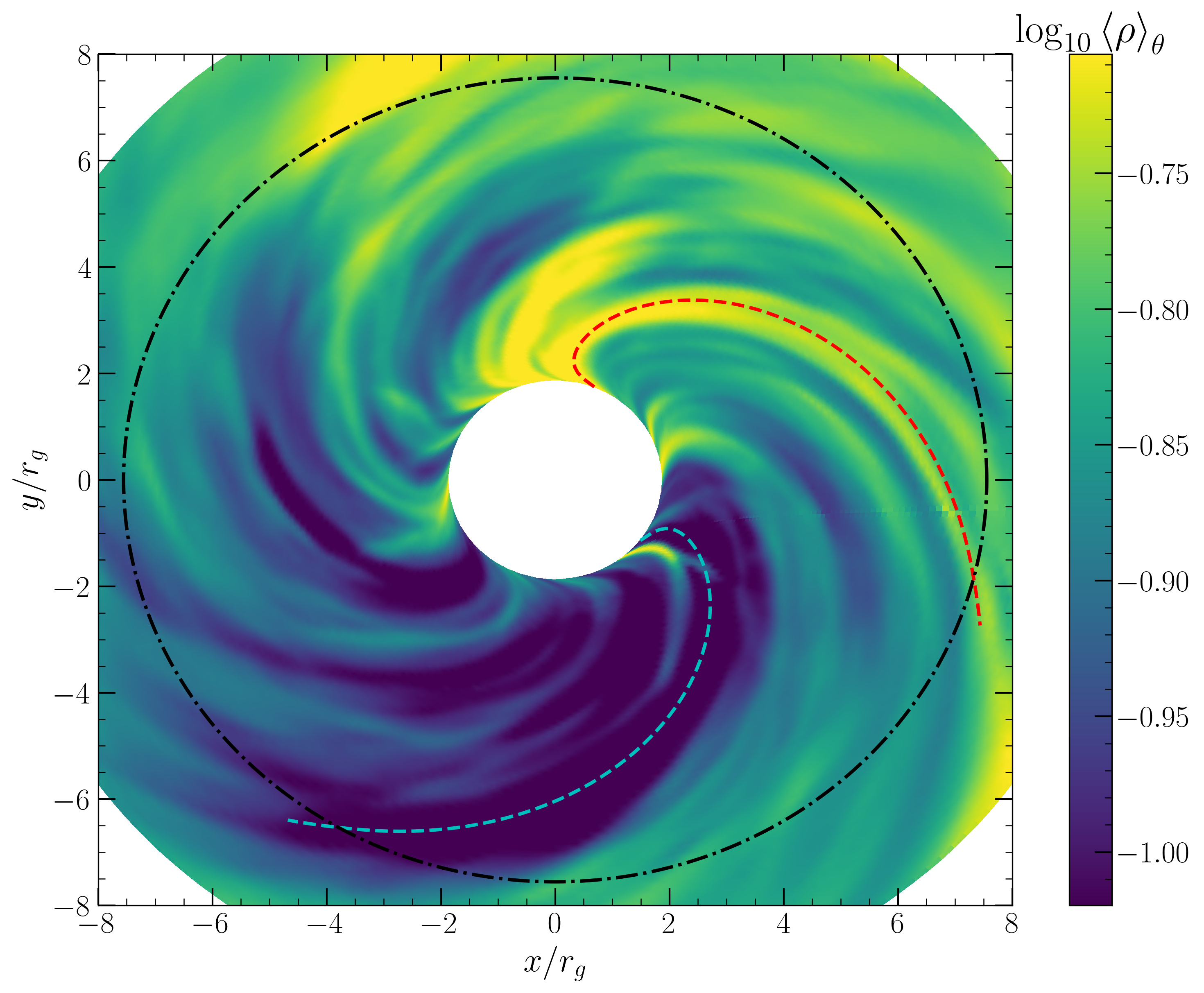}
        \caption{The azimuthal structure of the $\theta$-averaged density $\rhobar_{\theta}$ extracted from the Kerr black hole {\tt AthenaK} simulations, plotted against Boyer-Lindquist $x_{\rm BL}-y_{\rm BL}$ coordinates {(see text)}, chosen to highlight frame dragging effects. This figure is for a black hole spin $a = -1/2$.   By dashed curves we highlight how the over (red) and under (blue) densities in the inner flow follow the  characteristic spirals derived in this paper.  }
    \label{fig:kerr_BL}
\end{figure}

One interesting property of these retrograde inspirals not highlighted by Kerr-Schild coordinates is the effects of frame-dragging which occur when the fluid elements cross into the ergosphere of the black hole, at a radius $r_E = 2$ (in the equatorial plane). This property of the flow is best highlighted in Boyer-Lindquist coordinates  (see eq. \ref{blp} for the precise definition of the two systems), which correspond to the coordinate system associated with an observer at rest at infinity. In Figure \ref{fig:kerr_BL} we show the  vertically averaged disc density $\rhobar_{\theta}$ extracted from the Kerr black hole {\tt AthenaK} simulations, plotted against Cartesian Boyer-Lindquist $x_{\rm BL}-y_{\rm BL}$ coordinates {(defined by $x_{\rm BL} \equiv r \cos \varphi, y_{\rm BL} = r \sin \varphi$, where $\varphi$ is the Boyer-Lindquist azimuthal coordinate; equation \ref{blp})}, for $a = -1/2$. It is clear to see that all fluid elements are co-rotating with the black hole within the ergosphere, as they must \citep[see also][]{Ricarte22}. 

\subsection{ Discussion and summary } 
In this section we have analysed the properties of {\tt AthenaK} GRMHD simulations of Schwarzschild, and $a = \pm 1/2$ Kerr black hole accretion flows. We have focussed on the properties of the accretion disc density, as this traces both the dynamic (through mass conservation) and thermodynamic (through pressure balance in the vertical direction) properties of the fluid. Remarkably, we have found that analytical theory which uses some thin disc approximations provides an excellent qualitative (and in many cases quantitative) description of the flow. These results highlight that the fact that the plunging region is a particularly simple region of black hole accretion to describe analytically.

The fact that some thin disc assumptions hold up in this region, despite the thickness of the disc, warrants additional scrutiny. In effect what we learn from this analysis is that the accretion flow acts only weakly like a ``fluid'' in this region, and is mainly behaving (certainly in terms of its dynamics) as a collection of test particles. While this is clearly true for the so-called SANE configuration considered here (i.e., pressure forces are unable to overcome gravity), this appears less likely to be the case in so-called MAD dynamics, where non-geodesic magnetic forces will substantially change the physical picture. The leading order solutions of the equations of vertical hydrostatic equilibrium (equation \ref{height}), used as part of this intra-ISCO theory, explicitly neglect terms of order ${\cal O}(H/r)^2$. Clearly in a gravitationally dominated regime these corrections remain small even for order unity aspect ratios. 

It is interesting to contrast the success of the intra-ISCO theory in the three different directions considered here (namely the radial, vertical and azimuthal structure). Perhaps surprisingly (given the thin disc assumptions) the vertical structure of the disc is very well described by the theory. The azimuthal structure (spiral accretion flows) is also very well described by this theory, while the largest deviations occur in the radial direction (particularly for more negative spins). What is likely occurring here is that the vertical and azimuthal structure of the disc principally probe only one of the thermodynamic (vertical) or dynamic (azimuthal) behaviour of the flow, while the radial structure probes both. Clearly there is room to improve upon the theory of \cite{MummeryBalbus2023} in the radial description, particularly at radii closest to the ISCO. 

\section{An extended model for the internal energy of the flow within the plunging region} \label{pressec}
The internal energy (or pressure for our simple constant-$\Gamma$ equation of state systems) of accretion flows is an important and potentially observable property of these systems, particularly for the purposes of black hole imaging, as the ratio $P/\rho$ sets the temperature of the flow. The assumption of an adiabatic plunge (used to derive the theory tested above) would suggest that the pressure of these systems within the plunging region is simply given by 
\begin{equation}
{P \over P_I} = {K \over K_I} \left({ \rho \over \rho_I}\right)^{\Gamma} ,
\end{equation}
with $K/K_I = 1$. However, this theory neglects turbulent heating of the flow over its final inspiral, an effect we find is non-negligible in our simulations. 

This at first appears a surprising result, especially as the adiabatic theory reproduces the properties of the density and scale height so well. However, the key point is that while turbulent heating rapidly changes the internal energy of the flow, it takes a finite time for this extra heat to be communicated to the disc density. The timescale for this communication is given by the time it would take a sound wave to propagate over the disc thickness
\begin{equation}
\Delta t_{\rm s} \simeq {H\over c_s} ,
\end{equation}
while the infall time within the ISCO is 
\begin{equation}
\Delta t_{\rm ff} \simeq {r\over U^r} . 
\end{equation}
As we are considering thicker discs in this work, we have $H/r \sim 1$, and therefore 
\begin{equation}
{\Delta t_{\rm s} \over \Delta t_{\rm ff} } \simeq {H\over r} {U^r \over c_s} \sim {c \over c_s} \gg 1 .
\end{equation}
Therefore the fluid has insufficient time to communicate this additional heating to the disc density, and the density evolves on an effective adiabat. Note that this would not necessarily be the case for a much thinner disc which is heated by turbulent dissipation over the plunging region, which may well have sufficient time to communicate over its (shorter) vertical extent. 

We found numerically that the increase in heating throughout the plunging region was well approximated by a simple power-law $K \sim r^{\beta}$, and so a modified theory for the intra-ISCO internal energy which includes this turbulent heating (and lack of communication time) is 
\begin{equation}
{P \over P_I} = \left({r_I \over r} \right)^{6\Gamma/(1 + \Gamma)+\beta} \left[ \varepsilon^{-1} \left({r_I \over r} - 1\right)^{3/2} + 1\right]^{-2\Gamma/(1 + \Gamma)} . \label{PP}\\
\end{equation}

\begin{figure}
    \centering
    \includegraphics[width=\linewidth]{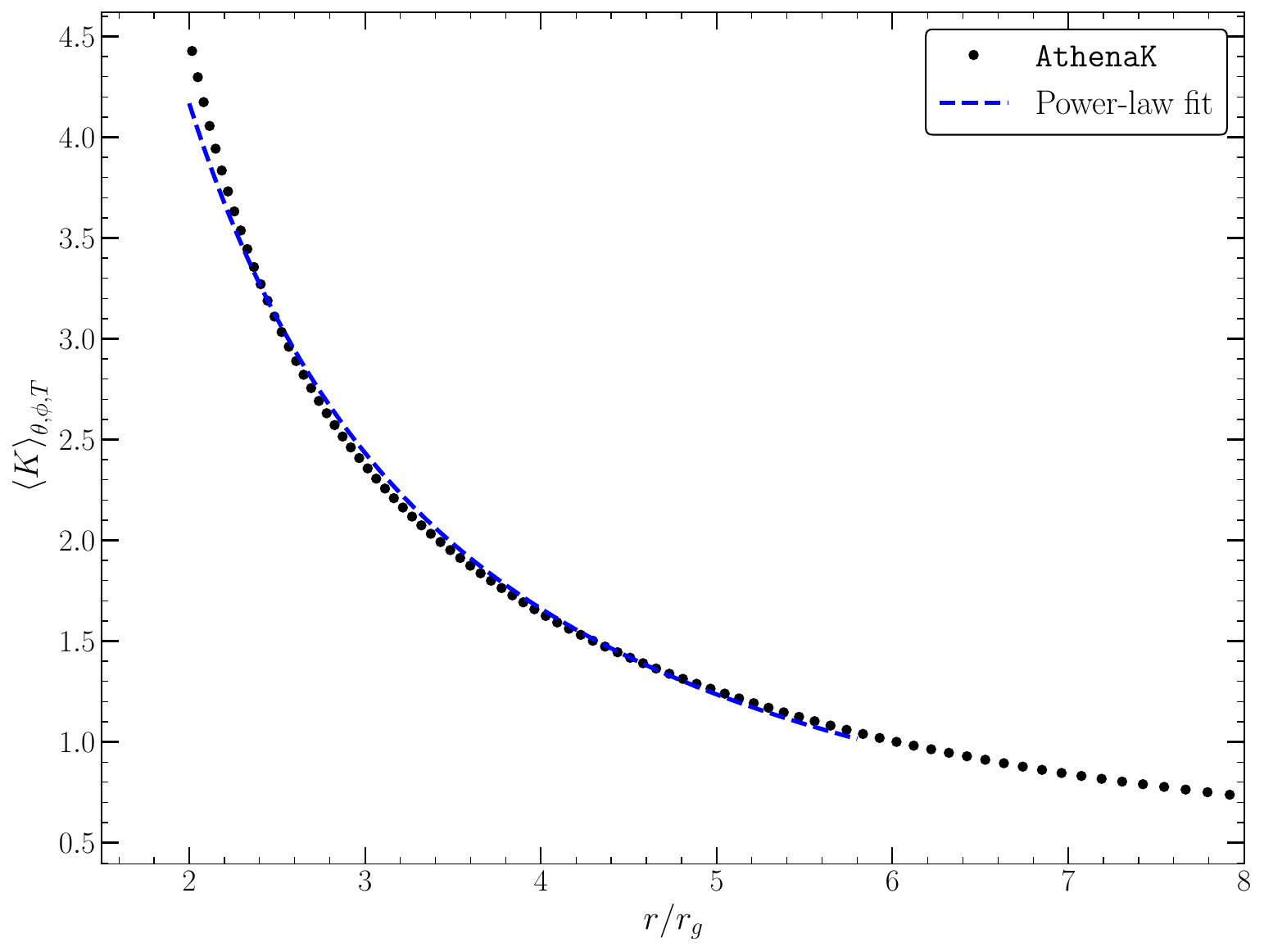}
    \includegraphics[width=\linewidth]{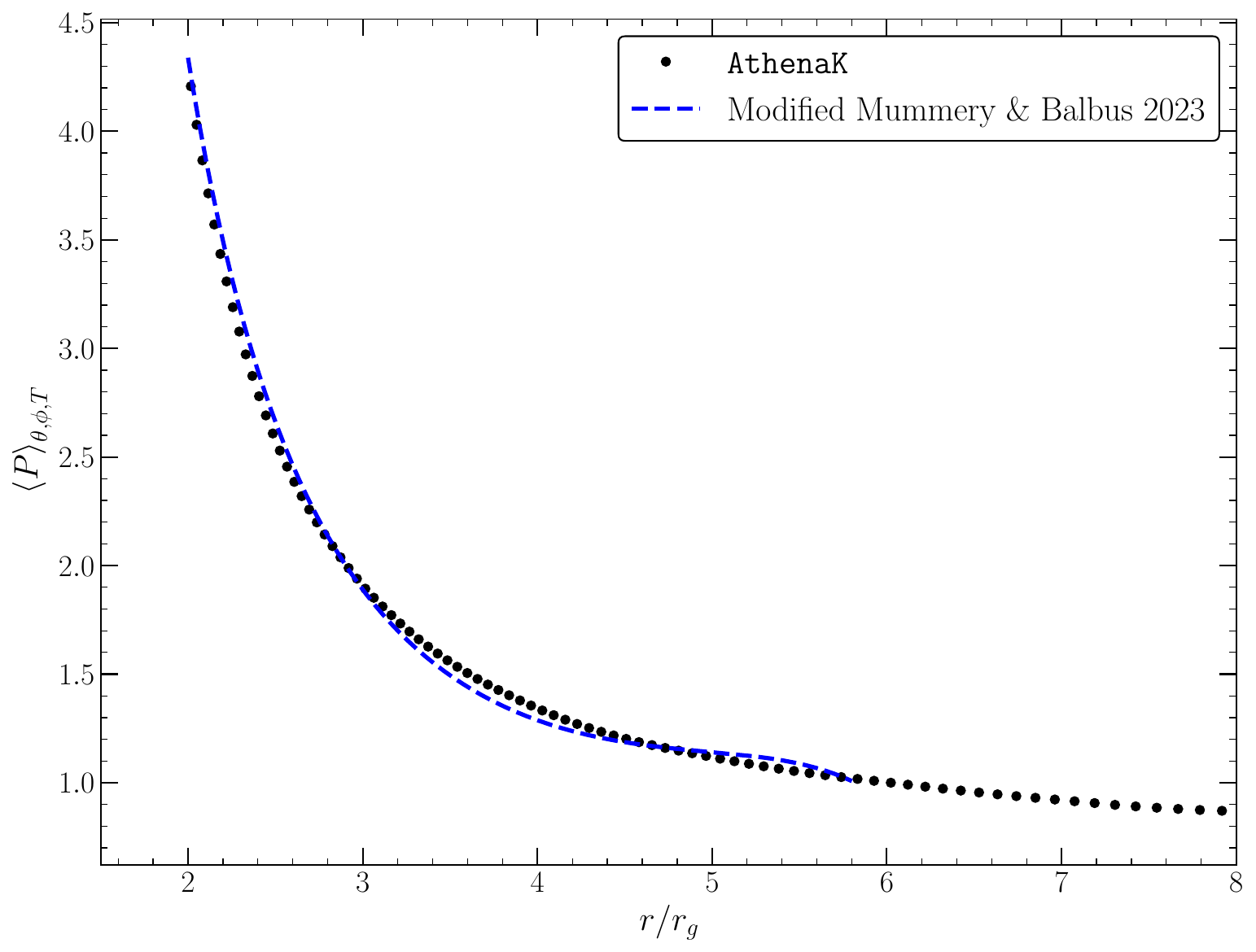}
    \includegraphics[width=\linewidth]{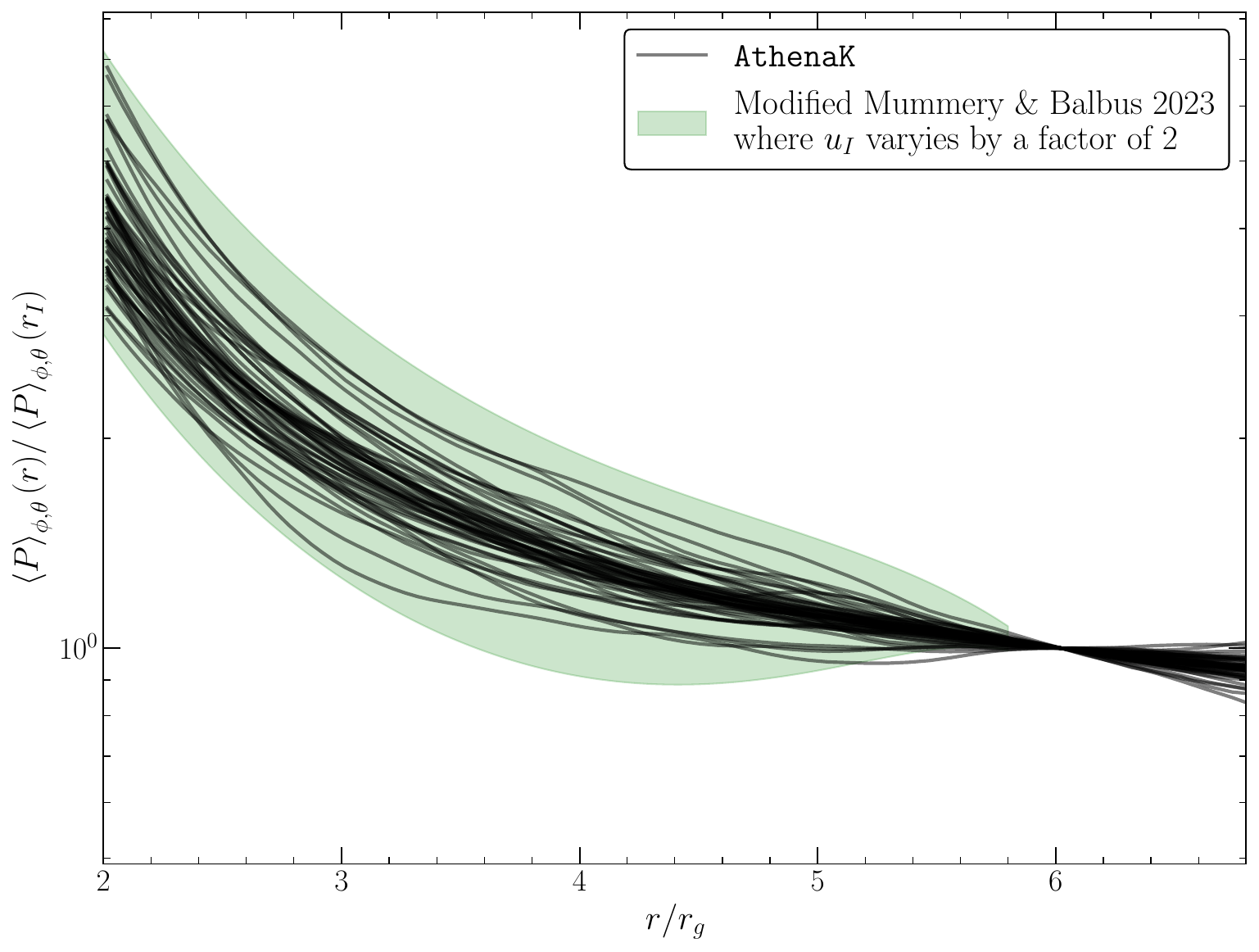}
    \caption{The radial structure of the accretion disc entropy and pressure extracted from the Schwarzschild black hole {\tt AthenaK} simulations. On the vertical axis of the upper  plot we display the angle and time-averaged entropy parameter $\left\langle K \right\rangle_{\theta, \phi, T} = \left\langle P \rho^{-\Gamma} \right\rangle_{\theta, \phi, T}$, showing that the disc flow continues to be turbulently heated over the plunging region. A simple modification of the \citealt{MummeryBalbus2023} theory to account for the turbulent heating (and short communication time, see text) in this region leads  to analytical pressure profiles which accurately reproduce the results of the simulation in both a time averaged (middle panel) and individual epoch (lowest panel) sense.  }
    \label{fig:pressure}
\end{figure}

A numerical fit to the time and angle averaged adiabatic parameter $K$ (for the simulation in the Schwarzschild spacetime) finds a best fit of $\beta \approx -1.327\dots$, suggesting that the entropy of the gas rises by a factor $\sim 4$ over the plunge (upper panel, Figure \ref{fig:pressure}). Taking this value of $\beta$, we can  compare the extended theory developed above to the pressure of the flow within the plunging region, both in a time averaged and independent snap-shot sense. We do this in the lower two panels of Figure \ref{fig:pressure}. This extended theory describes the numerical simulation extremely well. The variance between individual snapshots can once again be explained by turbulent variability in the trans-ISCO velocity $u_I$ at the factor 2 level. 

As for the disc density, we find exactly analogous behaviour for the pressure of the disc within the plunging region for each of the two Kerr simulations, with no dependence on black hole spin (except implicitly through the location of the ISCO).

\section{ Implications of the two dimensional nature of intra-ISCO accretion  }\label{obssec}
In this paper we have highlighted two novel results. Firstly,  theoretical simplifications usually associated with ``thin'' discs result in intra-ISCO thermodynamic expressions which accurately reproduce the vertical (e.g., Fig. \ref{fig:averaged_vertical}) and radial (e.g., Fig. \ref{fig:radial_density}) density of GRMHD simulations of thick discs. Secondly, accretion within the ISCO is fundamentally two-dimensional in character, with accretion mediated along spiral structures with a characteristic shape (which can be described analytically). In this section we discuss some observational implications of this second result. 

\subsection{ X-ray  continuum variability  }
The disc material within the ISCO remains hot, and can contribute significantly to the continuum emission observed from X-ray binaries at photon energies above $\sim 5$ keV. Continuum emission originating from within the ISCO was first speculated to have been detected in the X-ray binary MAXI J1820+070 by \cite{Fabian20}. This hypothesis  was recently confirmed by \mumetal\ who used continuum fitting models based on the analytical theory discussed here to perform a detailed analysis of four soft-state X-ray spectra, finding a requirement of non-zero intra-ISCO emission at extremely high statistical significance.  

One of the interesting results discussed by \cite{Fabian20} was that the fractional variability observed on short timescales in the MAXI J1820+070 light curve increased as the observing energy increased across the bandpass (see their Figure 10). As higher energies are associated with shorter length scales (owing to the increasing disc temperature at shorter disc radii), this suggests that the fractional disc variability increases as the disc fluid approaches the horizon.  While we have simulated thicker discs in this work, it seems likely that thinner discs (which is the limit in which the theory was constructed) will also show spiral-accretion in the innermost regions. 

Of course, MAXI J1820+070 is a stellar mass black hole (with mass roughly $M_\bullet = 8 M_\odot$), and so the dynamical timescale over which the spiral structures would evolve in this system is rather short $t_g \sim GM_\bullet/c^3 \sim 10^{-3}$ s, and so it is unlikely that emission from individual spiral structures will be temporally resolved in even short-timescale X-ray observations of Galactic X-ray binaries.  Nevertheless, even when integrated over many spiral evolution timescales, the geodesic stretching of turbulent fluctuations in the inner disc (the process through which the spiral structures are formed) seems likely to amplify the observed variability of this region, which may well be an observational signature of this region.  The fact that the analytical properties of this region are particularly simple, with each fluctuation resulting in a conserved quantity along a characteristic spiral, suggests that this may be a particularly interesting theoretical and observational route to probing this highly relativistic region. 

While most supermassive black hole discs in active galactic nuclei (AGN) are insufficiently hot to produce observable thermal X-ray emission, on the rare occasion in which an unfortunate star is tidally disrupted by the central black hole and then accreted (a so-called TDE) disc densities (or equivalently accretion rates) can get high enough to produce observable continuum emission, which is typically well described by relativistic disc models \citep[e.g.][]{MumBalb20a, Wen20}. This continuum emission then evolves on timescales appropriate for the evolutionary timescale of a supermassive black hole disc, which is significantly longer ($t \sim$ months) than for an X-ray binary disc. For these systems variability occurs on $\sim$ day timescales, and may well be significantly more enhanced, as it is possible that each observation probes a single realisation of the intra-ISCO density/temperature profile of the flow. 

\subsection{ Iron lines in active galactic nuclei  }
While active galactic nuclei do not typically produce significant thermal X-ray continuum emission, they are routinely observed to display prominent iron line features in addition to the non-thermal continuum \citep[see e.g.,][for a review]{Reynolds13}. These spectral features result from the irradiation of the accretion flow by hard X-rays sourced from a hot plasma located close to the central object. This process is typically referred to as `reflection', although it is more accurately described as reprocessing and re-emission \citep{Bambi2021}. The iron K$\alpha$ fluorescence line at $\sim 6.4$ keV is a prominent feature of the reflection spectrum, which also includes emission lines and absorption edges from all astrophysically abundant elements and a broad Compton scattering feature at $\sim 20-30$ keV known as the `Compton hump' \citep{Matt1991,Ross2005,Garcia2013}. The iron line is a particularly powerful diagnostic because it is narrow in the emission rest frame, whereas the observed line profile is heavily distorted by a combination of the relativistic orbital motion of the emitting material, and the gravitational energy-shifting of the emitted photons over their trajectory to the observer \citep{Fabian1989,Laor1991,Dauser2010}. 

In contrast to Galactic X-ray binaries, active galactic nuclei house significantly more massive black holes $M_\bullet \sim 10^6 - 10^9 M_\odot$, with dynamical timescales which can be easily probed with X-ray observations on human timescales $t_g \sim 10^2 - 10^5$ s. Iron line observations appear to be a very promising observational avenue for probing the two-dimensional nature of the plunging region, for the following reasons.  Firstly, the variability contained within a typical observational epoch (usually of order $\sim 10^4$ s) may allow for the individual observational signal of (e.g.,) a rotating $m=1$ spiral feature.  Secondly, and potentially more simply, the relativistic signatures of an asymmetric plunging region in black hole accretion flows may be imprinted on the time-averaged iron-line profile observed from these sources. 

Turbulent variability in iron line features is not a new concept \citep[see e.g.,][for iron line flux extracted from simulations produced in a modified Newtonian potential]{Armitage03}, but the simple analytical models developed here should allow controlled studies of these effects to be performed. 

\subsection{ High-frequency quasi-periodic oscillations }
Accreting stellar mass black hole X-ray binaries regularly show quasi-periodic oscillations (QPOs) in their X-ray emission \citep[e.g.,][for a review]{IngramQPO19}. Most of these observed QPOs are so-called low-frequency QPOs, but on rarer occasions so-called high-frequency QPOs (with frequencies around $\sim 100$ Hz) have been detected  from these systems \cite{Belloni12}. High-frequency QPOs are also relatively common in accreting neutron star systems. 

In a popular class of models which seek to explain observed QPO phenomenology, known broadly as the ``relativistic precession models'', high frequency QPOs are associated with the orbital timescale of the inner disc edge (with lower frequency QPOs associated with Lense--Thirring precession). A natural question to ask, if this broad phenomenology is correct, is ``what is orbiting at this inner edge''? Speculatively, it may well be that the spiral features discussed in this paper satisfy many of the properties required for the source of these high-frequency QPOs. Given their turbulent origin, random initial angle and finite lifetime, these spiral features will never produce strictly periodic signals. These structures do however all orbit with similar frequencies, that of the bulk fluid motion at the ISCO (see section \ref{spiralsec}). Of course significant extensions of the work carried out in this paper must be performed before this hypothesis can be tested, but it appears to be an intriguing possibility. 

\subsection{ Sagittarius A$^*$ }
The supermassive black hole at the center of our Galaxy, Sagittarius A*, is surrounded by a hot accretion flow comprising of a  relativistic plasma that emits bright synchrotron radiation while accreting at an extremely sub-Eddington rate (estimated to be around $L \sim 10^{-8} \, L_{\rm edd}$). This emission peaks in the millimeter band, and is thought to originate in the immediate vicinity of the black hole \citep[e.g.,][]{EHT22}.  One of the most interesting properties of this sources is that Sagittarius A$^*$ is known to display flaring behaviour on the top of its base emission, during which the observed flux can increase by a factor of up to $\sim$ 100 (in the near infrared). 

The flaring events observed in the Sagittarius A* supermassive black hole system are typically attributed to the assumed non-homogeneous nature of the near-horizon accretion flow. Brightness anisotropies,  typically referred to as  ``bright" or ``hot" ``spots”, are hoped to explain this variability features, and are typically modelled with orbiting Gaussian components \citep[e.g.,][]{Vos22, Yfantis23}. These models do a (potentially) surprisingly good job of describing the observed Sagittarius A$^*$ variability.    As we have argued here, however, it may be more natural (particularly if these anisotropies are in the innermost regions of the accretion flow) for these models to consider rotating spiral arms, rather than Gaussian components which will be rapidly sheared by gravitational effects (see Fig. \ref{fig:charac_spiral}). Analytical models, such as those developed here, may have practical utility in the world of event horizon telescope data analysis as they are significantly cheaper to compute and have parameter spaces which are therefore much simpler to exhaustively explore (when compared to GRMHD simulations).

\section{ Conclusions }\label{conc}
In this paper we have studied a series of {\tt AthenaK} GRMHD simulations of thick discs evolving with a so-called SANE magnetic field structure. We have shown that despite the large aspect ratio of the simulated discs, analytical models derived in the ``thin disc'' limit do a remarkably good job at reproducing the quantitative and qualitative properties of the GRMHD simulations within the plunging region. 

Extending these analytical models, we have derived a simple description for the asymmetric $r-\phi$ structure of accretion within the plunging region. We highlight how the rapid gravitational acceleration of the plunging region results in the formation of spiral accretion channels, through which turbulent over densities are ultimately accreted. This means that accretion within the ISCO is fundamentally two-dimensional in character, which may well have interesting observational implications. This is likely to be particularly relevant for supermassive black hole disc systems such as active galactic nuclei and event horizon telescope sources, for which the near-horizon dynamical timescale (which is of order the timescale over which the spirals propagate) approaches the observing timescale of astrophysical sources. 

We encourage the extension of existing analytical models of accretion to include asymmetric features at small radii, which are a signature of the highly relativistic regime of gravity and may well carry strong-field signatures which can be leveraged to probe in detail the properties of astrophysical compact objects. 

\section*{Acknowledgments} 
AM gratefully acknowledges support and hospitality from the Institute for Advanced Study. AM is grateful to Chris White for extremely helpful conversations regarding the analysis of {\tt AthenaK} data, and to Adam Ingram and Peter Jonker for discussions regarding the possible signatures of this region. This research used resources of the National Energy Research Scientific Computing Center (NERSC), a Department of Energy Office of Science User Facility using NERSC award FES-ERCAPm4307.
 This work was supported by a Leverhulme Trust International Professorship grant [number LIP-202-014]. For the purpose of Open Access, AM has applied a CC BY public copyright licence to any Author Accepted Manuscript version arising from this submission. 
 
\section*{Data availability }
No observational data was used in producing this manuscript. Numerical results will be shared upon reasonable request with the corresponding author. 

\bibliographystyle{mnras}
\bibliography{andy}

\appendix
\section{The Kerr-Schild coordinate system }\label{KerrApp}
{The {\tt AthenaK} code solves the governing GRMHD equations in Cartesian Kerr-Schild coordinates $(T, x, y, z)$, with line element given by   }
\begin{multline}
 g_{\mu\nu}{\rm d}x^\mu {\rm d}x^\nu = -{\rm d}T^2 + {\rm d}x^2 + {\rm d}y^2 + {\rm d}z^2 \\ + {2r^3 \over r^4 + a^2 z^2} \left[{\rm d}T + {r (x {\rm d}x + y{\rm d} y) \over r^2 + a^2} +  {a (y {\rm d}x - x{\rm d} y) \over r^2 + a^2} + {z\over r} {\rm d}z \right]^2 . 
\end{multline}
{where $r$ is the function defined via}
\begin{equation}
r^2 = {x^2 + y^2 + z^2 - a^2 \over 2} + \sqrt{\left({x^2 + y^2 + z^2 - a^2 \over 2}\right)^2 + a^2 z^2 }. 
\end{equation}
{While {\tt AthenaK} solves the equations in cartesian coordinates, it is more convenient for analysis purposes to transform back to so-called ``spherical Kerr-Schild coordinates'' $(r, \theta, \phi)$, which are related to cartesian $(x, y, z)$ coordinates by  } 
\begin{align}
x &= (r \cos\phi - a\sin \phi) \sin \theta, \\
y &= (r \sin\phi + a\cos \phi) \sin \theta, \\
z &= r \cos\theta.
\end{align}
{The timelike coordinate $T$ is identical in the two systems. }

The Kerr metric in spherical Kerr-Schild coordinates has the following non-zero coefficients 
\begin{align}
g_{TT} &= -1 + 2r/(r^2 + a^2 \cos^2\theta), \\
g_{rr} &= 1 + 2r/(r^2 + a^2 \cos^2\theta) \\
g_{\phi\phi} &=  \sin^2\theta (r^2 + a^2 + 2 a^2 r \sin^2\theta/(r^2 + a^2 \cos^2\theta))  , \\
g_{\theta\theta} &=r^2 + a^2 \cos^2\theta , \\
g_{T\phi} &= g_{\phi T} =  - 2ar\sin^2\theta/(r^2 + a^2 \cos^2\theta)  , \\ 
g_{Tr} &=  g_{rT} =  2r/(r^2 + a^2 \cos^2\theta)  , \\
g_{r\phi} &=  g_{\phi r} = - a\sin^2\theta (1 + 2r/(r^2 + a^2 \cos^2\theta)).
\end{align}
 In the equatorial plane ($\theta=\pi/2$, relevant for computing the characteristic spirals), this becomes
\begin{align}
g_{TT} &= -1 + 2/r, \\
g_{rr} &= 1 + 2/r \\
g_{\phi\phi} &=  r^2 + a^2 + 2 a^2 / r , \\
g_{\theta\theta} &= r^2 , \\
g_{T\phi} &=  g_{\phi T} = - 2a/r  , \\ 
g_{Tr} &= g_{rT} =  2/r  , \\
g_{r\phi} &=  g_{\phi r} = - a (1 + 2/r) .
\end{align}

\label{lastpage}
\end{document}